ΕΛΛΗΝΙΚΗ ΔΗΜΟΚΡΑΤΙΑ
Εθνικό και Καποδιστριακό
Πανεπιστήμιο Αθηνών

ΠαιΤΔΕ-ΕΚΠΑ

# «Η Ιστορία στην ψηφιακή εποχή»

Πανεπιστημιακές παραδόσεις Δεκ2020-Ιαν2021

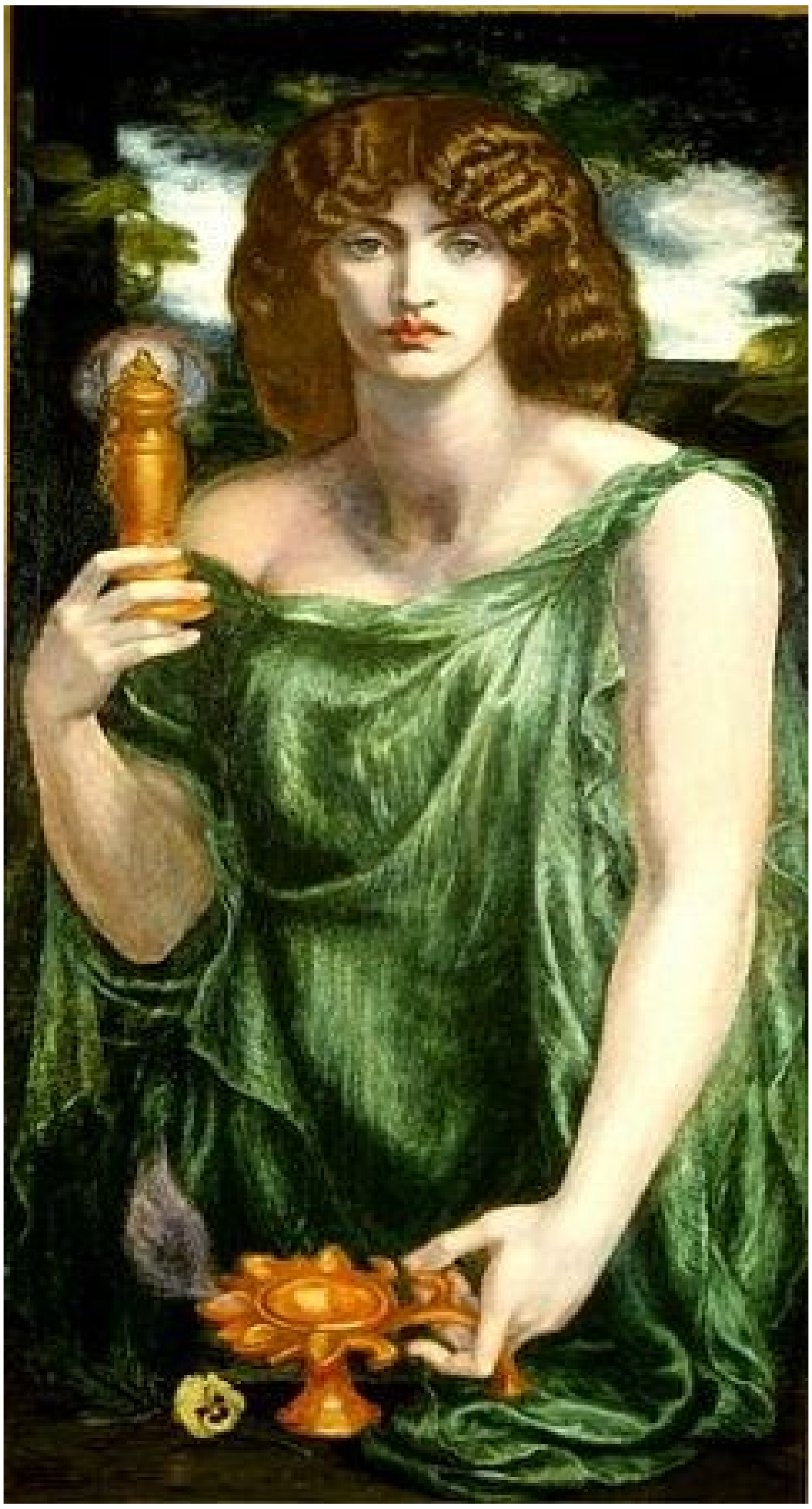

*Μνημοσύνη ή Η Λάμπα της Μνήμης (1881)*
*Ψηφιακό Αρχείο Rosseti http://www.rossettiarchive.org/*
*© Delaware Art Museum, Samuel and Mary R. Bancroft Memorial*

**Δρ. Μαρία Παπαδοπούλου-Αν. Καθ. Ζαχαρούλα Σμυρναίου**

**ΤΕΚ 301 - ΠΑΙΔΑΓΩΓΙΚΑ Ι**
**ΔΙΔΑΚΤΙΚΗ ΤΗΣ ΙΣΤΟΡΙΑΣ ΘΕΩΡΙΑ & ΠΡΑΞΗ**

# Περιεχόμενα

Clio loves those who bred them better horses
Found answers to their questions, made their things

W. H. Auden, Makers of History (1955)

Την εποχή που γράφτηκαν αυτοί οι στίχοι, η Κοινωνική Ιστορία γεννιόταν στην Αγγλία. Η λέξη «makers» στον τίτλο του ποιήματος παραπέμπει στους «δημιουργούς της Ιστορίας», τα μέλη της εργατικής τάξης. Η εποχή μας είναι εποχή των makerspaces, χώρων όπου γεννιούνται τολμηροί πειραματισμοί με την τεχνολογία. Τα τελευταία χρόνια τέτοιοι χώροι αποτελούν εκκολαπτήρια ψηφιακών καινοτομιών που «γράφουν ιστορία».

Επικοινωνία: maria.papadopoulou@univ-savoie.fr

# Η Ιστορία στην Ψηφιακή Εποχή

Ψηφιακές τεχνολογίες, όπως αυτές του Διαδικτύου και της Τεχνητής Νοημοσύνης αποτελούν μέρος της καθημερινότητάς μας επηρεάζοντας ευρύτερες όψεις του τρόπου ζωής μας, καθώς και τον τρόπο με τον οποίο αλληλεπιδρούμε με το παρελθόν. Έχοντας μεταβάλει καθοριστικά τους τρόπους παραγωγής και κατανάλωσης της γνώσης, η αλγοριθμική εποχή έχει επίσης αλλάξει ριζικά τη σχέση που έχει το ευρύ κοινό με την Ιστορία προς όφελος πεδίων της Ιστορίας όπως είναι η Δημόσια[1] και η Προφορική Ιστορία.[2]

Πώς επηρεάζει ο ψηφιακός μας πολιτισμός τον τρόπο που σκεφτόμαστε, μελετάμε, ερευνούμε και διδάσκουμε το παρελθόν, καθώς ιστορικά τεκμήρια διαχέονται με ταχείς ρυθμούς στην δημόσια σφαίρα; Πώς προάγουν οι ψηφιακές τεχνολογίες τη μελέτη, συγγραφή και διδασκαλία της Ιστορίας; Τι πρέπει να έχει υπόψη του ο μελετητής και ο σπουδαστής της Ιστορίας, ενώ ψηφιοποιημένο ή εγγενώς ψηφιακό πολιτισμικό περιεχόμενο, που ολοένα αυξάνεται σε όγκο, τον βομβαρδίζει με καταιγισμό πληροφοριών, μέσω του διαδικτύου; Και ενώ οι αλλαγές αυτές είναι πλέον ορατές σε παγκόσμιο επίπεδο, ποια είναι η θέση της Ιστορίας στην προσπάθεια για ψηφιακή σύγκλιση που προχωρά με ταχείς ρυθμούς στη χώρα μας; Ποιες, τέλος, είναι οι συνέπειες των αλλαγών αυτών για τη λεγόμενη ‘σχολική ιστορία’; Αυτά είναι μερικά μόνο από τα ζητήματα που βρίσκονται στο επίκεντρο των συζητήσεων σχετικά με την Ιστορία στην ψηφιακή εποχή και που εξετάζονται στο κείμενο που ακολουθεί.

# 1. Τι είναι «Ψηφιακή Ιστορία»;

## 1.1 Ορισμός

Τι είναι Ψηφιακή Ιστορία; Ακολουθούν τέσσερις ορισμοί από ιστορικούς που ασχολούνται με την Ψηφιακή Ιστορία.

Ορισμός 1:

Ψηφιακή Ιστορία είναι το σχετικά νέο διεπιστημονικό πεδίο που προέκυψε από την συνάντηση της επιστήμης της Ιστορίας με τις σύγχρονες ψηφιακές τεχνολογίες. Αντικείμενό της είναι η εξέταση και επίλυση ερωτημάτων που θέτει η Ιστορία ως επιστήμη αξιοποιώντας τις ψηφιακές τεχνολογίες, τα ψηφιακά περιβάλλοντα και τα ψηφιακά εργαλεία. Αξιοποιεί τις εξελίξεις στις Επιστήμες της Πληροφορίας και των Υπολογιστών, υιοθετεί νέες μεθόδους και

[1] Σε αντίθεση με την Ακαδημαϊκή Ιστορία, που διαχέεται μέσω των ακαδημαϊκών ιστορικών μονογραφιών, η Δημόσια Ιστορία διαχέεται εντός της δημόσιας σφαίρας αξιοποιώντας τις δυνατότητες που παρέχουν οι μουσειακές συλλογές, και τα αρχεία που πλέον διαθέτουν τις συλλογές τους όχι μόνο στον φυσικό τους χώρο, αλλά και, μέσω του διαδικτύου, στον κυβερνοχώρο.

[2] Η Προφορική Ιστορία ασχολείται με την παρουσίαση, διατήρηση και ερμηνεία της Ιστορίας ως βιώματος. Καταγράφει τις προφορικές μαρτυρίες των προσώπων για τα γεγονότα, όπως τα βίωσαν, και αξιοποιεί τα ψηφιακά μέσα και τις τεχνολογίες του διαδικτύου για την καταγραφή, διατήρηση και διάχυσή τους.

ροές εργασίας και δημιουργεί νέες ψηφιακές πλατφόρμες και εργαλεία που εξυπηρετούν τις ανάγκες της.

Ορισμός 2:[3]

«ΩΣ ΨΗΦΙΑΚΗ ΙΣΤΟΡΙΑ νοείται η ιστορική έρευνα η οποία κάνει χρήση πρωτογενών πηγών υπό μορφή ηλεκτρονικών δεδομένων και εφαρμόζει μεθόδους ανάλυσης και αναπαράστασης του παρελθόντος με τη χρήση ψηφιακών τεχνολογιών, λογισμικού και των πόρων του διαδικτύου.

Ο όρος, ο οποίος μαρτυρείται από τα τέλη της δεκαετίας του 1990 στις ΗΠΑ, γενικεύτηκε κατά την επόμενη δεκαετία και συνέβαλε στην ανάδυση των Ψηφιακών Ανθρωπιστικών Επιστημών (Digital Humanities) ως κλάδου των Ανθρωπιστικών Επιστημών.

Η ψηφιακή ιστορία καλύπτει ένα ευρύ φάσμα πρακτικών, μεθόδων και αντικειμένων που άπτονται της ακαδημαϊκής έρευνας, της δημόσιας ιστορίας και της διδασκαλίας της ιστορίας.»

Ορισμός 3:[4]

Ψηφιακή ιστορία είναι μια προσέγγιση για την εξέταση και αναπαράσταση του παρελθόντος που χρησιμοποιεί τις νέες τεχνολογίες της επικοινωνίας και του διαδικτύου και τα συστήματα λογισμικού. Σε ένα πρώτο επίπεδο, η ψηφιακή ιστορία είναι μια ανοιχτή αρένα επιστημονικής παραγωγής και επικοινωνίας που περιλαμβάνει την ανάπτυξη νέου διδακτικού υλικού και τη συλλογή επιστημονικών δεδομένων. Σε ένα δεύτερο επίπεδο, είναι μια μεθοδολογική προσέγγιση που πλαισιώνεται από την υπερκειμενική δύναμη αυτών των τεχνολογιών να δημιουργούν, να ορίζουν, να κάνουν εφικτές αναζητήσεις και συσχετίσεις στο ψηφιακό αρχείο του παρελθόντος της ανθρωπότητας. Κάνεις ψηφιακή ιστορία, λοιπόν, όταν, μέσω της τεχνολογίας, δημιουργείς ένα πλαίσιο, ώστε οι άνθρωποι να μπορούν να βιώσουν, να διαβάσουν και να παρακολουθήσουν την επιχειρηματολογία που διατυπώνεις σχετικά με ένα ιστορικό πρόβλημα.

Ορισμός 4:

Οι όροι 'Ιστορική Πληροφορική' ('History Informatics' ή 'Historical Computing'), όπως και αντίστοιχοι των επίσης υβριδικών πεδίων 'Βιοπληροφορική' ('Bioinformatics'), 'Οικονομική Πληροφορική' ('Business Informatics'), 'Πολιτισμική Πληροφορική' ('Cultural Informatics') και 'Πληροφορική Ανθρωπιστικών Επιστημών' ('Humanities Computing'), χρησιμοποιούνται για τον καινούριο κλάδο που συνδυάζει την Ιστορία με την Πληροφορική.[5]

---

[3] «Ψηφιακή ιστορία», Journée d'étude, *Calenda*, 14/6/2018, https://calenda.org/445713 και http://www.archive.uoa.gr/fileadmin/archive.uoa.gr/uploads/PSIFIAKI_ISTORIA_FYLLADIO_final.pdf

[4] Daniel J. Cohen, Michael Frisch, Patrick Gallagher, Steven Mintz, Kirsten Sword, Amy Murrell Taylor, William G. Thomas III, and William J. Turkel, "Interchange: The Promise of Digital History," Journal of American History 95, 2 (2008), www.historycooperative.org/journals/jah/95.2/interchange.html

[5] Burkard, B., Vogeler, G., Gruner, S. 'Informatics for Historians: Tools for Medieval Document XML Markup, and their Impact on the History-Sciences'. Journal of Universal Computer Science, 14, 2 (2007), 193-210.

## 1.2 Πότε εμφανίστηκε

Ο όρος «ψηφιακή ιστορία» εμφανίστηκε για πρώτη φορά το 1997, με την ίδρυση του ερευνητικού κέντρου ψηφιακής ιστορίας στην πολιτεία Βιρτζίνια των ΗΠΑ.[6] Τρία χρόνια νωρίτερα, το 1994, στο Παν/μιο George Mason της ίδιας πολιτείας, ο διδάκτορας του Παν/μίου Harvard Roy Rosenzweig ιδρύει το Κέντρο Ιστορίας και Νέων Μέσων (Center for History and New Media-CHNM)[7] το οποίο διευθύνει ως το θάνατό του, το 2007. Στο διάστημα αυτό, το κέντρο προώθησε την έρευνα και τη διδασκαλία της Ιστορίας με ψηφιακά εργαλεία και λογισμικά ανοικτής πρόσβασης, όπως το ψηφιακό αποθετήριο Zotero[8] (2005) και το σύστημα διαχείρισης ψηφιακού περιεχομένου (Content Management System-CMS) Omeka[9] (2007).

Το 2009, το ίδιο κέντρο, που είχε στο μεταξύ μετονομαστεί σε Roy Rosenzweig Center for History and New Media (RRCHNM), ιδρύει το THAT CAMP. Η ονομασία THAT προέρχεται από το ακρωνύμιο The Humanities and Technology, https://thatcamp.org/. Τα THAT CAMP αυτοχαρακτηρίζονται ως "unconferences", ανοικτά συνέδρια για ειδικούς στις Ανθρωπιστικές Επιστήμες, αλλά και στην Τεχνολογία, γιατί δεν ακολουθούν τις πάγιες διαδικασίες πρόσκλησης συμμετοχών βάσει θεματικής, αλλά η ατζέντά τους διαμορφώνεται επί τόπου από εκείνους που μετέχουν. Το THAT CAMP που πραγματοποιήθηκε το 2010 στο Παρίσι δημοσίευσε το Manifesto for the Digital Humanities (https://tcp.hypotheses.org/411) το οποίο αποτελεί κείμενο-ορόσημο για το πεδίο των Ψηφιακών Ανθρωπιστικών Επιστημών (Digital Humanities).

Στους πρωτοπόρους μελετητές της Ψηφιακής Ιστορίας συγκατελέγονται οι παρακάτω:[10]

- Daniel Cohen, συνεργάτης (και διάδοχος του Rosenzweig στο πηδάλιο του CHMM). Σε αυτόν συνοφείλεται το εγχειρίδιο Ψηφιακής Ιστορίας *Digital History: A Guide to Gathering, Preserving, and Presenting the Past on the Web* (2006)[11] και η δημοσίευση

---

[6] Virginia Center for Digital History (VCDH), http://www.vcdh.virginia.edu/index.php?page=VCDH. Βλ. ενδεικτικά δύο πρωτοπόρα ερευνητικά προγράμματα ψηφιακής ιστορίας: το πρώτο εξετάζει τον τρόπο που βίωσαν τον Αμερικανικό εμφύλιο δύο κοινότητες από τα αντίπαλα στρατόπεδα Βορείων και Νοτίων https://valley.lib.virginia.edu/, το δεύτερο εξετάζει τη ζωή στην πολιτεία Βικτώρια του Καναδά τον καιρό της βασίλισσας Βικτωρίας https://web.uvic.ca/vv/.

[7] Center for History and New Media-CHNM, https://rrchnm.org/

[8] Zotero, https://www.zotero.org/

[9] Omeka, https://omeka.org/

[10] Μπορεί κανείς να αναζητήσει Ιστορικούς Ψηφιακής Ιστορίας από κάθε χρονολογικό και θεματικό εύρος στα πανεπιστημιακά ιδρύματα των ΗΠΑ χρησιμοποιώντας το ακόλουθο ευρετήριο: http://digitalhistory.unl.edu/database/index.php

[11] Ανοικτά προσβάσιμο εδώ: https://chnm.gmu.edu/digitalhistory/

συζήτησης στρογγυλής τράπεζας με τον τίτλο *Interchange: The Promise of Digital History*.[12]
- Andreas Fickers, διευθυντής του ερευνητικού Κέντρου Σύγχρονης και Ψηφιακής Ιστορίας του Λουξεμβούργου (Luxembourg Centre for Contemporary and Digital History - C2DH).[13]

### 1.3 Βασικές αναφορές

Βασικές αναφορές για τη διδασκαλία και την έρευνα στο πεδίο της Ψηφιακής Ιστορίας αποτελούν:

- Η Λευκή Χάρτα[14] (2017) που προέκυψε από τη συνάντηση εργασίας είκοσι τεσσάρων ιστορικών στο Παν/μιο George Mason με στόχο να εντοπιστούν οι λόγοι για τους οποίο καθυστέρησε η ενθάρρυνση και προώθηση του διαλόγου για την ψηφιακή Ιστορία, δεδομένου ότι σχετικές πρωτοβουλίες χρονολογούνται εδώ και σχεδόν τρεις δεκαετίες.
- Το βιβλίο των Jack Dougherty & Kristen Nawrotzki. 2003. Writing History in the Digital Age. Series: Digital Humanities. University of Michigan Press, Digitalculturebooks (σε ανοικτή πρόσβαση https://doi.org/10.2307/j.ctv65sx57 ).
- Το άρθρο του G. Zaagsma, «On digital history». BMGN-Low Countries Historical Review, 2013 (προσβάσιμο εδώ) προσφέρει μια κατατοπιστική εισαγωγή στην Ιστορία και κυριώτερες πρακτικές της Ψηφιακής Ιστορίας.
- Τα επιστημονικά περιοδικά Current Research in Digital History (ετήσια ψηφιακή έκδοση του Roy Rosenzweig Center for History and New Media από το 2018-)[15] και Journal of Digital History (JDH) (ψηφιακή έκδοση του Luxembourg Centre for Contemporary and Digital History),[16] που προορίζεται να χρησιμεύσει ως φόρουμ για κριτική συζήτηση στον τομέα της Ψηφιακής Ιστορίας, προσφέροντας μια καινοτόμο πλατφόρμα που θέτει νέα πρότυπα στη δημοσίευση της ιστορίας με βάση μια νέα πολυεπίπεδη προσέγγιση. Κάθε άρθρο που θα δημοσιεύεται πρέπει να περιλαμβάνει τρία 'στρώματα': το αφηγηματικό (narration layer), το ερμηνευτικό (hermeneutic layer) που διερευνά τις μεθοδολογικές επιπτώσεις των ψηφιακών/αλγοριθμικών μεθόδων και εργαλείων και, τέλος, και το στρώμα των δεδομένων (data layer) που θα παρέχει πρόσβαση στα δεδομένα και στον κώδικα.

---

[12] Daniel J. Cohen, Michael Frisch, Patrick Gallagher, Steven Mintz, Kirsten Sword, et al. Interchange: The Promise of Digital History, Journal of American History, Volume 95, Issue 2, September 2008, Pages 452–491, https://doi.org/10.2307/25095630

[13] Σχετικά με τον A. Fickers βλ. την ιστοσελίδα του Κέντρου Σύγχρονης και Ψηφιακής Ιστορίας του Λουξεμβούρου εδώ https://wwwfr.uni.lu/c2dh/people/andreas_fickers, καθώς και σχετική ομιλία του A. Fickers στο Εθνικό Κέντρο Τεκμηρίωσης (ΕΚΤ) στις 6/2/2019, στο πλαίσιο του κύκλου ανοιχτών διαλέξεων «Μεγάλα Δεδομένα, Νέα Μέσα, Ζητήματα Τεκμηρίωσης: Μαθαίνοντας από πρωτοπόρα εγχειρήματα», βίντεο και διαφάνειες: http://helios-eie.ekt.gr/EIE/handle/10442/16128).

[14] "Digital History and Argument" white paper, Roy Rosenzweig Center for History and New Media (November 13, 2017). Η Λευκή Βίβλος είναι προσβάσιμη εδώ https://rrchnm.org/argument-white-paper/, https://rrchnm.org/wordpress/wp-content/uploads/2017/11/digital-history-and-argument.RRCHNM.pdf

[15] Current Research in Digital History https://crdh.rrchnm.org/volume/

[16] Journal of Digital History https://journalofdigitalhistory.org/en/about

- Ο ιστότοπος https://shsulibraryguides.org/digitalhistory που περιλαμβάνει, μεταξύ άλλων, υπερσυνδέσμους για ψηφιακά σύνολα δεδομένων, ψηφιακά εργαλεία και για τη Διδακτική της Ψηφιακής Ιστορίας και ο ιστότοπος https://maehr.github.io/awesome-digital-history/ που περιλαμβάνει, μεταξύ άλλων, υπερσυνδέσμους για ψηφιακές συλλογές ιστορικού ενδιαφέροντος, όπως η Europeana για την οποία γίνεται λόγος παρακάτω. Μια περιήγηση στους δύο παραπάνω ιστοτόπους καταδεικνύει την υβριδικότητα του πεδίου της Ψηφιακής Ιστορίας που καταργεί το νοητό σύνορο ανάμεσα στην Ιστορία (και τα λοιπά πεδία των Ανθρωπιστικών Επιστημών) και τα πεδία της Πληροφορικής και της Επιστήμης των Υπολογιστών.

Ακολουθούν τέσσερις ελληνόγλωσσες αναφορές σχετικά με την Ψηφιακή Ιστορία (Βαλατσού 2016, Παπακώστα 2014) και τα ζητήματα που θέτει το Διαδίκτυο για την ελληνική ιστορική εκπαίδευση (Γιακουμάτου 2006, Τσιβάς κ.ά. 2018):

- ✓ Βαλατσού, Δ. (2016) Ανάδυση νέων μνημονικών τόπων στο διαδίκτυο. Διδ. διατριβή. Τμήμα Ιστορίας και Αρχαιολογίας, Φιλοσοφική Σχολή, ΕΚΠΑ. http://repository.edulll.gr/edulll/retrieve/9398/3034_1.77_%CE%94%CE%94_7_10_14.pdf
  Αντικείμενο της διατριβής είναι η ιστορία της ψηφιακής ιστορίας, τα βασικά χαρακτηριστικά της, τα πλεονεκτήματα και τα μειονεκτήματα της ψηφιακής τεχνολογίας για τους ιστορικούς, καθώς και βασικά ζητήματα θεωρίας και μεθοδολογίας, τα οποία προκύπτουν από την ψηφιακή στροφή στην ιστορία.
- ✓ Γιακουμάτου, Τ. (2006) Διδάσκοντας Ιστορία την εποχή του διαδικτύου, Φιλολογική, Τεύχος 97, Δεκέμβριος, 69-75, http://www.netschoolbook.gr/epimorfosi/conferences/p9_dec2006_dig_history.pdf
  Το άρθρο επικεντρώνεται σε ζητήματα και προκλήσεις που θέτει η στροφή στο Διαδίκτυο για την ελληνική ιστορική εκπαίδευση.
- ✓ Παπακώστα, Κ. (2014) Ψηφιακή ιστορία. Πρακτικά ΙΕ΄ Διεθνούς Συνεδρίου της Παιδαγωγικής Εταιρείας Ελλάδος, 152-162, http://www.pee.gr/wp-content/uploads/eRA8_1-481.pdf
- ✓ Τσιβάς Αρμόδιος, Ανδρεάδου Χαρά, Κασκαμανίδης Γιάννης, Σαλβάνου Αιμιλία Παιδαγωγικές και διδακτικές αρχές σχεδιασμού και ανάπτυξης ψηφιακών πόρων για το μάθημα της Ιστορίας στο πλαίσιο του «Ψηφιακού σχολείου». Στ. Δημητριάδης, Β. Δαγδιλέλης, Θρ. Τσιάτσος, Ι. Μαγνήσαλης, Δ. Τζήμας (επιμ.), Πρακτικά 11ου Πανελλήνιου και Διεθνούς Συνεδρίου «Οι ΤΠΕ στην Εκπαίδευση», ΑΠΘ – ΠΑΜΑΚ, Θεσσαλονίκη, 19-21 Οκτωβρίου 2018, 73-77, https://www.diva-portal.org/smash/get/diva2:1287344/FULLTEXT01.pdf

# 2. Ζητήματα & Προκλήσεις

## 2.1 Η Αφθονία της Πληροφορίας

Κύριο έργο της Ιστορίας είναι η, κατά το δυνατόν «αμερόληπτη», «αξιόπιστη» και εμπεριστατωμένη αφήγηση (*εξιστόρηση*) των γεγονότων του παρελθόντος. Ο ρόλος αυτός της Ιστορίας τείνει να καταλυθεί στην εποχή του Διαδικτύου, που μπήκε στις ζωές μας δυναμικά από το 2000 και μετά; Πώς έχει αλλάξει η καινούρια ιστορική πραγματικότητα του 21ου αιώνα τον τρόπο εργασίας των ιστορικών, τον τρόπο που προσλαμβάνεται η Ιστορία από το ευρύ κοινό, καθώς και τη διδακτική της Ιστορίας;

Χαρακτηριστικό για το πώς αντιλαμβάνονται το ρόλο τους ως ιστορικοί της «Ψηφιακής Ιστορίας» είναι το ακόλουθο απόσπασμα από το άρθρο του A. Fickers (2012):

> Σύμφωνα με τον Αμερικανό ιστορικό Roy Rosenzweig, που υπήρξε ένας από τους λίγους που ασχολήθηκαν με την «Ψηφιακή ιστορία» ήδη από τη δεκαετία του 1990, οι περισσότεροι από τους συναδέλφους του τείνουν να αποφεύγουν συζητήσεις σχετικά με την ψηφιοποίηση εκλαμβάνοντάς την ως ζήτημα «τεχνικής φύσεως» και κάνουν τον ακόλουθο καταμερισμό εργασίας μεταξύ ιστορικών και αρχειονόμων: οι μεν αρχειονόμοι είναι αρμόδιοι να αντιμετωπίσουν τα ζητήματα ψηφιοποίησης των πηγών, καθώς άπτονται της διατήρησης και της διατήρησης των πηγών, ενώ οι ιστορικοί δεν είναι αρμόδιοι γι'αυτά, αλλά επικεντρώνονται σε ζητήματα αυθεντικότητας και αξιοπιστίας.[17]

Λίγο πιο κάτω, στο ίδιο κείμενο, ο Fickers συμμερίζεται το ερώτημα που θέτει ο Rosenweig (2011, 7),[18] εάν η τεράστια αλλαγή ως προς το πλήθος και τη διαθεσιμότητα των πηγών στην ψηφιακή εποχή όντως διευκολύνει το έργο του/της ιστορικού, αν και, κανονικά, δεν θα έπρεπε παρά να αποτελεί εκπλήρωση των προσδοκιών του/της, καθώς η σπανιότητα των πηγών ανέκαθεν δυσχέραινε το έργο του/της.

Ο Fickers παραλληλίζει τη σύγχρονη έμφαση στην ψηφιοποίηση και διάδοση της πολιτιστικής κληρονομιάς με την έκρηξη του αριθμού κριτικών εκδόσεων πηγών στα τέλη του 19ου αιώνα και συνεχίζει:

> «Εάν υποθέσουμε ότι το Διαδίκτυο θα είναι το κύριο αρχείο στο μέλλον, τι είδους κριτικές ικανότητες πρέπει να αποκτήσουν οι ιστορικοί, για να είναι σε θέση να εξακριβώσουν την αυθεντικότητα μιας διαδικτυακής πηγής; Εάν οι μελλοντικές γενιές ιστορικών θέλουν να διατηρήσουν αυτή τη βασική ικανότητα εντός της δικαιοδοσίας της επιστήμης τους, θα πρέπει να αναπτύξουν δεξιότητες στην επιστήμη των υπολογιστών, την ψηφιακή ανάλυση εικόνας και τις τεχνολογίες δικτύου.»[19]

Από την άλλη πλευρά, το Διαδίκτυο διαθέτει πλήθος πηγών στη διάθεση του τελικού χρήστη. Αυτή η 'πληροφοριακή αφθονία', όμως, αυξάνει την ανάγκη για τα παραδοσιακά εργαλεία του ιστορικού: κριτική αξιολόγηση της ποιότητας του ψηφιακού υλικού, διασταύρωση

---

[17] Fickers, A. «Towards A New Digital Historicism? Doing History In The Age Of Abundance». *Journal of European History and Culture*, Vol. 1, 1, 2012. Προσβάσιμο εδώ: http://158.64.76.181/bitstream/10993/7615/1/4-4-1-PB.pdf

[18] Rosenzweig, Roy, 'Scarcity or Abundance? Preserving the Past', in Roy Rosenzweig, Clio Wired. The Future of the Past in the Digital Age, Columbia University Press, 2011, 3-27.

[19] «If we assume that the internet will be the main archive of the future, what kind of critical competences must historians acquire or possess to be able to ascertain the authenticity of an online source? If future generations of historians want to keep this key competence within the realm of their discipline and habitus, they will need to develop skills in computer science, digital image analysis and network technology.» (Fickers 2012).

πηγών, αξιολόγηση του πλαισίου δημιουργίας του ψηφιακού υλικού και αναζήτηση πιθανών κενών.

## 2.2 Οι Τεχνολογίες Τεχνητής Νοημοσύνης

Η προϊούσα έμφαση στο ψηφιακό μέσο και ειδικά το Διαδίκτυο ως μέσο επικοινωνίας, πληροφορίας, συνεργασίας και έρευνας οδηγεί στη γοργή ανάπτυξη της Ψηφιακής Ιστορίας και, γενικότερα, των Ψηφιακών Ανθρωπιστικών Επιστημών. Ψηφιακές Ανθρωπιστικές Επιστήμες είναι το πεδίο που συνδυάζει το ανθρωπιστικό, ως σύστημα παραγωγής ανοιχτά προσβάσιμων, μη δομημένων ή ημιδομημένων δεδομένων σχετικά με τις ανθρώπινες κοινωνίες και τον πολιτισμό, με την ψηφιακή τεχνολογία ως περιβάλλον εργασίας για τη δόμηση, επεξεργασία, ανάλυση, ερμηνεία και διάχυση των δεδομένων αυτών. Αν και το πρώτο ερευνητικό έργο των Ψηφιακών Ανθρωπιστικών Επιστημών χρονολογείται στα πρώτα χρόνια μετά το Δεύτερο Παγκόσμιο πόλεμο, οι Ψηφιακές Ανθρωπιστικές Επιστήμες γνωρίζουν ιδιαίτερη άνθηση κυρίως τα τελευταία είκοσι χρόνια, στην εποχή του Ιστού δεύτερης γενιάς.

Η πρώτη γενιά του Παγκόσμιου Ιστού (Web 1.0) χαρακτηριζόταν από τη στατικότητα των ιστοσελίδων. Η δεύτερη γενιά του Παγκόσμιου Ιστού (Web 2.0) χαρακτηρίζεται από τεχνολογίες νέφους (cloud), τη διαδραστικότητα και τα συνεργατικά περιβάλλοντα, τα οποία έκαναν εφικτή την ανάπτυξη της εξ αποστάσεως σύγχρονης και ασύγχρονης εκπαίδευσης. Η σταδιακή μετάβαση στον Ιστό τρίτης γενιάς (Web 3.0) ή Ιστό των (Ανοικτών Συνδεδεμένων) Δεδομένων έχει στόχο την αποτελεσματικότερη συνεργασία ανθρώπου-υπολογιστή και την προσωποποιημένη, και ταυτόχρονα διαφανή, πρόσβαση στην πληροφορία. Ο Ιστός τρίτης γενιάς δίνει έμφαση στα δομημένα δεδομένα, που θα επεκτείνουν την πρόσβαση στην πληροφορία μέσω των ιστοσελίδων και των υπερσυνδέσμων. Τα δομημένα δεδομένα δίνουν τη δυνατότητα αυτοματοποιημένης επεξεργασίας και ανάσυρσης, ώστε οι μηχανές αναζήτησης να μπορούν να 'κατανοήσουν' πλήρως τα δεδομένα που είναι αναρτημένα στο Διαδίκτυο και να επιστρέφουν μόνο όσα σχετίζονται με την αναζήτηση.

Παραδοσιακά, σε τομείς όπως η Ιστορία, η Αρχαιολογία, η Φιλολογία, η Λογοτεχνία, οι ερευνητές/τριες περνούν πολύ χρόνο εξετάζοντας έγγραφα προκειμένου να βρουν σημαντικές συνδέσεις. Καθώς η Τεχνητή Νοημοσύνη, η Μηχανική Μάθηση (Machine Learning) και η Εξόρυξη Δεδομένων (Data Mining) έχουν ωριμάσει, ένα ευρύ φάσμα υπολογιστικών εργαλείων, μεθόδων και τεχνικών επιτρέπουν σε όσους ασχολούνται με τις Ανθρωπιστικές Επιστήμες να διεξάγουν έρευνα σε κλίμακα που δεν ήταν ως τώρα εφικτή σε πεδία όπως η δημιουργία ψηφιακών δεδομένων, η διαχείριση και ανάλυση δεδομένων (data analytics), η εξόρυξη κειμένων (text mining) και εικόνων (image mining), η οπτικοποίηση δεδομένων (data visualisation), η ανάλυση φυσικής γλώσσας (natural language processing) και τα γεωγραφικά πληροφοριακά συστήματα (GIS).

## 2.3 Ο Κριτικός Γραμματισμός στα Δεδομένα

Ο πρωταγωνιστικός ρόλος της Τεχνητής Νοημοσύνης κάνει το αίτημα για κριτικό γραμματισμό στα δεδομένα (critical data literacy) να αποκτά άμεση προτεραιότητα για όσους ασχολούνται με τις Ανθρωπιστικές Επιστήμες.

Ο γραμματισμός δεδομένων (data literacy) επικεντρώνεται στις ικανότητες που απαιτούνται όταν κανείς χρησιμοποιεί δεδομένα: η ικανότητα ανάγνωσης, κατανόησης, δημιουργίας και επικοινωνίας των δεδομένων ως πληροφοριών. Οι σημερινές ανάγκες προσδιορίζουν την ανάγκη όχι για απλό γραμματισμό δεδομένων (όπως είναι λ.χ. η δεξιότητα χρήσης δεδομένων), αλλά για κριτικό γραμματισμό δεδομένων (critical data literacy).που περιλαμβάνει την κριτική στάση απέναντι στους αλγορίθμους και τη διαφανή, δεοντολογική χρήση της Τεχνητής Νοημοσύνης για την ανεύρεση και επεξεργασία των δεδομένων. Στόχος του είναι να προωθήσει τη σύγκριση αλγορίθμων και τον εντοπισμό αυτών που δεν τηρούν τις αρχές της ηθικής δεοντολογίας και της διαφάνειας. Ο κριτικός γραμματισμός δεδομένων περιλαμβάνει την αναστοχαστική συνειδητοποίηση του τρόπου που λειτουργούν τα υπολογιστικά συστήματα και τους πολλούς και ποικίλους τρόπους με τους οποίους δημιουργούνται και χρησιμοποιούνται τα δεδομένα. Τέλος, καλλιεργεί την ικανότητα κατανόησης των ευρύτερων συστημάτων στα οποία τα δεδομένα παίζουν ουσιώδη ρόλο.

Το «Σχέδιο Δράσης για την Ψηφιακή Εκπαίδευση (2021-2027) - Επαναπροσδιορίζοντας την Εκπαίδευση και Κατάρτιση για την Ψηφιακή Εποχή)»[20] της Ευρωπαϊκής Επιτροπής, το οποίο τέθηκε σε διαβούλευση από τον Ιούνιο ως το Σεπτέμβριο του 2020, περιλαμβάνει δύο βασικά σημεία:

1. Προώθηση της ανάπτυξης ενός οικοσυστήματος ψηφιακής εκπαίδευσης υψηλής απόδοσης. Απαιτείται:
   - υποδομή, συνδεσιμότητα και ψηφιακός εξοπλισμός
   - αποτελεσματικός σχεδιασμός και ανάπτυξη ψηφιακών ικανοτήτων
   - ψηφιακά ικανοί δάσκαλοι, καθηγητές και προσωπικό εκπαίδευσης και κατάρτισης
   - υψηλής ποιότητας περιεχόμενο μάθησης, φιλικά προς το χρήστη εργαλεία και ασφαλείς πλατφόρμες που σέβονται την ιδιωτικότητα και θα τηρούν τις αρχές της δεοντολογίας

- 2. Ενίσχυση των ψηφιακών δεξιοτήτων και ικανοτήτων για τον ψηφιακό μετασχηματισμό. Για το σκοπό αυτό απαιτούνται:
  βασικές ψηφιακές δεξιότητες και ικανότητες από νεαρή ηλικία
- κριτικός ψηφιακός γραμματισμός,
- καταπολέμησης της παραπληροφόρησης
- καλή γνώση και κατανόηση τεχνολογιών δεδομένων, όπως η τεχνητή νοημοσύνη
- προηγμένες ψηφιακές δεξιότητες που παράγουν περισσότερους ειδικούς στις ψηφιακές τεχνολογίες και διασφαλίζουν ότι τα κορίτσια και οι νέες γυναίκες εκπροσωπούνται εξίσου στις ψηφιακές σπουδές και σταδιοδρομία

Απαντώντας στο «Σχέδιο Δράσης για την Ψηφιακή Εκπαίδευση (2021-2027) Επαναπροσδιορίζοντας την Εκπαίδευση και Κατάρτιση για την Ψηφιακή Εποχή)» της Ευρωπαϊκής Επιτροπής, η Ευρωπαϊκή Ένωση Εκπαιδευτικών Ιστορίας EuroClio - European Association of History Educators τονίζει την ανάγκη για: [21]

---

[20] Digital Education Action Plan (2021-2027) Resetting education and training for the digital age. Βλ. το Σχέδιο Δράσης εδώ.

[21] Βλ. το κείμενο της Ευρωπαϊκής Εταιρείας Εκπαιδευτικών Ιστορίας Euroclio εδώ.

- ✓ Ανάπτυξη ή βελτίωση εύχρηστων εργαλείων που μπορούν να χρησιμοποιήσουν οι εκπαιδευτικοί για να δημιουργήσουν, να μοιραστούν και να προσαρμόσουν δικούς τους ανοιχτούς εκπαιδευτικούς πόρους.
- ✓ Ανάπτυξη ανοικτής εκπαίδευσης υψηλής ποιότητας.
- ✓ Έρευνα για τον προσδιορισμό και την κοινή χρήση στρατηγικών για τη χρήση αποτελεσματικών ψηφιακών τεχνολογιών για τη διδασκαλία και τη μάθηση στις Ανθρωπιστικές Επιστήμες.

# 3. Η Ιστορία στον Παγκόσμιο Ιστό

## 3.1 Παράγοντες που ευνοούν την στροφή προς την Ψηφιακή Ιστορία

**1.** Μια σειρά από τεχνολογικές εξελίξεις, όπως:
- ✓ Η δυνατότητα χρήσης βάσεων δεδομένων (databases) και βάσεων γνώσεων (knowledge bases).[22]
- ✓ Ο συνδυασμός χάρτη και ιστορικής αφήγησης (γεω-αφήγηση) μέσω ψηφιακών προγραμμάτων GIS (Geographic Information Systems).[23]
- ✓ Η δυνατότητα τρισδιάστατης απεικόνισης και εκτύπωσης (3D imaging & printing).
- ✓ Η δημιουργία βιντεοπαιχνιδιών με ιστορικό περιεχόμενο. Λειτουργεί κατά κάποιο τρόπο όπως μια προσομοίωση. Σημειωτέον ότι οι προσομοιώσεις (simulations) και το παιχνίδι ρόλων (role play) είναι τεχνικές που χρησιμοποιούνται εδώ και χρόνια από τους εκπαιδευτικούς στο μάθημα της Ιστορίας στην τυπική εκπαίδευση.

**2.** Η ψηφιοποίηση και ανάρτηση στο Διαδίκτυο τεράστιου σε ποσότητα ιστορικού/πολιτισμικού ψηφιακού περιεχομένου:
- ✓ Η δημιουργία πλούσιων σε υλικό ψηφιακών αρχείων και πολιτισμικών συλλογών, όπως η Europeana <https://www.europeana.eu/en>.

---

[22] Βάση δεδομένων (database) είναι ένα σύνολο δομημένων δεδομένων (structured dataset) που έχουν μορφοποιηθεί με συγκεκριμένο τρόπο ώστε ο υπολογιστής να μπορεί εύκολα να βρει τις επιθυμητές πληροφορίες. Βάση γνώσεων (knowledge base) δεν είναι απλώς ένας χώρος αποθήκευσης δεδομένων, αλλά μπορεί να είναι ένα εργαλείο Τεχνητής Νοημοσύνης και της Αναπαράστασης της Γνώσης (Knowledge Representation) για τη λήψη έξυπνων αποφάσεων με βάση τα δομημένα δεδομένα που περιέχει.

[23] Το σύστημα γεωγραφικών πληροφοριών (GIS) είναι ένα υπολογιστικό σύστημα για τη συλλογή, την αποθήκευση, τον έλεγχο και την προβολή δεδομένων που σχετίζονται με τοποθεσίες στην επιφάνεια της Γης (spatial data). Μπορεί να εμφανίσει πολλά διαφορετικά είδη δεδομένων σε έναν χάρτη, και να ενσωματώσει δεδομένα σε μορφή πίνακα. Η σύζευξη των γεωγραφικών δεδομένων με άλλους τύπους δεδομένων μπορεί να οδηγήσει στο να απαντηθούν σημαντικά ιστορικά ερευνητικά ερωτήματα. Βλ. ενδεικτικά Knowles, Anne Kelly, Amy Hillier, επιμ. 2008. Placing History: How Maps, Spatial Data, and GIS Are Changing Historical Scholarship. 1st έκδ. ESRI Press. Leidwanger, Justin. 2013. “Modeling Distance with Time in Ancient Mediterranean Seafaring: A GIS Application for the Interpretation of Maritime Connectivity.” *Journal of Archaeological Science* 40 (8): 3302–8. Πρόσβαση εδώ: https://www.sciencedirect.com/science/article/abs/pii/S0305440313001064.

- ✓ Η διάδοση πολύγλωσσων ψηφιακών πλατφορμών όπως: Wikipedia <https://www.wikipedia.org/>και Wikidata <https://www.wikidata.org/>.
- ✓ Η δημιουργία ενός continuum μεταξύ των λεγόμενων ιδρυμάτων μνήμης (Galleries Libraries Archives Museums- GLAM).

**3.** Η ύπαρξη ενός πολυπληθούς και διεθνούς κοινού που είναι πρόθυμο να αναποκριθεί στα αιτήματα και τις μεθόδους της Ψηφιακής Ιστορίας:

- ✓ Η διάδοση της χρήσης του Διαδικτύου σε δισεκατομμύρια χρήστες παγκοσμίως.[24]
- ✓ Η δυνατότητα πληθοπορισμού (crowdsourcing) των πηγών δημιουργεί νέες ευκαιρίες για τους ιστορικούς να αλληλεπιδράσουν με ένα κοινό που ενδιαφέρεται σαφώς για το νόημα του παρελθόντος. Μέσω του πληθοπορισμού οι ιστορικές πηγές συλλέγονται με διαδικασία «από κάτω προς τα πάνω» (bottom up), καθώς τα μέλη μιας κοινότητας μπορούν εύκολα να συνεισφέρουν υλικό.

**4.** Η δυνατότητα χρηματοδότησης ερευνητικών προγραμμάτων και πρωτοβουλιών στο πεδίο της Ψηφιακής Ιστορίας

- ✓ Τα πρώτα κέντρα ψηφιακής Ιστορίας στις ΗΠΑ προέκυψαν από χρηματοδότηση του National Endowment for the Humanities (NEH). Ανάλογη πρωτοβουλία στην Ευρώπη είναι το ίδρυμα Europeana (Europeana Foundation) που τελούν υπό την αιγίδα της Ευρωπαϊκής Επιτροπής.[25]

**5.** Η στροφή, συνολικά, της εκπαίδευσης στο ψηφιακό.

- ✓ Συνέπεια της στροφής αυτής είναι αφενός η αύξηση του αριθμού των αποφοίτων πανεπιστημιακών προγραμμάτων Ψηφιακών Ανθρωπιστικών Επιστημών, αφετέρου η αναμόρφωση των αναλυτικών προγραμμάτων στα εκπαιδευτικά συστήματα όλων των βαθμίδων στις χώρες τις ΕΕ, ώστε να καλυφθούν οι νέες ανάγκες.
- ✓ Η στροφή αυτή είναι αισθητή και στην Ελλάδα, όπου το ψηφιακό χάσμα διευρύνεται μεταξύ εκπροσώπων της γενιάς του διαδικτύου που γεννήθηκαν τον 21$^{ο}$ αιώνα και όσων γεννήθηκαν τον 20$^{ο}$ αι. Οι πρώτοι χαρακτηρίζονται ως ψηφιακοί αυτόχθονες (digital natives), ενώ οι δεύτεροι ως ψηφιακοί μετανάστες (digital immigrants).[26]

---

[24] Παρά το ότι το λεγόμενο ψηφιακό χάσμα (digital divide) εξακολουθεί να υφίσταται, ειδικά σε χώρες με χαμηλό κατά κεφαλήν εισόδημα, η κοινότητα των χρηστών του Διαδικτύου άγγιξε το 59% του παγκόσμιου πληθυσμού κατά το έτος 2020 (Στατιστικά στοιχεία εδώ).

[25] NEH: https://www.neh.gov/, Europeana Pro: https://pro.europeana.eu/about-us/foundation

[26] Οι όροι ανήκουν στον Marc Prensky. 2001. Digital Natives Digital Immigrants. On the Horizon (MCB University Press, Vol. 9 No. 5, October 2001) https://www.marcprensky.com/writing/Prensky%20-%20Digital%20Natives,%20Digital%20Immigrants%20-%20Part1.pdf

**6**. Η στροφή προς την ανοικτότητα των εργαλείων, των δεδομένων, του κώδικα και των ερευνητικών αποτελεσμάτων που εκπορεύτεται τόσο από θεσμικά κέντρα αποφάσεων, όσο και από ad hoc κοινότητες.[27]

## 3.2 Η Ψηφιακή Ιστορία χρησιμοποιεί μη συμβατικές πηγές

Βασική πηγή των ιστορικών ήταν, ανέκαθεν, τα κείμενα. Από την άλλη, οι υπολογιστές, ενώ ξεκίνησαν την ιστορική τους ύπαρξη ως μηχανές υπολογισμού αριθμητικών δεδομένων, τις τελευταίες δεκαετίες, έχουν άρρηκτη σχέση με τα κείμενα καθώς τεράστιος όγκος κειμένων έχει ψηφιοποιηθεί σε διαφορετικό βαθμό ψηφιοποίησης και μορφότυπους, από την απλή, ‘φωτογραφική’ ψηφιοποίηση, ως την μετατροπή σε δεδομένα που μπορεί να επεξεργαστεί το λογισμικό κάθε υπολογιστή (όπως είναι η μετατροπή σε αρχεία XML). Η χρήση του ψηφιακού μέσου χωρίς τον κατάλληλο ψηφιακό γραμματισμό, οδηγεί στην άκριτη αποδοχή των όρων (περιορισμών, μεροληψιών) που θέτει το ψηφιακό μέσο, σχεδόν να γίνονται αντιληπτοί από τους τελικούς χρήστες.

### 3.2.1 Από την ψηφιοποίηση στη διαλειτουργικότητα

Πολλές από τις ψηφιακές πηγές δεν παρέχουν τις δυνατότητες χρήσης και επομένως δεν καλύπτουν τις ανάγκες που έχουν οι ιστορικοί, αλλά και το ευρύ κοινό. Η δυνατότητα χρήσης του ψηφιακού περιεχομένου εξαρτάται, μεταξύ άλλων, από το είδος ψηφιοποίησης του περιεχομένου που διατίθεται ψηφιακά.

Η ψηφιοποίηση (digitization) ενός εγγράφου είναι συνήθως μια διαδικασία δύο σταδίων:

Το πρώτο στάδιο είναι σχετικά απλό, πραγματοποιείται με τη χρήση σαρωτή ή κάμερας και, εάν γίνει σωστά, οδηγεί σε αποτέλεσμα με σχετικά μικρές αφαιρέσεις από το πρωτότυπο (το αποτέλεσμα είναι αντίγραφο τηλεομοιοτυπίας). Πρώτα μια ψηφιακή εικόνα του εγγράφου δημιουργείται ως bitmap και μετά το περιεχόμενο του κειμένου κωδικοποιείται και γίνεται αναγνώσιμο από τη μηχανή. Η διαδικασία αυτή παρέχει στον τελικό χρήστη τη δυνατότητα να πληκτρολογήσει έναν όρο αναζήτησης και, εάν ο όρος αυτός βρίσκεται στο κείμενο, στον χρήστη θα εμφανιστεί η κατάλληλη εικόνα της σελίδας.

Το δεύτερο στάδιο είναι πιο σύνθετο: περιλαμβάνει είτε πληκτρολόγηση του κειμένου με το χέρι είτε με το λογισμικό οπτικής αναγνώρισης χαρακτήρων (OCR) που χρησιμοποιείται για την αυτόματη αναγνώριση γραμμάτων από την εικόνα bitmap. Και οι δύο επιλογές είναι χρονοβόρες και επιρρεπείς σε σφάλματα. Το λογισμικό OCR τείνει πλέον να χρησιμοποιείται σε μεγάλη κλίμακα, γιατί είναι ταχύτερο και φθηνότερο.

---

[27] Όπως η Unesco (βλ. σχετικά τη διακήρυξη για ανοικτούς εκπαιδευτικούς πόρους: Open Educational Resources – OER, https://en.unesco.org/oer/paris-declaration), η πρωτοβουλία του Ευρωπαϊκού δικτύου Παν/κών και Ερευνητικών Ιδρυμάτων OPERAS-Open Scholarly Communication in the European Research Area for Social Scienes and Humanities (https://www.operas-eu.org/), η διεθνής πρωτοβουλία για την ανοικτή πρόσβαση στη έρευνα και στη γνώση (https://creativecommons.org/) και η αντίστοιχη ελληνική πρωτοβουλία ΕΛΛΑΚ (https://ellak.gr/ και https://creativecommons.ellak.gr/).

Οι υπολογιστές μπορούν να χρησιμοποιούν έναν αριθμό προτύπων κωδικοποίησης χαρακτήρων: το πιο κοινό πρότυπο είναι το UTF-8 που υπάρχει στο 98% των ιστοσελίδων του Παγκόσμιου Ιστού. Το UTF-8 είναι το συνιστώμενο πρότυπο κωδικοποίησης κειμένου, γιατί είναι σε θέση να αποδίδει χαρακτήρες σε πολλά σύγχρονα αλφάβητα, Λατινικά και μη (όπως είναι π.χ. το Ελληνικό αλφάβητο, το Αραβικό abjad και οι Κινεζικοί χαρακτήρες. Ο όρος ‘digitalization’, από την άλλη, συνδέεται με την έννοια του εγγενούς ψηφιακού (born-digital), δηλαδή έγγραφα ή πληροφορίες που δεν υπέστησαν μετατροπή από την αναλογική τους μορφή σε ψηφιακή μορφή, αλλά δημιουργήθηκαν απευθείας σε ψηφιακό περιβάλλον (ΕΙΚΟΝΑ 1). Ένα παράδειγμα ψηφιοποίησης είναι η φωτογραφία μιας σελίδας βιβλίου, που μπορούμε να τραβήξουμε με το κινητό μας. Παράδειγμα ψηφιακά εγγενούς εγγράφου είναι το μήνυμα ηλεκτρονικού ταχυδρομείου που πληκτρολογεί κανείς στον υπολογιστή.

**ΕΙΚΟΝΑ 1**

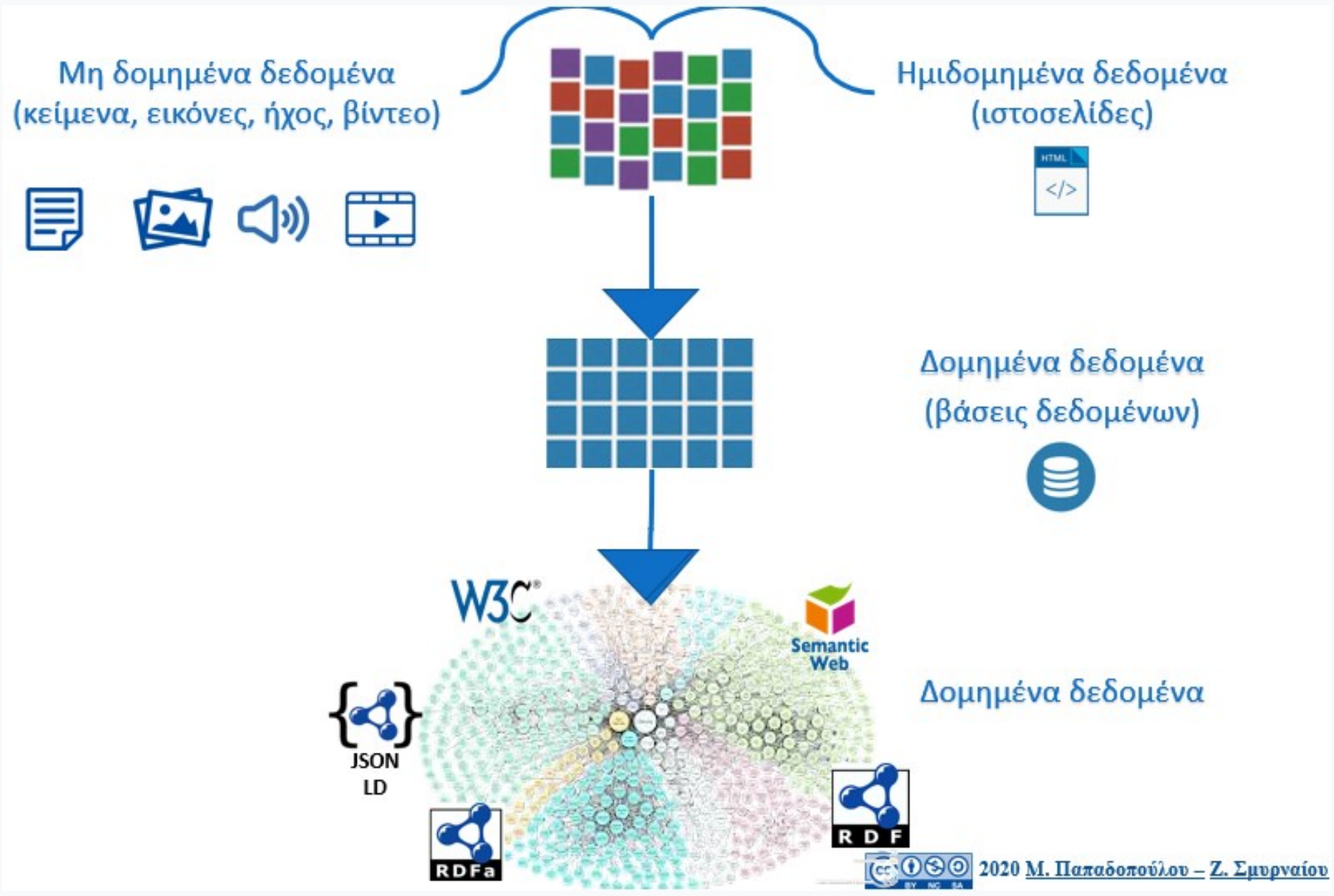

Όταν αντί για μορφότυπους που μπορεί διαθέτουμε πόρους σε μορφή που μπορεί να ‘κατανοήσει’ ο υπολογιστής (machine processible, machine tractable), όχι απλώς να ‘διαβάσει’ (machine readable), τότε μπορούμε να μιλάμε για διαλειτουργικότητα (interoperability) και διαλειτουργικά δεδομένα. Η διαλειτουργικότητα διευκολύνει τις μηχανές αναζήτησης βελτιώνοντας τη συνεργασία μεταξύ ανθρώπου και υπολογιστή. Γνωστοί φορείς που τηρούν προδιαγραφές διαλειτουργικότητας είναι η Βιβλιοθήκη του Κογκρέσου και η ελεύθερα προσβάσιμη βάση γνώσεων Wikidata, η οποία εμπλουτίζεται διαρκώς μέσω πληθοπορισμού.[28]

[28] Βλ. τις Προδιαγραφές διαλειτουργικότητας και ποιότητας για τη διαδικτυακή διάθεση ψηφιακού πολιτιστικού περιεχομένου. Εθνικό Κέντρο Τεκμηρίωσης-ΕΚΤ. 1η έκδοση Μάρτιος 2019, Ανοικτή πρόσβαση εδώ.

ΕΙΚΟΝΑ 2

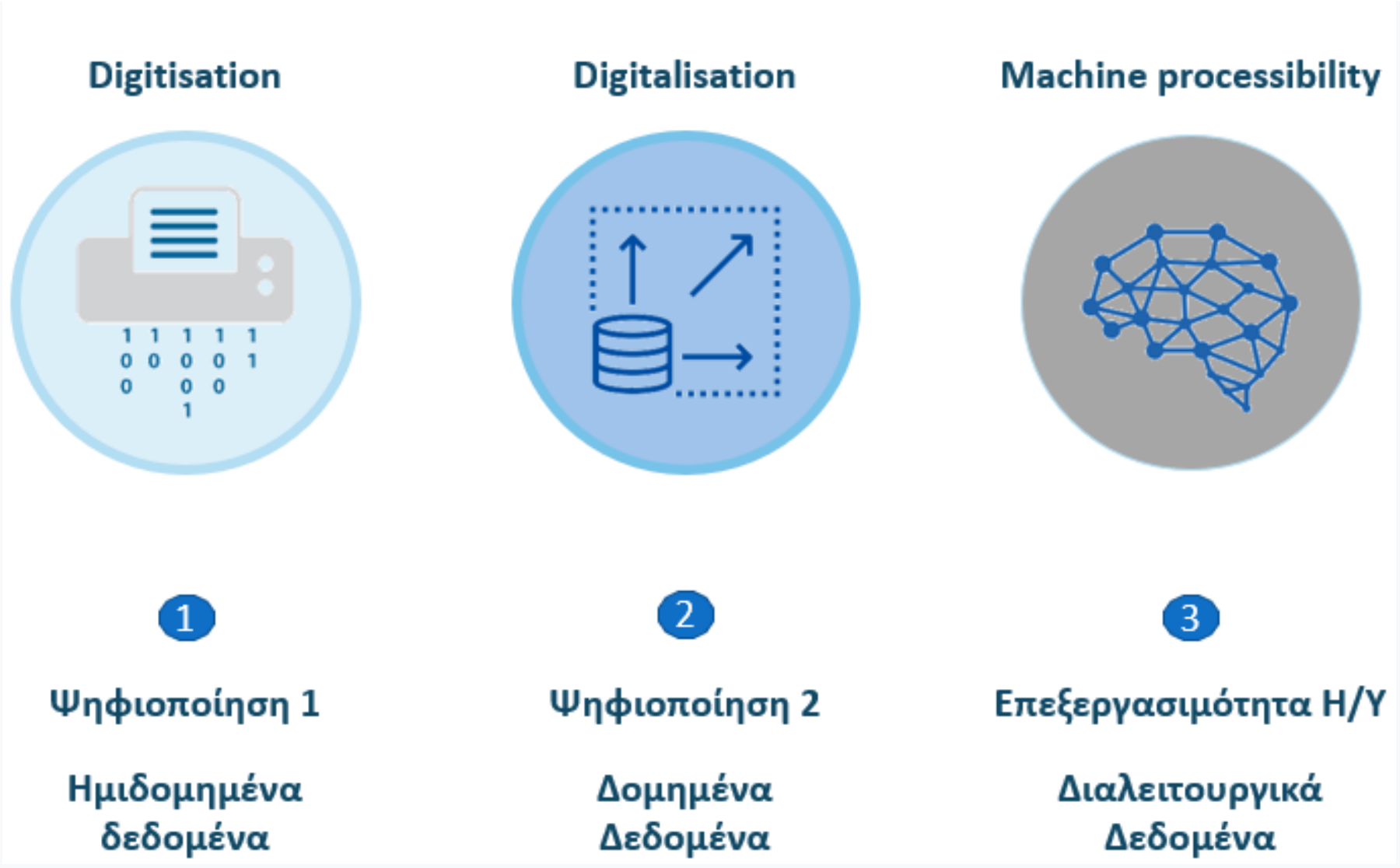

Τα διαλειτουργικά δεδομένα είναι κατ'ανάγκη δομημένα δεδομένα. Τα ανοικτά συνδεδεμένα δεδομένα είναι εγγενώς διαλειτουργικά (ΕΙΚΟΝΕΣ 2-3). Ο εθνικός συσσωρευτής πολιτισμικού περιεχομένουτ ΕΚΤ (Εθνικού Κέντρο Τεκμηρίωσης), για τον οποίο γίνεται εκτενής λόγος πιο κάτω, παρέχει τα δεδομένα του ως αναγνώσιμα από τους ανθρώπους-χρήστες, αλλά και ως δομημένα διαλειτουργικά δεδομένα (δηλαδή ανοικτά συνδεδεμένα δεδομένα), κατανοητά από τους πράκτορες λογισμικού (software agents) και συνεπώς αξιοποιήσιμα για ομογενοποίηση και επανάχρηση από τον ευρωπαϊκό συσσωρευτή πολιτισμικού περιεχομένου, Europeana.

ΕΙΚΟΝΑ 3

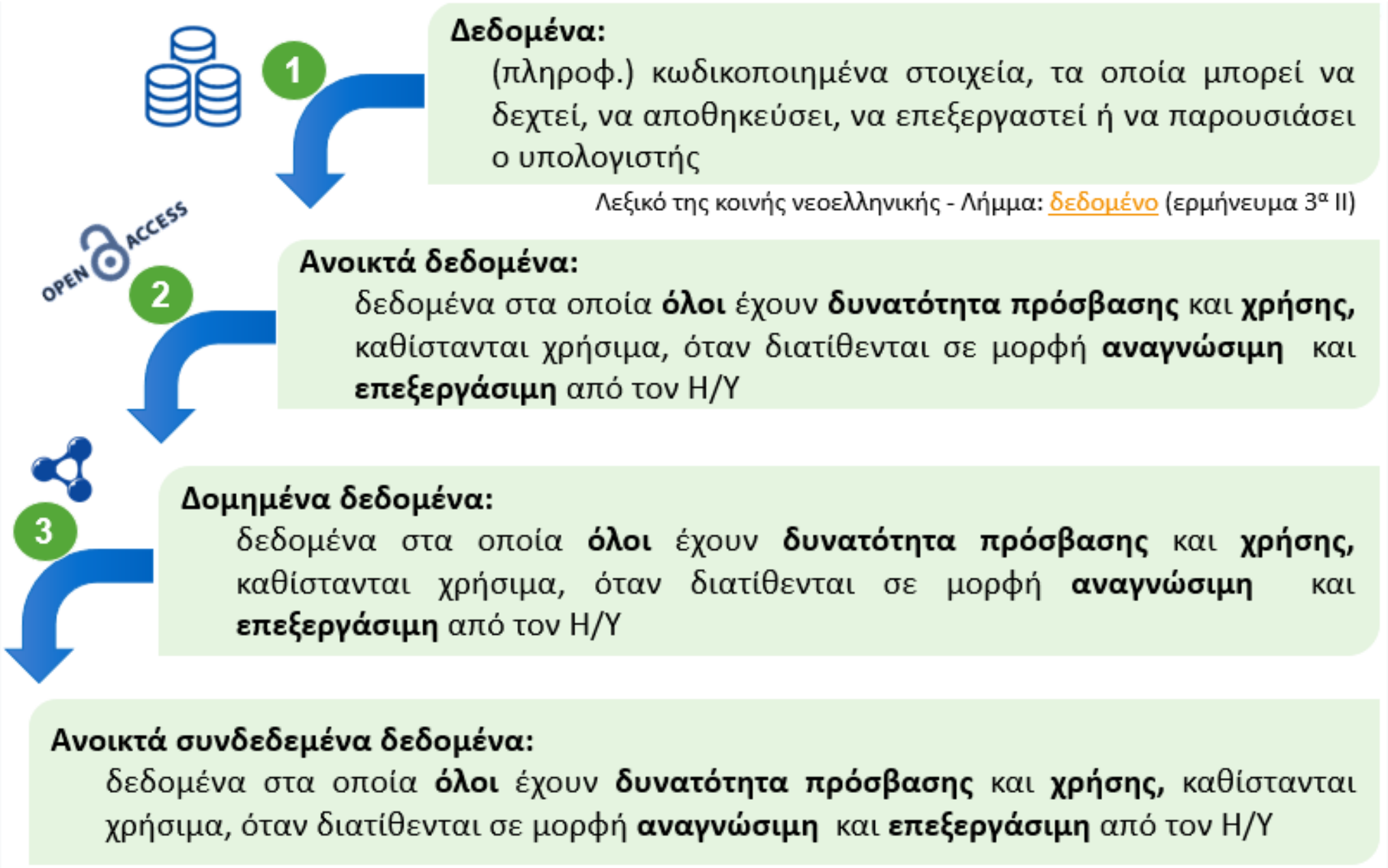

Η λεγόμενη «πυραμίδα της γνώσης» ή (Data, Information, Knowledge, Wisdom-DIKW Pyramid)[29] αποτελεί ένα εννοιολογικό πλαίσιο για να κατανοήσει κανείς πώς τα σήματα αποθηκεύονται ως δεδομένα και μετατρέπονται πρώτα σε πληροφορίες, μετά σε γνώση και τέλος σε σοφία. Στη βάση της πυραμίδας βρίσκονται τα δεδομένα. Η δόμηση των δεδομένων τα καθιστά πληροφορίες. Η ικανότητα των υπολογιστικών συστημάτων για απόκτηση της πολύτιμης ‘γνώσης’ από τις πληροφορίες και τα δεδομένα, είναι δυνατό να αναβαθμιστεί από το ένα στάδιο στο άλλο, ανάλογα με την κωδικοποίηση των δεδομένων και την εξάλειψη του λεγόμενου ‘θορύβου’ (noise).

Τα δεδομένα είναι μια συλλογή στοιχείων σε πρωτογενή ή μη οργανωμένη μορφή, όπως μια ακολουθία χαρακτήρων. Ωστόσο, χωρίς πλαίσιο (context), τα δεδομένα δεν είναι ιδιαίτερα χρήσιμα. Για παράδειγμα, το 15012021 είναι απλώς μια ακολουθία αριθμών χωρίς εμφανή σημασία. Αλλά αν το δούμε στο πλαίσιο «πρόκειται για ημερομηνία», μπορούμε εύκολα να αναγνωρίσουμε την 15η Ιανουαρίου 2021. Με τη σύνδεση με άλλα δεδομένα ή την προσθήκη περιβάλλοντος (context), τα δεδομένα αποκτούν αξία και νόημα. Αφού τα δεδομένα «καθαριστούν», γίνονται πληροφορίες που μπορούν να αναλυθούν και να οπτικοποιηθούν. Για τη μετάβαση από την πληροφορία στη γνώση πρέπει τα δεδομένα, να εμπλουτιστούν σημασιολογικά, να καθορισθεί με σαφήνεια ο τύπος τους. Η σοφία, τέλος, προϋποθέτει τη δυνατότητα λήψης αιτιολογημένων αποφάσεων. Η μετάβαση από τα δεδομένα στην πληροφορία είναι συντακτικού τύπου (συνάρθρωση), από την πληροφορία στη γνώση είναι σημασιολογικού τύπου (σημασιοδότηση) και από τη γνώση στη σοφία είναι συστημικού τύπου (αφορά το όλο σύστημα των συναρθωμένων και σημασιοδοτημένων δεδομένων).

Ένα παράδειγμα: Πώς μπορεί ο υπολογιστής να ‘αναγνωρίσει’ ότι το εικονιζόμενο αντικείμενο είναι π.χ. γάτα; Το σήμα, δηλαδή το επίπεδο τάσης σε μια κυψέλη πυκνωτή στο τσιπ RAM γίνεται δεδομένα, δηλαδή bits που αποτελούν το αρχείο εικόνας. Στο στάδιο αυτό δεν υπάρχει ακόμη δομή, μόνο bytes. Ο υπολογιστής έχει πληροφορίες, όταν υπάρχει δομή στα δεδομένα (π.χ. κεφαλίδα, pixel, διάταξη). Ο υπολογιστής αποκτά γνώση, όταν μπορεί να αναγνωρίσει ότι πρόκειται για την εικόνα μιας γάτας και όταν μπορεί να κάνει προβλέψεις για κάθε αντίστοιχη εικόνα. Ο υπολογιστής αποκτά σοφία, όταν μπορεί να απαντήσει ερωτήσεις του τύπου «γιατί αυτό που υπάρχει στην εικόνα είναι γάτα»;

### 3.2.2 Ψηφιακή αναζήτηση δεδομένων

Μόλις δημιουργηθούν οι ψηφιακές πηγές, οι χρήστες τους συχνά ανατρέχουν σε αυτές χρησιμοποιώντας τεχνικές που δεν τους είναι κατανοητές σε βάθος. Ένα απλό παράδειγμα είναι η αναζήτηση με βάση λέξεις-κλειδιά ή φασέτες (facets)[30] σε μια βάση δεδομένων ή στον Παγκόσμιο ιστό, η οποία επιστρέφει μια σειρά από αποτελέσματα ταξινομημένα κατά «συνάφεια».

---

[29] Ackoff, Russell (1989). "From Data to Wisdom". Journal of Applied Systems Analysis. 16: 3–9. Αποδίδεται στον Ackoff (1989, παρόλο που στο άρθρο του αυτό δεν υπάρχει σχετική εικόνα

[30] Ο όρος εξελληνίζει τον Αγγλικό όρο facet που χρησιμοποιείται στην επιστήμη της Βιβλιοθηκονομίας και σημαίνει «όψη, ως προς την οποία ομαδοποιούνται έννοιες της ίδιας εγγενούς κατηγορίας». Καπιδάκης, Σ., Λαζαρίνης, Φ., Τοράκη, Κ. 2015. Θησαυροί. Στο Καπιδάκης, Σ., Λαζαρίνης, Φ., Τοράκη, Κ. 2015. Θέματα βιβλιοθηκονομίας και επιστήμης των πληροφοριών. Αθήνα: Σύνδεσμος Ελληνικών Ακαδημαϊκών Βιβλιοθηκών. κεφ 8, σελ. 261. Διαθέσιμο στο: http://hdl.handle.net/11419/1680).

Ο Hitchcock (2013)[31] επισημαίνει σωστά, ότι οι περισσότεροι ιστορικοί που χρησιμοποιούν ψηφιακές πηγές το κάνουν χωρίς να έχουν καμία ιδέα ούτε για τις επιπτώσεις του τρόπου με τον οποίο ψηφιοποιήθηκε η αρχική πηγή για να δημιουργηθεί το ψηφιακό της 'υποκατάστατο' (digital surrogate) ούτε του τρόπου με τον οποίο γίνονται οι ψηφιακές αναζητήσεις των ψηφιοποιημένων δεδομένων. Ούτε, πώς οι μηχανές αναζήτησης 'αποφασίζουν' τι είναι και τι δεν είναι «σχετικό» κι έτσι επιστρέφουν ό,τι θεωρείται σχετικό ως αποτέλεσμα της αναζήτησης, ενώ ό,τι δεν θεωρείται σχετικό δεν το επιστρέφουν. Είναι κοινός τόπος ότι οι μηχανές αναζήτησης είναι δυνατό να φανούν μεροληπτικές και αναξιόπιστες. Όμως παραμένουν το μόνο ψηφιακό εργαλείο που χρησιμοποιούν οι περισσότεροι ιστορικοί για να κάνουν αναζητήσεις στον μεγάλο όγκο ψηφιακού υλικού που είναι διαθέσιμο, παρόλο που υπάρχουν εναλλακτικοί τρόποι αναζήτησης δεδομένων στον Ιστό, όπως με χρήση των δημόσιων Διεπαφών Προγραμματισμού Εφαρμογών (Application Programming Interface- API).[32]

### 3.2.3 Δεδομένα και Μεταδεδομένα

Οι έννοιες «δεδομένα» και «μεταδεδομένα» είναι βασικές στις Ψηφιακές Ανθρωπιστικές Επιστήμες και την Ψηφιακή Ιστορία. Τα δεδομένα διακρίνονται σε αδόμητα (π.χ. κείμενο), ημιδομημένα (π.χ. ιστοσελίδες) και δομημένα (αρχεία XML).

Τα μεταδεδομένα είναι κατηγορίες ή πεδία που περιέχουν συγκεκριμένες πληροφορίες, όπως τίτλος, ημερομηνία ή δημιουργός, για κάθε πηγή. Είναι δυνατό να είναι αδόμητα ή δομημένα. Οι ετικέτες που προσθέτει κανείς σε φωτογραφίες σε πλατφόρμες κοινωνικής δικτύωσης όπως π.χ. στο Pinterest είναι ο πιο απλός, όχι, όμως ο πιο αποτελεσματικός, τρόπος να μοιραστεί κανείς ψηφιακά το υλικό που επιθυμεί συνοδεύοντάς το με αδόμητα μεταδεδομένα. Πολλές από τις λύσεις που παρέχουν οι ψηφιακοί συσσωρευτές Europeana και SearchCulture (βλ. παρακάτω ενότητα 3.4.2) αφορούν την τυποποίηση των μεταδεδομένων με σκοπό την οργάνωση, περιγραφή, κοινή χρήση και διατήρηση του ψηφιακού περιεχομένου.

Η Επεκτάσιμη Γλώσσα Σήμανσης (Extensible Markup Language - XML), το Πλαίσιο Περιγραφής Πόρων (Resource Description Framework – RDF), η Πρωτοβουλία Κωδικοποίησης Κειμένου (Text Encoding Initiative TEI), το πρότυπο Dublin Core (DC), το πρότυπο Open Archives Initiative Protocol for Metadata Harvesting (OAI-PMH), παρέχουν τη δυνατότητα δόμησης δεδομένων και μεταδεδομένων με τρόπο επεξεργάσιμο από τον υπολογιστή.

---

[31] Hitchcock, T. 2013. Confronting the digital or how academic history writing lost the plot. Culture and Social History 10:9–23. http://sro.sussex.ac.uk/id/eprint/58045/

[32] Π.χ. το API του ΕΚΤ SearchCulture προσβάσιμο εδώ (μετά από συμπλήρωση της ηλεκτρονικής αίτησης για χορήγηση 'κλειδιού' (API key):
https://www.searchculture.gr/aggregator/portal/interoperability#rest_api

## 3.3 Πόροι της Ψηφιακής Ιστορίας: Ιδρύματα Ιστορικής Μνήμης

Η Ψηφιακή Ιστορία χρησιμοποιεί μη συμβατικές πηγές. Πέρα από τις παραδοσιακές πηγές της Ιστορίας, αρχειακές πηγές και κείμενα, οι ιστορικοί Ψηφιακής Ιστορίας στρέφονται όλο και περισσότερο σε οπτικές, προφορικές, ηχητικές και πολυμεσικές πηγές. Οι προκλήσεις που θέτουν αυτές οι πηγές απαιτούν από τους ιστορικούς να αξιοποιήσουν όλα τα μέσα (media), να εξετάσουν εναλλακτικές θεωρητικές και πρακτικές προσεγγίσεις (π.χ. σχετικά με το πότε και πώς μπορούν να χρησιμοποιηθούν αυτές οι λιγότερο παραδοσιακές πηγές) καθώς και να αποκτήσουν νέες δεξιότητες για την ερμηνεία τους.

Τα λεγόμενα ιδρύματα (ιστορικής) μνήμης (Πινακοθήκες-Βιβλιοθήκες-Αρχεία-Μουσεία) είναι κύριοι πόροι για την Ψηφιακή Ιστορία και χάρη στο διαδίκτυο αποτελούν ένα continuum γνωστό με το ακρωνύμιο LAM ή GLAM (Gallery-Library-Archive-Museum). Στην εποχή της πανδημίας COVID-19 προσφέρουν τις πολυμεσικές συλλογές τους σε ανοικτή, δωρεάν πρόσβαση, καθιστώντας εφικτή την επίσκεψή τους από το κοινό.

### 3.3.1 Ψηφιακά Αρχεία

Η έννοια του ψηφιακού αρχείου (digital archive), είναι κομβική για την Ψηφιακή Ιστορία και αναφέρεται σε ιστότοπους όπου συγκεντρώνονται ψηφιοποιημένες αρχειακές πηγές. Στα Ελληνικά η λέξη «αρχείο» συμπυκνώνει τις σημασίες «έγγραφο» (document) και «αρχείο» (αγγλικό ‘archive’ και έχει γίνει τετριμμένη, αφού χρησιμοποιείται κατά την καθημερινή μας αλληλεπίδραση με τους υπολογιστές σε φράσεις όπως «σώζω, αποθηκεύω, κατεβάζω το αρχείο».

Στην παρούσα χρονική στιγμή, οι ιστορικοί έχουν πρόσβαση σε έναν συνεχώς αυξανόμενο πλούτο ψηφιοποιημένων εκδόσεων, ή ψηφιακών υποκατάστατων (digital surrogates) επιλεγμένων πρωτογενών πηγών μέσω διαδικτυακών συλλογών. Ταυτόχρονα, παρατηρείται μια έκρηξη εγγενώς-ψηφιακού υλικού που παράγεται και συλλέγεται σε πρωτοφανή κλίμακα (ιστότοποι, μέσα κοινωνικής δικτύωσης, ψηφιακά βίντεο και φωτογραφίες κ.λπ.). Γνωστά ψηφιακά αρχεία είναι οργανισμοί όπως το Internet Archive, το Ψηφιακό Αρχείο της 11ης Σεπτεμβρίου, και το Ψηφιακό Αρχείο Rossetti.[33] Παρά την ύπαρξη διεθνών προτύπων του Διεθνούς Οργανισμού Προτυποποίησης ISO,[34] σπανίως ακολουθούνται, με αποτέλεσμα κάθε ψηφιακό ιστορικό αρχείο να αντιπροσωπεύει ένα κάπως διαφορετικό όραμα για τη φύση και την έννοια ενός ψηφιακού αρχείου. Την τελευταία εικοσαετία η ερευνητική και επαγγελματική αρχειονομική πρακτική έχει στραφεί ανεπιστρεπτί προς το ψηφιακό.[35]

Βασικές ψηφιακές πηγές στη διάθεση του ελληνόφωνου κοινού και της ερευνητικής κοινότητας είναι οι εξής:

---

[33] Internet Archive https://archive.org/, September 11 Digital Archive https://911digitalarchive.org/, Rossetti Digital Archive http://www.rossettiarchive.org/

[34] Σχετικά με τα Πρότυπα του Διεθνούς Οργανισμού Προτυποποίησης ISO (International Organization for Standardization) για τα ψηφιακά αρχεία, βλ. ISO Standards for Digital Archives: https://gfbio.biowikifarm.net/wiki/ISO_Standards_for_Digital_Archives

[35] Bountouri Lina 2017. Archives in the Digital Age. Standards, Policies and Tools. Chandos, Elsevier. Βλ. τον Πρόλογο.

- www.openarchives.gr η διαδικτυακή πύλη αναζήτησης και πλοήγησης σε ελληνόγλωσσο επιστημονικό περιεχόμενο του Εθνικού Κέντρου Τεκμηρίωσης (ΕΚΤ).
- http://arxeiomnimon.gak.gr/index.html η ψηφιακή πολιτιστική συλλογή των Γενικών Αρχείων του Κράτους. Περιλαμβάνει 7 εκατομμύρια ψηφιοποιημένες σελίδες.
- https://anemi.lib.uoc.gr/ η ψηφιακή Βιβλιοθήκη Νεοελληνικών Σπουδών. Περιλαμβάνει 2 εκατομμύρια ψηφιοποιημένες σελίδες. Δημιουργήθηκε με χρηματοδότηση από το πρόγραμμα Κοινωνία της Πληροφορίας το 2006-2008.
- http://efimeris.nlg.gr/ns/main.html η ψηφιακή βιβλιοθήκη εφημερίδων της Εθνικής Βιβλιοθήκης. Περιλαμβάνει μεταξύ άλλων: Ακρόπολη (1883-1884), Εμπρός (1896-1969), Ταχυδρόμος (1958-1977).
- http://e-library.iep.edu.gr/iep/index.html η συλλογή σχολικών βιβλίων του Παιδαγωγικού Ινστιτούτου.
- https://www.searchculture.gr/ ο εθνικός συσσωρευτής ελληνικού πολιτισμικού περιεχομένου του ΕΚΤ (βλ. ενότητα 3.4.2.2).

### 3.3.2 Ψηφιακές Πολιτισμικές Συλλογές

Πώς συνδέονται οι ψηφιακές συλλογές πολιτισμικής κληρονομιάς[36] με την Ιστορία και την ιστορική εκπαίδευση;

Η πολιτισμική κληρονομιά (υλική και άυλη) ανέκαθεν συνδεόταν με την ιστορική εκπαίδευση. Με την ίδια λογική, η ψηφιακή πολιτισμική κληρονομιά είναι δυνατό να χρησιμεύσει ως όχημα για την κατανόηση ιστορικών εννοιών όπως η αιτιότητα, η συνέχεια, η αλλαγή.

Καθένας μπορεί να είναι ιστορικός στοχαστής με την ευρεία έννοια. Τα προγράμματα Ψηφιακής Ιστορίας και εμπλουτισμού ψηφιακών συλλογών με πληθοπορισμό συχνά καλούν το ευρύ κοινό να αναζητήσει τεκμήρια του παρελθόντος (φωτογραφίες, παλιά ημερολόγια, γράμματα, κάθε είδους χειρόγραφο ή άλλο αντικείμενο) ξεχασμένα στο πατάρι ή στο σεντούκι της γιαγιάς. Αξιοποιώντας τέτοιου είδους ανοικτές προσκλήσεις το ευρύ κοινό αναπτύσσει όχι μόνο συνείδηση της πολιτισμικής κληρονομιάς, αλλά και την ιστορική συνείδηση. Η λεγόμενη «Επιστήμη των Πολιτών» (citizen science) αποτελεί πεδίο αξιοποιήσιμο για την ιστορική εκπαίδευση στο πλαίσιο της καλλιέργειας ιστορικής συνείδησης και ενσυναίσθησης.

> «Στο παρελθόν η επιστήμη ήταν θέμα μιας μικρής ομάδας ατόμων, των εξειδικευμένων επιστημόνων και λάμβανε χώρα σε ειδικά διαμορφωμένα εργαστήρια, ενώ το ευρύτερο κοινό μόνο επιφανειακά ή ίσως και καθόλου μπορούσε να κατανοήσει τις έννοιες και τη διαδικασία. […] Η Επιστήμη των Πολιτών (Citizen Science), όπως ονομάζεται η νέα τάση, αποτελεί ένα παγκόσμιο φαινόμενο που αναφέρεται στη συμβολή των πολιτών στην επιστημονική γνώση και πληροφορία. Συμπεριλαμβάνει ένα ευρύ πεδίο δραστηριοτήτων, από την κινητοποίηση των πολιτών για τη συλλογή δεδομένων έως τη συμμετοχή εκπαιδευμένων εθελοντών στην ερμηνεία δεδομένων, ή ακόμα και την πλήρη συμμετοχή στη διαμόρφωση πολιτικών για την επιστήμη.

[36] Ο όρος «πολιτισμική κληρονομιά» είναι ευρύτερος του όρου «πολιτιστική κληρονομιά», καθώς περιλαμβάνει κάθε είδους υλική (κινητή ή μη) και άυλη έκφανση ή εκδήλωση του πολιτισμού μιας ομάδας, κοινότητας σε μια δεδομένη χρονική περίοδο.

Το ενδιαφέρον του κοινού για την Επιστήμη των Πολιτών έχει αυξηθεί ραγδαια τα τελευταία χρόνια, είτε για τη διεξαγωγή ερευνών, που διαφορετικά θα ήταν αδύνατον να υλοποιηθούν, είτε για τη βελτίωση της διαδικασίας λήψης αποφάσεων για πολιτικές βασισμένες σε στοιχεία (evidence based policies). Ταυτόχρονα, δίνεται η μοναδική ευκαιρία για την ανάπτυξη και βιωματική εμπειρία σε επιστημονικά πεδία που μπορεί να ενδιαφέρουν το εκάστοτε κοινό, με αποτέλεσμα να αναπτύσσεται απο κοινού η κατανόηση των κρυφών προβλημάτων και των πιθανών λύσεων.»[37]

Είναι ανάγκη (και ευκαιρία) να αναπτυχθούν μέθοδοι για την εκμετάλλευση ψηφιακών πηγών. Η ανάγνωση των πηγών είναι πάντα ένα σημαντικό στοιχείο της εργαλειοθήκης του ιστορικού και οι απλές αναζητήσεις λέξεων-κλειδιών παρέχουν έναν τρόπο προσπέλασης πολύ μεγάλου όγκου υλικού επιτρέποντας την ταχεία αναγνώριση των αποσπασμάτων που αξίζουν προσεκτικότερη ανάγνωση. Ωστόσο, απαιτούνται και άλλες προσεγγίσεις, όπως οι παρακάτω μέθοδοι που αφορούν στην ποσοτική ανάλυση των δεδομένων: n-gram, corpus linguistics, distant reading, network analysis (ανάλυση δικτύων), GIS. Καθώς οι ιστορικοί ασχολούνται πάντα με την ανάλυση και την κατανόηση των κειμένων, το να μπορούν να αναπτύξουν ή να εφαρμόσουν τεχνικές που θα τους βοηθήσουν σε αυτό, τους είναι ιδιαίτερα χρήσιμο, σε έναν κόσμο με όλο και μεγαλύτερη ποσότητα κειμένων σε ψηφιακή μορφή.

Η ψηφιακή ιστορία είναι μια προσέγγιση για την έρευνα και την ερμηνεία του παρελθόντος που βασίζεται σε τεχνολογίες υπολογιστών και επικοινωνιών για τη συλλογή, ποσοτικοποίηση, ερμηνεία και ανοικτή πρόσβαση και χρήση ιστορικού υλικού και αφηγήσεων τόσο από τους ειδικούς, όσο και από το ευρύ κοινό, για την έρευνα, την διδασκαλία-μάθηση και την ψυχαγωγία. Δίνει τη δυνατότητα σε άτομα και οργανισμούς να συμμετέχουν ενεργά στη διατήρηση και την αφήγηση ιστοριών από το παρελθόν και ξεκλειδώνει μοτίβα ενσωματωμένα σε διάφορα σώματα πηγών που χωρίς το ψηφιακό εργαλείο ή μέσο δεν θα ήταν ορατά, ώστε να μελετηθούν. Το να γίνει η τεχνολογία αναπόσπαστο στοιχείο της τέχνης του ιστορικού ανοίγει νέους τρόπους ανάλυσης των πιθανών μοτίβων που αναδύονται καθώς τα δεδομένα μελετώντας με τρόπους, εργαλεία και μέσα που επιτρέπουν την οπτικοποίησή τους, εμπλουτίζοντας έτσι την ιστορική έρευνα.

#### *3.4.2.1 Europeana*

Με την ενίσχυση της ΕΕ, από το 2008, η Europeana αποτελεί τον ευρωπαϊκό διαδικτυακό συσσωρευτή πολιτισμικού περιεχομένου (cultural content aggregator) συγκεντρώνοντας μεταδεδομένα από περισσότερα από 3000 ιδρύματα από όλα τα κράτη μέλη της ΕΕ. Αποστολή της Europeana είναι να ενισχύσει τον ψηφιακό μετασχηματισμό του τομέα της πολιτιστικής κληρονομιάς.

Όπως διαβάζουμε στον ιστότοπό της:

---

[37] Βλ. σχετικό άρθρο στον Ιστότοπο του ΕΚΤ με τίτλο «Νέα Πλατφόρμα για την Επιστήμη των Πολιτών από το Κοινό Κέντρο Ερευνών της Ε»Ε https://www.ekt.gr/el/news/21435, καθώς και το Αφιέρωμα του περιοδικού του ΕΚΤ «Καινοτομία, Έρευνα και Ψηφιακή Οικονομία» στην Επιστήμη των Πολιτών: Οι πολίτες παίρνουν την επιστήμη στα χέρια τους Αφιέρωμα, Τεύχος 104, Ιούνιος 2016-Αύγουστος 2016 http://www.ekt.gr/el/magazines/features/20429

«Αποστολή:

Αναπτύσσουμε εξειδίκευση, εργαλεία και πολιτικές για την υιοθέτηση της ψηφιακής αλλαγής και την ενθάρρυνση συνεργασιών που προωθούν την καινοτομία. Διευκολύνουμε τους ανθρώπους να χρησιμοποιούν την πολιτιστική κληρονομιά για εκπαίδευση, έρευνα, δημιουργία και αναψυχή. Το έργο μας συμβάλλει σε μια ανοιχτή, γεμάτη γνώση και δημιουργική κοινωνία.

Όραμα:

Η Europeana οραματίζεται έναν τομέα πολιτιστικής κληρονομιάς που τροφοδοτείται από τις ψηφιακές τεχνολογίες και μια Ευρώπη που τροφοδοτείται από τον πολιτισμό, προσφέροντάς του μια ανθεκτική, αναπτυσσόμενη οικονομία, αυξημένη απασχόληση, βελτιωμένη ευημερία και αίσθηση της ευρωπαϊκής ταυτότητας.» [38]

Στην πράξη, η Europeana συγκεντρώνει από τα ιδρύματα μνήμης ψηφιακά τεκμήρια σε μορφή επεξεργάσιμη από τα λογισμικά των υπολογιστών που ακολουθεί το ίδιο μοντέλο δεδομένων EDM (Europeana Data Model),[39] ώστε να είναι συμβατά μεταξύ τους, εξασφαλίζοντας τη διαλειτουργικότητά τους.

Εθνικοί συσσωρευτές από κάθε χώρα-μέλος της ΕΕ συγκεντρώνουν τα τεκμήρια και μαζί με τα μεταδεδομένα τους τα διαθέτουν και μέσω του ιστοχώρου της Europeana ή σε θεματικές ψηφιακές εκθέσεις που κατά καιρούς οργανώνει η Europeana. Χιλιάδες ευρωπαϊκά αρχεία, βιβλιοθήκες και μουσεία συνεργάζονται με την Europeana μέσω των συσσωρευτών πολιτισμικού περιεχομένου για να κάνουν τις συλλογές τους ανοιχτά προσβάσιμες σε χρήστες σε όλο τον κόσμο.

Η κοινή χρήση των δεδομένων μιας συλλογής με την Europeana εξασφαλίζει πρόσβαση σε μεγάλη μερίδα κοινού και σε ειδικούς στη μοντελοποίηση δεδομένων, σε ζητήματα πνευματικών δικαιωμάτων και αδειοδότησης των τεκμηρίων.

Σε αντίθεση με την πρώτη γενεά ψηφιακών συλλογών και αρχείων της πρώτης δεκαετίας του 21$^{ου}$ αιώνα, η Europeana προωθεί την επαναχρησιμοποίηση του ψηφιακού υλικού, τόσο από τους ανθρώπους-χρήστες (το ευρύ και το ακαδημαϊκό κοινό), αλλά και από τα λογισμικά των υπολογιστών. Για τον λόγο αυτό, τα τεκμήρια των συλλογών που περιλαμβάνει είναι σε μορφότυπους εναρμονισμένους προς τις απαιτήσεις της δημοσίευσης δεδομένων στην Τρίτη γενιά του ιστού, που δεν βασίζεται στο υπερκείμενο και στους υπερσυνδέσμους αλλά στο μοντέλο των Ανοικτών Συνδεδεμένων Δεδομένων.

[38] Europeana Mission https://pro.europeana.eu/about-us/mission

[39] Βλ. το μοντέλο Δεδομένων της Europeana EDM εδώ: https://pro.europeana.eu/files/Europeana_Professional/Share_your_data/Technical_requirements/EDM_Documentation/EDM_Mapping_Guidelines_v2.4_102017.pdf

### *3.4.2.2 SearchCulture.gr: η ψηφιακή υποδομή του Εθνικού Κέντρου Τεκμηρίωσης*

Στην Ελλάδα, υπεύθυνος φορέας για τη συγκέντρωση των τεκμηρίων από τους ελληνικούς φορείς και την δημοσίευση των μεταδεδομένων τους σύμφωνα με τις προδιαγραφές τεκμηρίωσης της Europeana είναι το Εθνικό Κέντρο Τεκμηρίωσης και Ηλεκτρονικού Περιεχομένου (ΕΚΤ), που έχει διαθέσει πάνω από 320.000 τεκμήρια στη Europeana. Εδώ και περίπου ενάμιση χρόνο το ΕΚΤ αποτελεί διαπιστευμένο εθνικού συσσωρευτή πολιτιστικών δεδομένων για τη Europeana μέσω της υποδομής SearchCulture.gr[40] και των σημασιολογικών λεξιλογίων Semantics.gr[41] τα οποία αποτελούν ένα πιλοτικό σύστημα που ανέπτυξε το ΕΚΤ. Το σύστημα αυτό διαθέτει ένα εργαλείο σημασιολογικού εμπλουτισμού που μπορεί να χρησιμοποιηθεί για τον εμπλουτισμό και την ομογενοποίηση των μεταδεδομένων διαφορετικών φορέων που μπορούν έτσι να καθιερώνουν εύκολα τα δικά τους λεξιλόγια όρων. Τα λεξιλόγια αυτά διατίθενται ελεύθερα ως Ανοικτά Διασυνδεμένα Δεδομένα (Linked Open Data). Στην ΕΙΚΟΝΑ 4 φαίνεται παράδειγμα τεκμηρίου που προέρχεται από τη.συλλογή του Κέντρου Ερεύνης της Ιστορίας του Ελληνικού Δικαίου της Ακαδημίας Αθηνών αναρτημένο στον ιστότοπο του SearchCulture (πάνω δεξιά) και εκφρασμένο διαλειτουργικά σύμφωνα με το πρότυπο EDM της Europeana.

**ΕΙΚΟΝΑ 4** *Ψηφιακό τεκμήριο ως ιστοσελίδα (πάνω δεξιά) και ως διαλειτουργικά δεδομένα-SearchCulture-ΕΚΤ.*

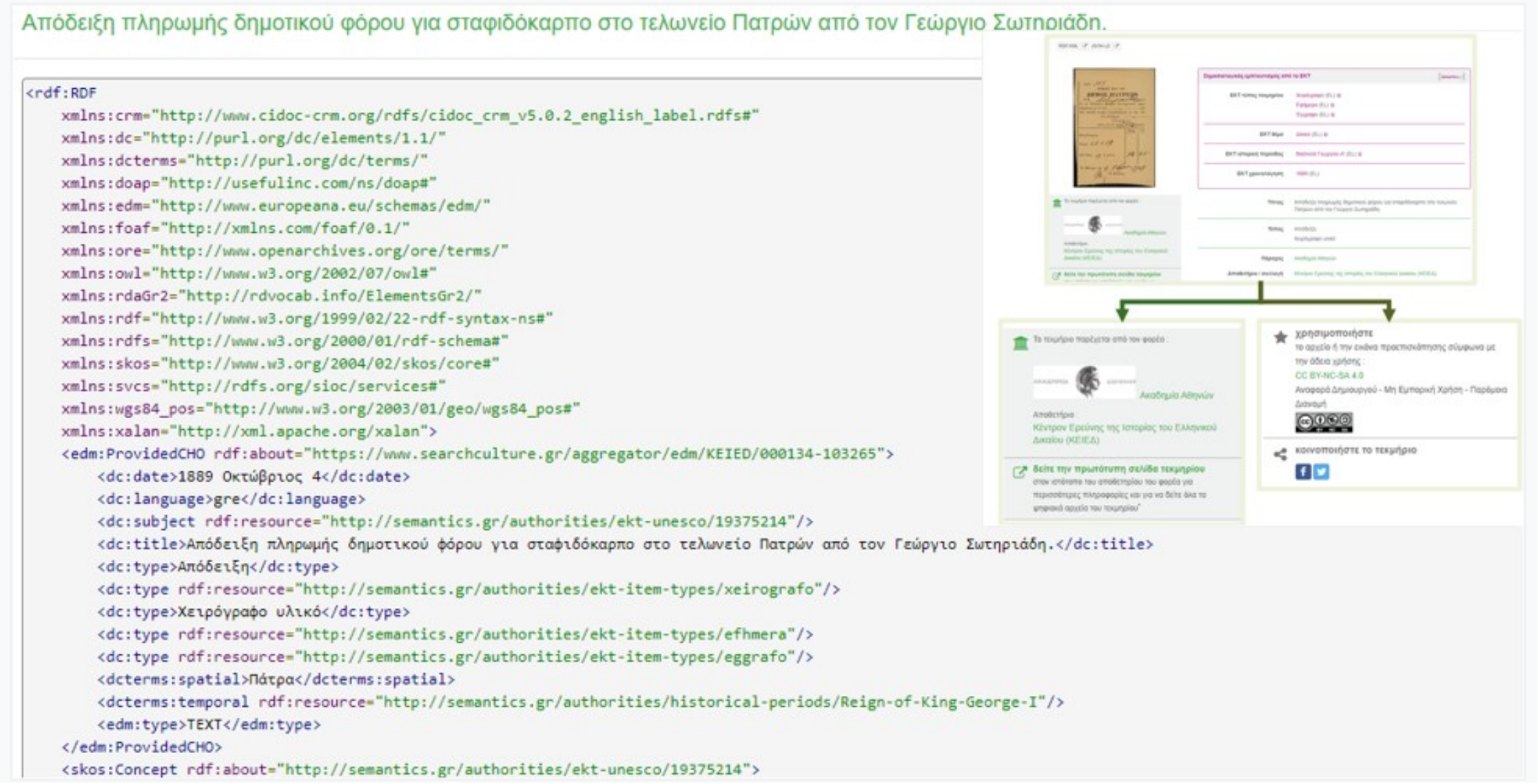

Η Ιστορία εκδημοκρατίζεται, γίνεται αντικείμενο πληθοπορισμού από το ευρύ κοινό και συναντά τον κόσμο των ψηφιακών πολιτισμικών συλλογών. Χαρακτηριστικό παράδειγμα είναι η ανοικτή πρόσκληση για ελληνική συνεισφορά (μέσω πληθοπορισμού) στη θεματική ψηφιακή συλλογή που ετοιμάζει η Europeana.[42] Την πρόσκληση απηύθυνε το 2019 το ΕΚΤ σε φορείς και πολίτες «να μοιραστούν την ιστορία τους και να συνεισφέρουν φωτογραφίες

---

[40] SearchCulture https://www.searchculture.gr/

[41] Semantics.gr https://www.semantics.gr/

[42] Βλ. τον κατάλογο των εθνικών συσσωρευτών (μεταξύ των οποίων και το ΕΚΤ) εδώ https://pro.europeana.eu/page/aggregators?utm_source=share-your-data%2Fprocess&utm_medium=Find%20an%20aggregator&utm_campaign=internal_link

και σχετικό αρχειακό υλικό από την Ελλάδα στο πανευρωπαϊκό την Βιομηχανική Επανάσταση της Europeana.[43]

## 4. Ο Ρόλος του/της Ιστορικού στην Ψηφιακή Εποχή

Πριν την έλευση της ψηφιακής εποχής, οι ιστορικοί βασίζονταν κυρίως σε επαγγελματίες αρχειονόμους, βιβλιοθηκονόμους ή επιμελητές μουσείων για να αποκτήσουν πρόσβαση στα ίχνη του παρελθόντος. Σήμερα, οι ιστορικοί έρχονται αντιμέτωποι με ψηφιακές πρωτογενείς πηγές που, διατεταγμένες κατά ‘συνάφεια’ με τη βοήθεια αλγορίθμων, διατίθενται από ένα ευρύ φάσμα φορέων και ατόμων διαμεσολαβώντας το παρελθόν στον κυβερνοχώρο. Απολαμβάνουν άνευ προηγουμένου πρόσβαση σε μια σειρά πρωτογενών πηγών, ελεύθερα προσβάσιμων ή όχι και έχουν τη δυνατότητα να συγκεντρώσει και να εξετάσει μεγάλο όγκο από αυτό τό ψηφιακό υλικό.

Το Διαδίκτυο αποτελεί πλέον το παγκόσμιο ψηφιακό αρχείο της ανθρώπινης ιστορίας, παγκόσμιας, εθνικής, τοπικής, προφορικής, δημόσιας, προερχόμενης από πληθοπορισμό (crowdsourcing) ή από τις επίσημες πηγές και κρατικά αρχεία. Παρόλο που η ψηφιοποίηση και η παρουσίαση ιστορικών πηγών στο διαδίκτυο είναι η μακροβιότερη μορφή ψηφιακής ιστορίας, συνηθισμένο λάθος κατά τη χρήση ψηφιακών υποκατάστατων των πρωτογενών πηγών είναι να γίνεται αναφορά μόνο την ιστοσελίδα στην οποία βρίσκεται μια ψηφιοποιημένη πρωτεύουσα πηγή και όχι και στο αρχικό χαρτώο τεκμήριο. Η αδυναμία να ληφθεί υπόψη η διαφορά μεταξύ ψηφιοποιημένου αντιγράφου και πρωτοτύπου δείχνει πως η μετάβαση από το αναλογικό στο ψηφιακό γίνεται, πολλές φορές, ασυναίσθητα.

Η συγκρότηση της ιστορικής μνήμης δεν παύει να είναι επιλεκτική είτε το περιβάλλον είναι αναλογικό είτε ψηφιακό. Όταν μια ψηφιακή συλλογή αναπαράγει μια μοναδική συλλογή αρχείων, η επιλογή αυτής της συλλογής είναι ένα επιχείρημα υπέρ της σημασίας της και της σημασίας της εξέτασής της. Το ίδιο ισχύει, εάν η ψηφιακή συλλογή συγκεντρώνεται από πολλά αρχεία και συλλογές. Το ψηφιακό αρχείο μπορεί να κάνει ορατές ορισμένες ομάδες και άτομα και άρα να θέτει υπό το πέπλο της ανωνυμίας κάποιες άλλες ομάδες ή άτομα. Όταν ο/η ιστορικός επιλέγει το υλικό (αρχειακό ή άλλο) που θα χρησιμοποιήσει διατυπώνει έμμεσα ένα επιχείρημα για το ποιες πηγές είναι σημαντικές για την κατανόηση ενός θέματος (και ποιες όχι). Ο τρόπος οργάνωσης, κατηγοριοποίησης και περιγραφής των πηγών, καθώς και ο σχεδιασμός μιας διεπαφής για την παρουσίασή τους και την πρόσβαση σε αυτές είναι μια ‘θέση’ (υπέρ ή κατά) που παίρνει ο/η ιστορικός. Η πράξη παρέμβασης σε μια συλλογή μέσω π.χ. του ψηφιακού εμπλουτισμού των μεταδεδομένων της συλλογής με τη χρήση μια γλώσσας επισημείωσης, (markup language),[44] της σχεδίασης ενός μοντέλου μηχανικής μάθησης που βασίζεται σε γνωστούς αλγορίθμους (π.χ. word2vec), ή του σχεδιασμού μιας

---

[43] Βλ. «Η Ιστορία της Εργασίας και η βιομηχανική ευρωπαϊκή κληρονομιά στο νέο αφιέρωμα της Europeana “Europe at Work”» εδώ: http://www.ekt.gr/el/news/23443 (26/09/2019)

[44] Π.χ. η γλώσσα RDF (Resource Description Framework) η οποία χρησιμοποιείται για τον σημασιολογικό εμπλουτισμό ψηφιακών πόρων. Τις προδιαγραφές της γλώσσας αυτής έχει δημοσιεύσει η Επιτροπή Παγκόσμιου Ιστού (W3C).

διεπαφής (interface) αποτελούν βασικές πτυχές που επηρεάζουν την ιστορική σκέψη και ερμηνεία.

## 4.1 Μέθοδοι και πρακτικές

Η ταξινόμηση των ερευνητικών πρακτικών των (Ψηφιακών) Ανθρωπιστικών Επιστημών παρουσιάζει ενδιαφέρουν για την Ψηφιακή Ιστορία και τον τρόπο διδασκαλίας της. Στόχος της είναι να κατανοηθεί καλύτερα ως διαδικασία και σύνθετη νοητική διεργασία προκειμένου να αναπτυχθούν εργαλεία που να διευκολύνουν τους ερευνητές και σπουδαστές των ανθρωπιστικών πεδίων.να εργασθούν σε διαδικτυακό περιβάλλον.

Σύμφωνα με τον John Unsworth (Unsworth 2000),[45] κοινές πρωτογενείς μέθοδοι στις Ανθρωπιστικές Επιστήμες είναι οι εξής: ανακάλυψη, επισημείωση, σύγκριση, αναφορά, δειγματοληψία, επεξήγηση, αναπαράσταση (Discovering, Annotating, Comparing, Referring, Sampling, Illustrating, and Representing). Σύμφωνα με τους Palmer κ. ά. (2009)[46] είναι οι ακόλουθες:

1. Αναζητώ (Searching)
    - 1.1 Εκτελώ μια Αναζήτηση Απευθείας Direct searching
    - 1.2 Εκτελώ Αλυσίδα Αναζητήσεων Chaining
    - 1.3 Ξεφυλλίζω Browsing
    - 1.4 Διερευνώ Probing
    - 1.5 Αποκτώ πρόσβαση Accessing
2. Συλλέγω Collecting
    - 2.1 Συγκεντρώνω Gathering
    - 2.2 Οργανώνω Organizing
3. Διαβάζω Reading
    - 3.1 Διαβάζω 'λοξά' Scanning
    - 3.2 Αξιολογώ Assessing
    - 3.3 Ξαναδιαβάζω Rereading
4. Συγγράφω Writing
    - 4.1 Συναρμολογώ Assembling
    - 4.2 Συγγράφω ως μέλος συγγραφικής ομάδας Co-authoring
    - 4.3 Διαχέω Disseminating
5. Συνεργάζομαι Collaborating
    - 5.1 Συντονίζω Coordinating
    - 5.2 Δικτυώνομαι Networking
    - 5.3 Συμβουλεύω-ομαι Consulting
6. Εγκάρσιες πρωτογενείς δραστηριότητες Cross-cutting Primitives
    - 6.1 Παρακολουθώ Monitoring
    - 6.2 Κρατώ σημειώσεις Notetaking
    - 6.3 Μεταφράζω Translating

---

[45] Unsworth, J. Scholarly Primitives: what methods do humanities researchers have in common, and how might our tools reflect this? Symposium on Humanities Computing: formal methods, experimental practice sponsored by King's College, London, 13 May 2000. http://www.iath.virginia.edu/~jmu2m/Kings.5-00/primitives.html.

[46] Palmer, C.L., L.C. Teffeau and C.M. Pirmann. Scholarly Information Practices in the Online Environment: Themes from the Literature and Implications for Library Service Development. 2009. www.oclc.org/programs/publications/reports/2009-02.pdf.

6.4 Πρακτικές για τα δεδομένα Data Practices

Ανάλογη προσπάθεια αποτελεί η TaDiRAH - Taxonomy of Digital Research Activities in the Humanities.[47] Αποτελεί πρωτοβουλία της ευρωπαϊκής υποδομής για τις Ανθρωπιστικές Επιστήμες Dariah[48] και στοχεύει στο να διευκολύνει τη δόμηση και εύρεση πληροφοριών σχετικά με τις μεθόδους των Ψηφιακών Ανθρωπιστικών Επιστημών.

## 4.2 Ο «πάγκος εργασίας» του Ιστορικού

Ενώ η Ιστορία και οι Ανθρωπιστικές Επιστήμες, γενικότερα, χρησιμοποιούν μεθόδους ποιοτικές, όπως η ερμηνευτική του κειμένου και η ανάγνωση εκ του σύνεγγυς (close reading), η μετάβασή τους στο ψηφιακό περιβάλλον ευνοεί τη χρήση, εκτός των ποιοτικών, και ποσοτικών μεθόδων. Μερικές από τις πιο συνηθισμένες παρουσιάζονται συνοπτικά στην ενότητα αυτή.

### 4.2.1 Ο Ιστορικός-Προγραμματιστής (Programming Historian)

Ο ιστότοπος Programming Historian (Ιστορικός-Προγραμματιστής)[49] είναι μια πρωτοβουλία στο πεδίο των Ψηφιακών Ανθρωπιστικών Επιστημών και μεθοδολογίας ψηφιακής ιστορίας. που σκοπό έχει τη δημοσίευση σεμιναριακών μαθημάτων (tutorials) που βοηθούν τους ανθρωπιστές να μάθουν ένα ευρύ φάσμα ψηφιακών εργαλείων, τεχνικών και ροών εργασίας για τη διευκόλυνση της έρευνας και της διδασκαλίας. Περιλαμβάνει μαθήματα κυρίως στην Αγγλική γλώσσα, αλλά και στη Γαλλική και Ισπανική.

Τα μαθήματα είναι ταξινομημένα κατά δυσκολία. Μπορεί κανείς να ενταχθεί στην κοινότητα, συνεισφέροντας, μεταφράζοντας ή ζητώντας νέο μάθημα σε τεχνικό ζήτημα που κάποιος άλλος γνωρίζει και είναι σε θέση να το προσφέρει εθελοντικά στην κοινότητα με τη μορφή ενός επεξηγηματικού κειμένου που, στη συνέχεια, θα αναρτηθεί στον ιστότοπο.

Τα περισσότερα από τα υπάρχοντα μαθήματα αφορούν σε ποικίλα τεχνικά θέματα όπως η διαχείριση των δεδομένων (data manipulation),[50] η χρήση γλωσσών προγραμματισμού όπως η Python, η εκ του μακρόθεν ανάγνωση (distant reading) (βλ. ενότητα 4.2.3), η ανάλυση Κοινωνικών Δικτύων (Social Network Analysis) (βλ. ενότητα 4.2.3).

[47] TaDiRAH - Taxonomy of Digital Research Activities in the Humanities http://tadirah.dariah.eu/vocab/index.php?setLang=en, https://github.com/dhtaxonomy

[48] Dariah https://www.dariah.eu/

[49] The Programming Historian https://programminghistorian.org/

[50] Ο όρος data manipulation αφορά στη μετατροπή των δεδομένων ώστε να είναι πιο εύκολο να διαβαστούν ή να δομηθούν, ώστε να είναι αξιοποιήσιμα από τους υπολογιστές.

### 4.2.2 Αλγοριθμικές προσεγγίσεις

Η γλώσσα προγραμματισμού Python είναι ανοικτής πρόσβασης (open source)[51] και χρησιμοποιείται ευρέως στις Ψηφιακές Ανθρωπιστικές Επιστήμες, καθώς αποτελεί εξαιρετική επιλογή για κειμενικά δεδομένα. Βασικές γνώσεις και δεξιότητες προγραμματισμού στη γλώσσα Python για τον/την ιστορικό περιλαμβάνουν τις εξής:

- Βασικές προγραμματιστικές τεχνικές της Python.
- Επεξεργασία κειμένου (text processing): 'καθαρισμός' του κειμένου από τις πιο κοινές λέξεις (stop words), όπως π.χ. ο σύνδεσμος και, τα άρθρα και υπολογισμός της κατανομής συχνότητας των λέξεων.
- Προκαταρκτική επεξεργασία κειμένου (text preprocessing).
- Χρήση βιβλιοθηκών, λ.χ. NLTK (Natural Language Toolkit) https://www.nltk.org/.

Όπως η Python, έτσι και η γλώσσα προγραμματισμού R είναι open source και χρησιμοποιείται ευρέως στην Επιστήμη των δεδομένων (Data science). Τα τελευταία χρόνια έχει χρησιμοποιηθεί στο πεδίο της Ψηφιακής Ιστορίας.[52]

### 4.2.3 Μοντελοποίηση κειμένου

Η μοντελοποίηση θεμάτων κειμένου (topic modelling) έχει γίνει δημοφιλής διαδικασία για τη συγκέντρωση εγγράφων (document clustering) σε θεματικές ομάδες. Μια εφαρμογή θεματικής μοντελοποίησης κειμένου με ροή εργασίας φιλική προς το χρήστη είναι προσβάσιμη από τον ιστότοπο της ευρωπαϊκής υποδομής για τις Ανθρωπιστικές Επιστήμες DARIAH-DE.[53]

### 4.2.4 Απομακρυσμένη Ανάγνωση

Η απομακρυσμένη ανάγνωση[54] είναι μια μέθοδος λογοτεχνικής κριτικής που χρησιμοποιεί τεχνικές υπολογιστικής ανάλυσης δεδομένων για τον προσδιορισμό σημαντικών προτύπων σε μεγάλες συλλογές κειμένων. Σε αντίθεση με την ανάγνωση εκ του σύνεγγις (close reading), το αντικείμενο της ανάλυσης είναι συχνά μια συλλογή εκατοντάδων ή χιλιάδων ψηφιοποιημένων κειμένων. Ένα από διαθέσιμα εργαλεία απομακρυσμένης ανάγνωσης είναι το Ngram της Google με το οποίο μπορεί κανείς να πραγματοποιήσει συγκριτικές αναζητήσεις στατιστικής συχνότητας στην εμφάνιση ως πέντε λέξεων στις ακόλουθες γλώσσες: Αγγλικά, Γαλλικά, Γερμανικά, Εβραϊκά, Ρωσικά, Ισπανικά. Τα δεδομένα που επιστρέφονται αντλούνται από τα ψηφιοποιημένα βιβλία Google Books. Το χρονικό

---

[51] Περισσότερα για την Ανοικτή πρόσβαση βλ. http://www.openaccess.gr/el/oa-about

[52] Arnold, Taylor, Lauren Tilton. *Humanities Data in R*. Springer, 2015. http://link.springer.com/10.1007/978-3-319-20702-5.

Lincoln A. Mullen Computational Historical Thinking With Applications in R, 2018, προσβάσιμο εδώ: https://dh-r.lincolnmullen.com/

[53] DARIAH-DE https://dariah-de.github.io/TopicsExplorer/#the%20sample%20corpus

[54] Ο όρος ανήκει στον F. Moretti, καθηγητή Ιστορίας της Λογοτεχνίας του Παν/μίου Stanford: https://english.stanford.edu/people/franco-moretti

διάστημα που καλύπτεται είναι από το 1800 ως σήμερα. Στην ΕΙΚΟΝΑ 5 φαίνεται η συγκριτική αναζήτηση αναφορών για τους ποιητές Καβάφη, Σεφέρη, Ελύτη » με το εργαλείο Google Ngram Viewer (με τη ρύθμιση Smoothing 3, δηλαδή με υπολογισμούς μέσων όρων ±ετών για κάθε έτος).

**ΕΙΚΟΝΑ 5** *Αποτελέσματα αναζήτησης των όρων «Cavafy, Seferis, Elytis*

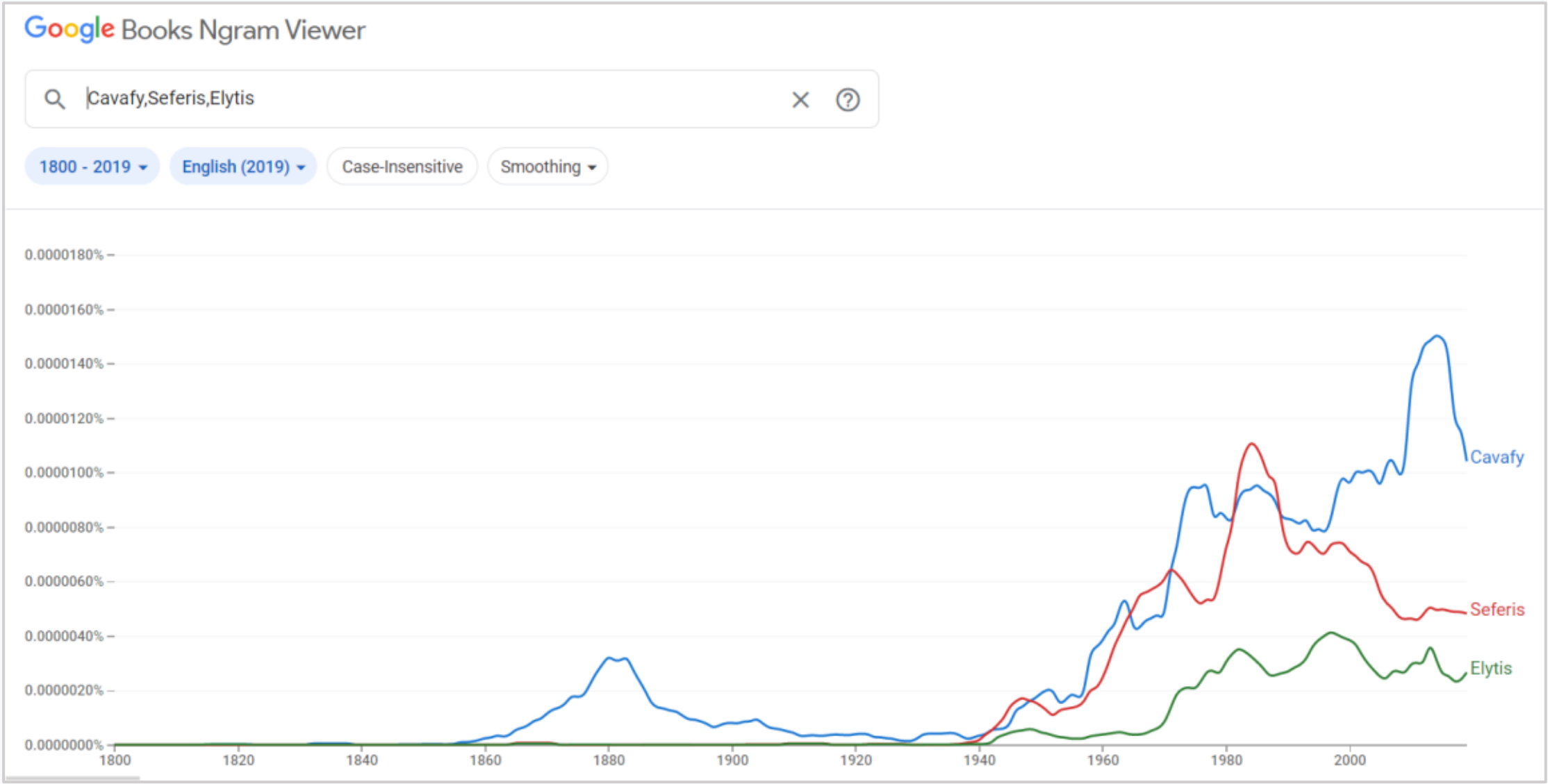

Η απομακρυσμένη ανάγνωση για την ευρωπαϊκή λογοτεχνική ιστορία (COST Action CA16204)[55] είναι ένα ερευνητικό έργο που στοχεύει στη δημιουργία σώματος κειμένων (corpus) που θα επιτρέπει καινοτόμες υπολογιστικές μεθόδους ανάλυσης λογοτεχνικού κειμένου σε τουλάχιστον 10 ευρωπαϊκές γλώσσες (συμπεριλαμβανομένης της Νέας Ελληνικής).

### 4.2.5 Ανάλυση Κοινωνικών Δικτύων

Η Ανάλυση Κοινωνικών Δικτύων μελετά άτομα (αποτελούν τις κορυφές ή κόμβους του δικτύου) και τις σχέσεις μεταξύ αυτών των ατόμων (αποτελούν τους συνδέσμους ή δεσμούς μετα ξύ των ατόμων του δικτύου).

Κατά την τελευταία δεκαετία, το πεδίο της αρχαίας ιστορίας και των κλασικών σπουδών γνώρισε μια αργή αλλά σταθερή αύξηση των δημοσιεύσεων που εφαρμόζουν στην ελληνορωμαϊκή ιστορία τις έννοιες της ανάλυσης κοινωνικού δικτύου (SNA).

Ενώ αρχικά εισήγαγε κυρίως την έννοια των δικτύων και της συνδεσιμότητας με μεταφορική έννοια, η πρόσφατη έρευνα στράφηκε όλο και περισσότερο στις πιο ποσοτικές πτυχές της ανάλυσης δικτύου χρησιμοποιώντας ψηφιακά εργαλεία ελεύθερου λογισμικού, όπως το Node XL (https://nodexl.com/) και το Gephi (https://gephi.org/).

[55] Distant Reading https://www.distant-reading.net/

Λογοτεχνικές πηγές, επιγραφές και πάπυροι προσφέρουν πλήθος πληροφοριών σχετικά με αυτοκρατορικές και άλλες ελίτ, και τους δεσμούς της οικογένειας, του γάμου, της φιλίας, της προστασίας και της δωροδοκίας που τους συνέδεαν. Όπως δείχνουν οι μελέτες περίπτωσης στο ειδικό τεύχος του περιοδικού The Journal of Historical Network Research, η SNA προσφέρει νέες προοπτικές σε ένα ερευνητικό πεδίο που κυριαρχείται από πιο παραδοσιακές προσωπογραφικές μελέτες παρέχοντας ένα ισχυρό εργαλείο για την ανάλυση και οπτικοποίηση κοινωνικών και πολιτικών συνδέσεων στις αρχαίες κοινωνίες.

Στο πεδίο της Αρχαίας Ιστορίας, πρωτοπόρες είναι οι έρευνες της Diana Harris Cline για το κοινωνικό δίκτυο του Περικλή (Harris Cline 2020a) και των κεραμέων στην Αρχαία Αθήνα (Harris Cline 2020β).[56] Στην ΕΙΚΟΝΑ 6 φαίνεται δραστηριότητα για μαθητές/τριες της Δευτεροβάθμιας εκπαίδευσης (στο μάθημα της Αρχαίας Ιστορίας ή των Αρχαίων Ελληνικών) που βασίζεται στην ανάλυση του κοινωνικού δικτύου του Περικλή από την Harris Cline.

**ΕΙΚΟΝΑ *6* Δειγματική δραστηριότητα αξιοποίησης στη Δευτεροβάθμια Εκπαίδευση των πορισμάτων της εφαρμογής της SNA (Ανάλυση Κοινωνικών Δικτύων) στην περίπτωση του Περικλή**

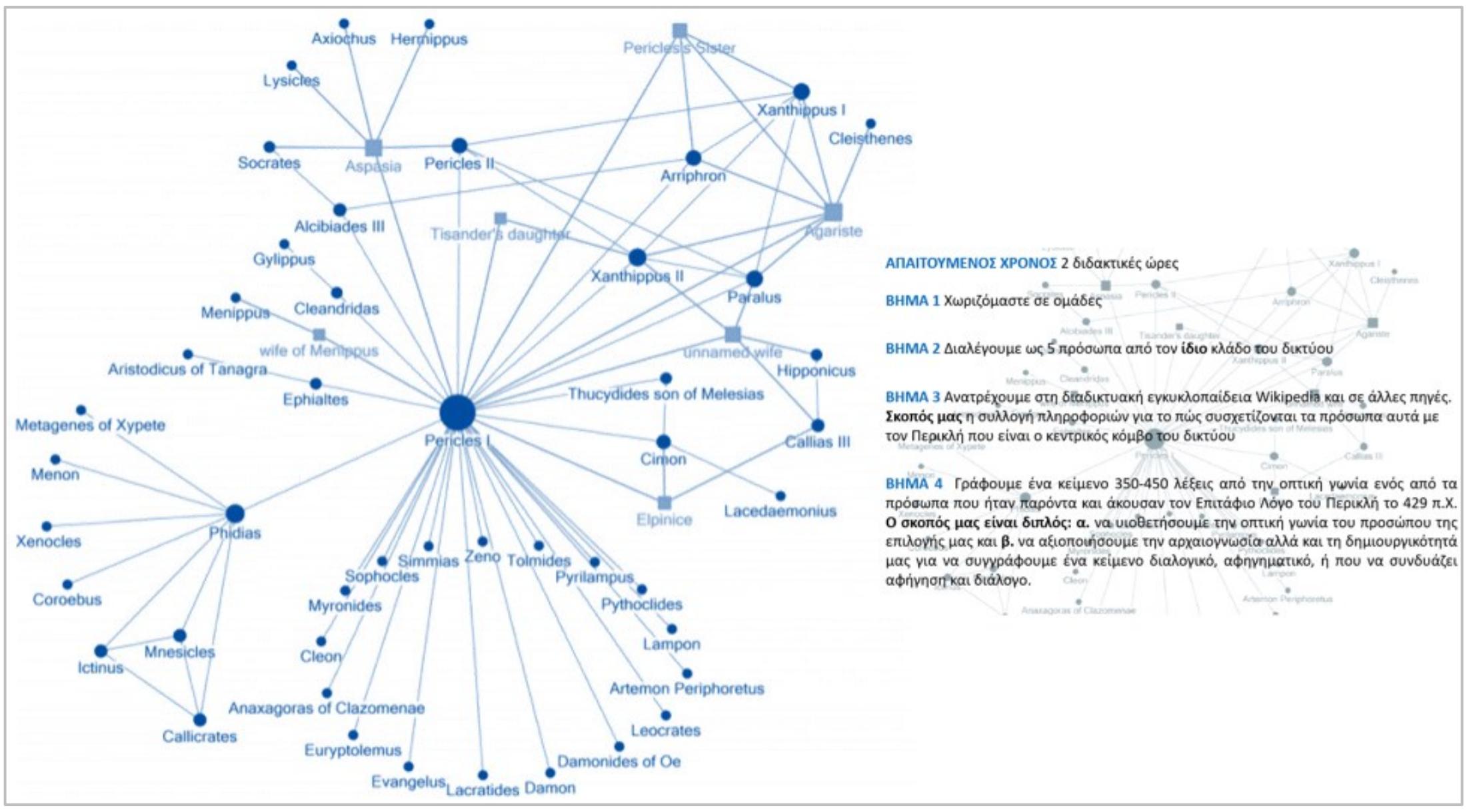

## 5. Αξιοποίηση της Ψηφιακής Ιστορίας στη Διδασκαλία & τη Μάθηση

Το βασικό υλικό της Ιστοριογραφίας του μέλλοντος πιθανότατα θα είναι κυρίως ψηφιακό, -είτε εγγενώς ψηφιακό, είτε ψηφιοποιημένο- και κάθε κατηγορία θα έχει συγκεκριμένα

---

[56] Harris Cline, Diane. 2020α. “Athens as a Small World.” Στο: *The Ties That Bind. Ancient Politics and Network Research. Journal of Historical Network Research*. Wim Broekaert, Elena Köstner, Christian Rollinger επιμ., 4:36–56. Luxembourg:Université du Luxembourg.

Harris Cline, Diane 2020β. “Beazley's Connoisseurship and Athenian Kerameikoi: A Social Network Analysis.” Στο: *Reconstructing Scales of Production in the Ancient Greek World: Producers, Processes, Products, People. Proceedings of the XIX Conference of Classical Archaeology, Archaeology and Economy in the Ancient World, Bonn 22-26 May 2018* edited by Eleni Hasaki and Martin Bentz, 59-80. Heidelberg: Propylaeum.

χαρακτηριστικά και ευρετικές ανάγκες. Οι σπουδαστές της Ψηφιακής Ιστορίας ασκούνται στο να προσεγγίσουν τα ψηφιακά ιστορικά έργα κριτικά συνειδητοποιώντας ότι κάθε απόφαση που αφορά τον ψηφιακό σχεδιασμό και το περιεχόμενο είναι μέρος της ψηφιακής ιστορικής ερμηνείας.

Η Ψηφιακή Ιστορία δεν διαφέρει από την (αναλογική) Ιστορία σε ένα βασικό σημείο: η ανάλυση της ψηφιακής ιστορικής αναπαράστασης απαιτεί ιστοριογραφική γνώση, παρόμοια με αυτή που απατείται για την συγγραφή και ανάλυση των ιστορικών επιστημονικών μονογραφιών.

Οι τρόποι διδασκαλίας και μάθησης έχουν αλλάξει με την αυξημένη χρήση ψηφιακών τεχνολογιών, ειδικά στις ανθρωπιστικές επιστήμες. Η ψηφιακή παιδαγωγική περιλαμβάνει ένα πολύ ευρύ φάσμα δραστηριοτήτων:

- διδασκαλία του πώς να είναι σε θέση οι μαθητές/τριες και φοιτητές/τριες να αλληλεπιδρούν με μεγάλα σύνολα δεδομένων
- διδασκαλία στους μαθητές πώς να πλοηγούνται και να χειρίζονται αποθετήρια πληροφοριών
- διδασκαλία στους μαθητές πώς να χρησιμοποιούν ψηφιακά εργαλεία.

Τα μεθοδολογικά μέσα, θα πρέπει να είναι σε θέση να θέτουν, κυρίως, αλλά και να απαντούν σύνθετα ερωτήματα σχετικά με τα ιστορικά ‘γεγονότα’ και τον τρόπο που οι θεωρούμενοι ως βασικοί ‘παράγοντες-συντελεστές’ σχετίζονται με αυτά και, επίσης να θέτουν ερωτήματα σχετικά με την ψηφιακή τους πραγμάτωση και παρουσίαση.

Η επεξεργασία των ιστορικών ερωτημάτων συχνά ακολουθεί την εξής πορεία:
1. διατύπωση ερωτήσεων
2. συλλογή και οργάνωση στοιχείων
3. ερμηνεία και ανάλυσή τους
4. αξιολόγηση και συναγωγή συμπερασμάτων
5. ανακοίνωση-επικοινωνία τους

Οι μαθητές έχουν τη δυνατότητα ηλεκτρονικής πρόσβασης σε ιστορικές πηγές χρησιμοποιώντας αρχειακό υλικό, τόσο κείμενα όσο και εικόνες, και έτσι εξοικειώνονται και εμπλέκονται σε αυθεντικές διαδικασίες έρευνας, που τους βοηθούν να οργανώνουν και να αναπαριστούν σύνθετες έννοιες, να ‘παρακολουθούν’ τα ιστορικά γεγονότα και να πραγματοποιούν εικονικές επισκέψεις ιστορικών χώρων ασκώντας την ιστορική τους ενσυναίσθηση. Μέσω των ψηφιακών τεκμηρίων βιώνουν το κλίμα μιας εποχής και αντιλαμβάνονται ότι τα τεκμήρια αποτελούν το μέσο προσέγγισης και διαλόγου με το παρελθόν.

Καθώς μαθητές/τριες καλούνται «να συγκρίνουν», «να περιγράψουν», «να αναλύσουν», «να επεξεργαστούν», «να εξηγήσουν», «να προβλέψουν», «να δικαιολογήσουν» και να φτάσουν από την πληροφορία στην ανασύσταση της γνώσης, ανακύπτουν τα ακόλουθα (ανοιτκά) ερωτήματα:
- Πόση έμφαση δίνεται στα περιεχόμενα της μάθησης και πόση στα ψηφιακά μέσα, περιβάλλοντα, εργαλεία;

- Ποιος είναι ο ελάχιστος βαθμός ιστορικού γραμματισμού που κρίνεται αναγκαίος και ποιος ο ελάχιστος βαθμός ψηφιακού γραμματισμού;
- Πώς ευθυγραμμίζεται η μαθησιακή δραστηριότητα ψηφιακής Ιστορίας με την επιδιωκόμενη ενεργοποίηση της ιστορικής κριτικής σκέψης;
- Πώς επιτυγχάνεται ο έλεγχος της αξιοπιστίας των πηγών στο ψηφιακό περιβάλλον;

Από την άλλη πλευρά, ο εκπαιδευτικός που διδάσκει Ψηφιακή Ιστορία καλείται να μεταδώσει και να προκαλέσει την ανάκληση και κατανόηση της ιστορικής γνώσης, αλλά και να εκπαιδεύσει τους μαθητές/τριες στα ψηφιακά περιβάλλοντα και εργαλεία.

## 5.1 Δεξιότητες Ψηφιακής Ιστορίας

Η συγγραφή των σχολικών εγχειριδίων της Ιστορίας στηρίζεται στις αρχές και τη φιλοσοφία του ισχύοντος Διαθεματικού Ενιαίου Πλαισίου Προγραμμάτων Σπουδών (Δ.Ε.Π.Π.Σ) και των Αναλυτικών Προγραμμάτων Σπουδών (Α.Π.Σ) και στις σύγχρονες παιδαγωγικές θεωρίες. Η διδακτική πράξη πρέπει να υπηρετεί τους σκοπούς του μαθήματος, δηλαδή την καλλιέργεια της ιστορικής συνείδησης και της ιστορικής σκέψης των μαθητών/τριών και τον ιστορικό τους εγγραμματισμό. Κάθε διδασκαλία του ιστορικού μαθήματος με ΤΠΕ σχεδιάζεται με βάση τρεις παραμέτους:

- το λογισμικό ή γενικά εργαλείο τεχνολογίας που θα χρησιμοποιηθεί
- τις έννοιες και δεξιότητες της ιστορικής σκέψης, οι οποίες θεωρούμε ότι μπορούν να καλλιεργηθούν αποτελεσματικά με τη χρήση του αντίστοιχου εργαλείου (π.χ. κατανόηση της συνέχειας και της αλλαγής)
- τις γενικές παιδαγωγικές αρχές και θεωρίες (π.χ. η έννοια της διαφοροποιημένης διδασκαλίας), ώστε η χρήση του ψηφιακού περιβάλλοντος να μην γίνεται ούτε αυτοσκοπός, ούτε ο πρώτιστος σκοπός.

### 5.1.1 Δεξιότητες ιστορικού εγγραμματισμού

Η σύνδεση των εννοιών που ακολουθούν με το ιστορικό υλικό αποτελεί τη βάση για το σχεδιασμό διδακτικών δραστηριοτήτων:

- η ιστορική συνείδηση
- η σημαντικότητα του ιστορικού γεγονότος
- ο ιστορικός χρόνος
- ο ιστορικός χώρος
- οι άνθρωποι
- τα γεγονότα
- τα αντικείμενα
- ο ιστορικός λόγος ως κατασκευή
- η ερμηνεία του ιστορικού περιεχομένου
- η ιστορική αιτιότητα (αίτια-αφορμές-συνέπειες)
- η ιστορική αλλαγή και συνέχεια
- τα ιστορικά τεκμήρια
- η ιστορική οπτική (perspective)

Βασικές δεξιότητες της ιστορικής σκέψης είναι (ΕΙΚΟΝΑ 7):

- η διατύπωση ιστορικών ερωτημάτων με βάση τις πρωτογενείς πηγές
- η σύνθεση ή ανασύνθεση των διαθέσιμων πληροφοριών

- η προσέγγισης της Ιστορίας ως Ιστοριογραφικής και διανοητικής κατασκευής με βάση τις δευτερογενείς ιστορι(ογραφ)ικές πηγές
- η διαμόρφωση ιστορικών αξιολογικών κρίσεων
- η σύγκριση και η σύνδεση του παρελθόντος με το παρόν (και το μέλλον)
- η διάκριση ιστορικών περιόδων
- η διάκριση τομέων του ιστορικού γίγνεσθαι (π.χ. πολιτικής, κοινωνίας, οικονομίας)
- η «μικροϊστορία» έναντι της «μακροϊστορίας» (longue durée)
- η συγκρότηση και διατύπωση επιχειρημάτων
- ο έλεγχος αξιοπιστίας των πηγών
- η χρήση ενός ιστορικού πλαισίου αναφοράς
- η κριτική εξέταση, ερμηνεία και αξιολόγηση πρωτογενών και δευτερογενών πηγών
- η προσέγγιση του ιστορικού παρελθόντος ως βιώματος
- ο κριτικός αναστοχασμός για τις χρήσεις (και στρεβλώσεις στη χρήση) του ιστορικού παρελθόντος

**ΕΙΚΟΝΑ 7** *Ιστορικές έννοιες Πηγή εικόνας: https://historicalthinking.ca/historical-thinking-concepts*

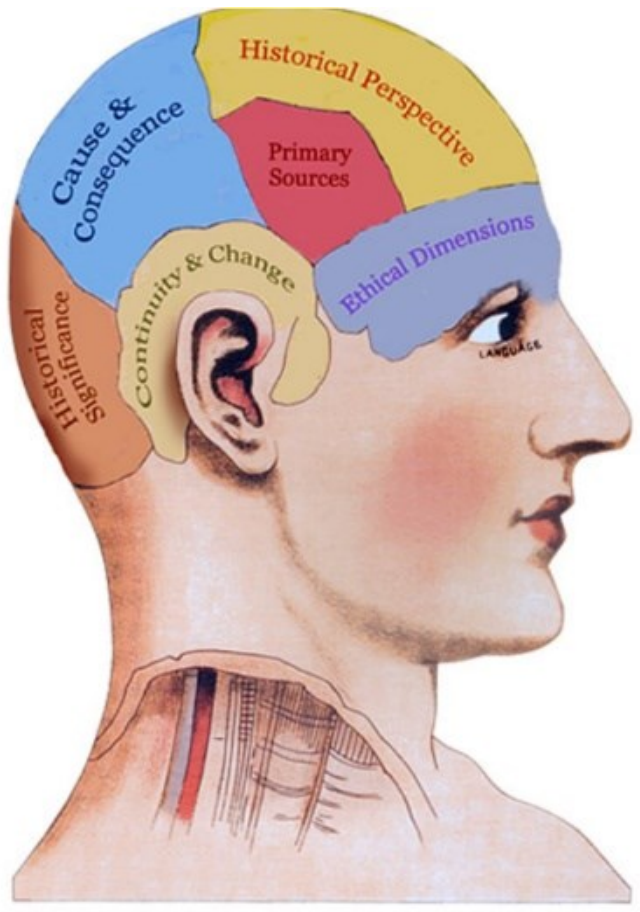

Στα (ανοικτά) ερωτήματα που τίθενται σχετικά με την κατάκτηση των δεξιοτήτων ιστορικής σκέψης συμπεριλαμβάνονται τα εξής:

- πόσο πληροφοριακό υλικό πρέπει να δίνεται στους μαθητές/τριες;
- πόσο απαραίτητη είναι η προϋπάρχουσα γνώση του ιστορικού και του χρονολογικού πλαισίου, προτού επιχειρηθεί η εμβάθυνση;
- πώς αποκτάται ιστορική οπτική (perspective);
- πώς αποκτάται ιστορική ενσυναίσθηση;
- πώς επιτυγχάνεται νοηματοδότηση στο σήμερα;

### 5.1.2 Δεξιότητες γλωσσικού εγγραμματισμού

Η ιστορική σκέψη συνυφαίνεται με την ικανότητα ανάγνωσης, αποκωδικοποίησης (πρόσληψη) και γραφής (παραγωγής), γεγονός που δυσχεραίνει τη μεταξύ τους διάκριση. Ζητείται, για παράδειγμα, από τους μαθητές-τριες να διαβάσουν μια γραπτή πηγή και, στη συνέχεια, τους ζητείται να επιδείξουν γραπτώς το βαθμό κατανόησή τους. Είναι δεδομένο ότι η κατανόηση μιας ιστορικής πηγής προαπαιτεί ένα βαθμό προϋπάρχοντος γλωσσικού εγγραμματισμού. Η διαπλοκή του γλωσσικού με τον ιστορικό εγγραματισμό αυξάνει τη δυσκολία του εγχειρήματος της αποτελεσματικής και ικανοποιητικής διδασκαλίας και εκμάθησης της Ιστορίας.

### 5.1.3 Δεξιότητες ψηφιακού εγγραμματισμού

Ψηφιακός εγγραμματισμός είναι η ικανότητα εύρεσης, αξιολόγησης, αξιοποίησης, κοινής χρήσης και δημιουργίας ψηφιακού περιεχομένου, καθώς και η ικανότητα υπολογιστικής/αλγοριθμικής σκέψης (επίλυσης προβλημάτων). Η Ευρωπαϊκή Επιτροπή ανέπτυξε το DigComp: το Ευρωπαϊκό Ψηφιακό Πλαίσιο Ικανότητας ως πλαίσιο αναφοράς του τι σημαίνει «ψηφιακά ικανός». Το DigComp επιδιώκει να υποστηρίξει την αυτοπεποίθηση του ατόμου, που θα το οδηγεί στην κριτική και υπεύθυνη χρήση της ψηφιακής τεχνολογίας. Περιλαμβάνει 5 τομείς και 21 ικανότητες (ΕΙΚΟΝΑ 8), οι οποίες συνθέτουν τον ψηφιακό γραμματισμό που επιδιώκεται (και απαιτείται) στο πλαίσιο της διδασκαλίας της Ιστορίας με τη χρήση ψηφιακών μέσων.

**ΕΙΚΟΝΑ 8** Ψηφιακές Ικανότητες DigComp 2019
https://ec.europa.eu/social/main.jsp?catId=738&langId=en&pubId=8203&furtherPubs=yes

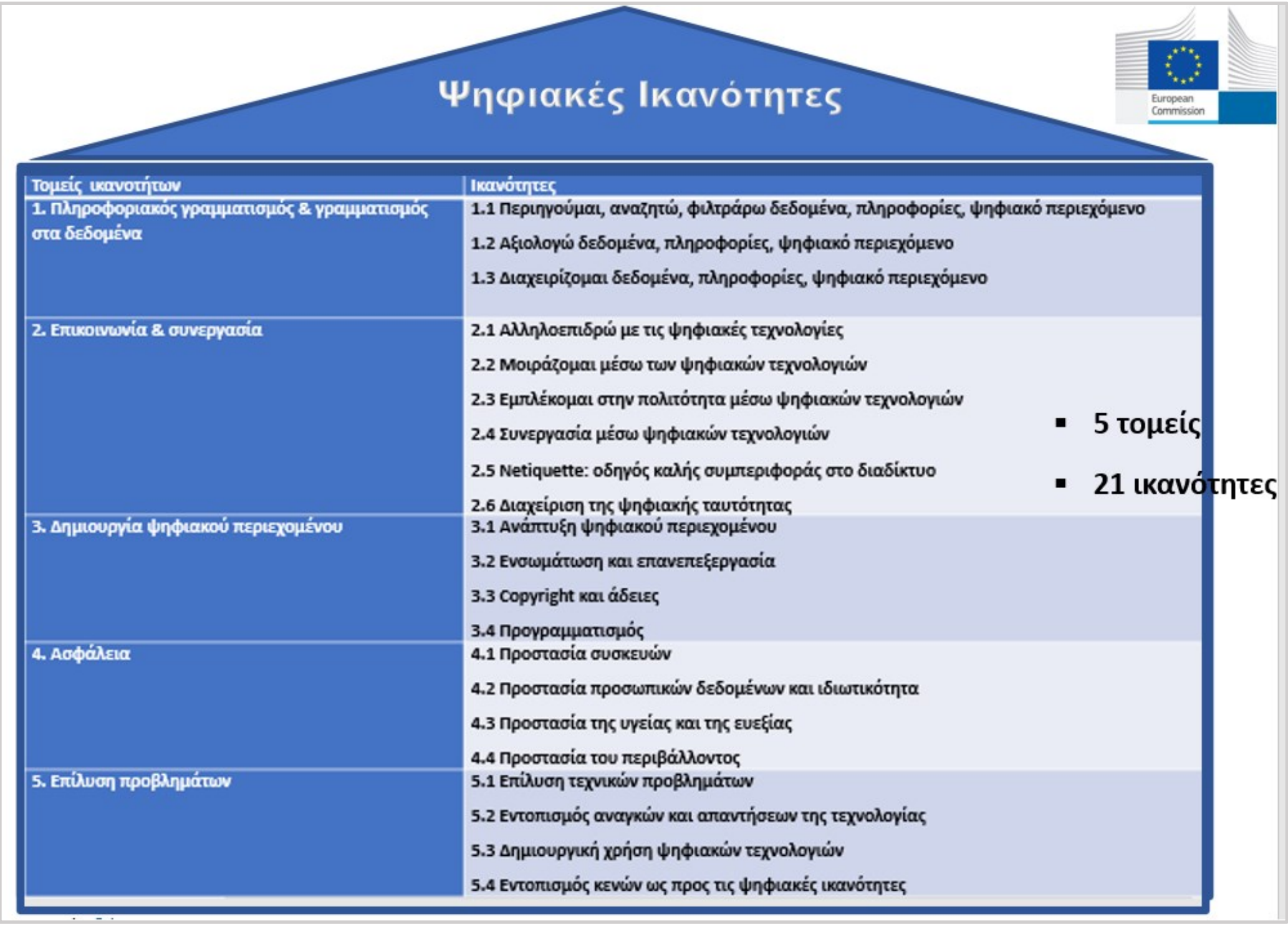

| Τομείς ικανοτήτων | Ικανότητες |
|---|---|
| 1. Πληροφοριακός γραμματισμός & γραμματισμός στα δεδομένα | 1.1 Περιηγούμαι, αναζητώ, φιλτράρω δεδομένα, πληροφορίες, ψηφιακό περιεχόμενο |
| | 1.2 Αξιολογώ δεδομένα, πληροφορίες, ψηφιακό περιεχόμενο |
| | 1.3 Διαχειρίζομαι δεδομένα, πληροφορίες, ψηφιακό περιεχόμενο |
| 2. Επικοινωνία & συνεργασία | 2.1 Αλληλοεπιδρώ με τις ψηφιακές τεχνολογίες |
| | 2.2 Μοιράζομαι μέσω των ψηφιακών τεχνολογιών |
| | 2.3 Εμπλέκομαι στην πολιτότητα μέσω ψηφιακών τεχνολογιών |
| | 2.4 Συνεργασία μέσω ψηφιακών τεχνολογιών |
| | 2.5 Netiquette: οδηγός καλής συμπεριφοράς στο διαδίκτυο |
| | 2.6 Διαχείριση της ψηφιακής ταυτότητας |
| 3. Δημιουργία ψηφιακού περιεχομένου | 3.1 Ανάπτυξη ψηφιακού περιεχομένου |
| | 3.2 Ενσωμάτωση και επανεπεξεργασία |
| | 3.3 Copyright και άδειες |
| | 3.4 Προγραμματισμός |
| 4. Ασφάλεια | 4.1 Προστασία συσκευών |
| | 4.2 Προστασία προσωπικών δεδομένων και ιδιωτικότητα |
| | 4.3 Προστασία της υγείας και της ευεξίας |
| | 4.4 Προστασία του περιβάλλοντος |
| 5. Επίλυση προβλημάτων | 5.1 Επίλυση τεχνικών προβλημάτων |
| | 5.2 Εντοπισμός αναγκών και απαντήσεων της τεχνολογίας |
| | 5.3 Δημιουργική χρήση ψηφιακών τεχνολογιών |
| | 5.4 Εντοπισμός κενών ως προς τις ψηφιακές ικανότητες |

## 5.2 Πόροι για τη διδασκαλία της Ψηφιακής Ιστορίας στη Δευτεροβάθμια Εκπαίδευση

Η Ψηφιακή Ιστορία προσφέρει στους/στις ιστορικούς πολλές οδούς για τη συλλογή πηγών, τη συνεργασία και τη δυνατότητα να μοιράζονται το έργο τους με ευρύ φάσμα κοινού. Τα ψηφιακά μέσα διευκολύνουν την πρόσβαση σε περιθωριοποιημένες ή αθέατες κοινωνικές ομάδες. Αυτό θα ήταν ανέφικτο, αν μοναδικό μέσο επικοινωνίας με το κοινό ήταν το έντυπο μέσο. Παράλληλα προσφέρουν πολλές δυνατότητες αξιοποίησης του ψηφιακού μέσου και του ψηφιακού ιστορικού υλικού για την ανάπτυξη και εμπέδωσης της ιστορικής γνώσης και την καλλιέργεια των ιστορικών δεξιοτήτων.

### 5.2.1 Ψηφιακοί Πόροι για την Ιστορία

Σημαντικοί ψηφιακοί πόροι, που θέτουν ψηφιακό υλικό και διαδραστικά εργαλεία στη διάθεση του/της εκπαιδευτικού, είναι οι ακόλουθοι:

- Η Historiana https://historiana.eu/, μια πρωτοβουλία της ένωσης EuroClio και της Europeana με χρηματοδότηση της Ευρωπαϊκής Ένωσης. Δίνει πρόσβαση σε συλλογές, πηγές και εργαλεία και απευθύνεται σε εκπαιδευτικούς που διδάσκουν Ιστορία.
- Η Μαθησιακή Ηλεκτρονική Τράπεζα Ιδιαίτερης αξίας Διδακτικών Αντικειμένων ΜΗΤΙΔΑ του ΕΚΤ, διαδικτυακή πλατφόρμα αναζήτησης εκπαιδευτικού, πολιτιστικού και επιστημονικού περιεχομένου του Εθνικού Κέντρου Τεκμηρίωσης και Ηλεκτρονικού Περιεχομένου (ΕΚΤ) http://www.mitida.gr/. Είναι ειδικά σχεδιασμένη για τη δημιουργία ψηφιακού εκπαιδευτικού υλικού και για την υποβοήθηση του έργου των εκπαιδευτικών, με περισσότερες από 1.200.000 πηγές περιεχομένου διαθέσιμες στο διαδίκτυο για εκπαιδευτική επανάχρηση.
- Ο ψηφιακός συσσωρευτής εκπαιδευτικών πόρων Φωτόδεντρο του Υπουργείου Παιδείας: http://photodentro.edu.gr/aggregator/ Περιλαμβάνει πόρους για την Ιστορία, και για μια σειρά από άλλα μαθησιακά αντικείμενα.

### 5.2.2 Ψηφιακοί Ιστορικοί Χάρτες

- Ο ιστορικός άτλας Centennia http://historicalatlas.com/. Διατίθεται για Windows και Mac, με διεπαφή (interface) στα Ελληνικά. Περιλαμβάνει τους Νεώτερους χρόνους 1500-σήμερα. Η έκδοση Nations 1789-1939 είναι δωρεάν.
- Ο ιστορικός άτλας Pleiades https://pleiades.stoa.org/, με διαλειτουργικές προδιαγραφές και δυνατότητα συμμετοχής της κοινότητας μέσω πληθοπορισμού. Το περιεχόμενό του οργανώνεται σε τέσσερις τύπους πληροφοριακών πόρων: τόποι (places), τοποθεσίες (locations), ονόματα (names) και συνδέσεις (connections)
- Ο ιστορικός άτλας της Ρωμαϊκής Αυτοκρατορίας https://imperium.ahlfeldt.se/ Οι αρχαίοι τόποι οργανώνονται ως χώροι και ως κτίρια. Τα κτίρια είναι οργανωμένα θεματικά (π.χ. αμφιθέατρα, θέατρα, ναοί κ.λπ.). Μεταδεδομένα για τοποθεσίες και κτίρια είναι διαθέσιμα παραπλεύρως. Αναζητήσεις για τους τόπους μπορούν να γίνουν βάσειτου αρχαίου τοπωνυμίου, αλλά και του σύγχρονου τοπωνυμίου.
- Ο ιστότοπος και η εφαρμογή ToposText https://topostext.org/. Η εφαρμογή είναι διαθέσιμη δωρεάν για Android και Apple. Μέσω της εφαρμογής και του ιστοτόπου, κάθε χρήστης, φοιτητής, ταξιδιώτης, ερευνητής, συνδέεται με τις αρχαίες πηγές που διαμόρφωσαν την ελληνική ιστορία. Ο ιστότοπος ToposText περιλαμβάνει:
  - Πάνω από 530 κείμενα της αρχαίας γραμματείας σε αγγλική μετάφραση
  - 5.350 αρχαίες τοποθεσίες, μουσεία και αρχαιολογικούς χώρους για τον αρχαίο ελληνικό κόσμο
  - Διαδραστικό χάρτη και ευρετήριο που συνδέει κάθε τοποθεσία με την αρχαία πηγή που αναφέρεται σε αυτήν
  - Ευρετήριο κύριων ονομάτων
  - Ακριβείς συντεταγμένες που επιτρέπουν στον χρήστη να «πλησιάσει» αρκετά στο χάρτη, ώστε να ‘δει’ τα αρχαία ερείπια.

### 5.2.3 Ψηφιακές Χρονογραμμές

- Χρονογραμμή του εργαστηρίου Knightlab: http://timeline.knightlab.com/. Εφαρμογή ελεύθερης πρόσβασης. Βλ. το Φύλλο Εργασίας 1.
- Χρονογραμμή Παγκόσμιας Ιστορίας: https://www.wdl.org/en/sets/world-history/timeline/
- Χρονογραμμή του Μουσείου Ακροπόλεως (Ελληνικά και Αγγλικά): https://xronologio.theacropolismuseum.gr/en/-645/-590

## 5.3 Δειγματικές Διδακτικές Προσεγγίσεις

### Φύλλο Εργασίας 1: Ψηφιακή Γεωαφήγηση με StoryMap Js

Τι είναι;

Η **Ψηφιακή Αφήγηση με χρήση Χάρτη** είναι ένας μικτός τρόπος αφήγησης μιας ιστορίας (πραγματικής ή φανταστικής, π.χ. τη ζωή μιας ιστορικής προσωπικότητας ή την πλοκή ενός λογοτεχνικού βιβλίου), συνδυάζοντας ψηφιακό κείμενο και εικόνα και, κυρίως, τοποθετώντας πρόσωπα και γεγονότα όχι σε ένα νοητό χώρο που συνάγεται κατά την αφήγηση από τον αναγνώστη, αλλά σε έναν οπτικοποιημένο χώρο που μετατρέπει τους γεωγραφικούς τόπους στους οποίους δρουν τα πρόσωπα ή/και εκτυλίσσονται τα γεγονότα της ιστορίας σε αναπόσπαστο τμήμα της αφήγησης.

Γιατί;

συνδυάζει πολλών ειδών γραμματισμούς: μαθηματικό, γλωσσικό, γεωγραφικό, ψηφιακό

- είναι πολυτροπική (κείμενο, χάρτης, πίνακες, εικόνα, βίντεο, εξοικείωση με τη γλώσσα html)
- ενθαρρύνει την προσέγγιση από πάνω προς τα κάτω (top-down)
- εξασκεί τη σχεδιαστική σκέψη (design thinking)
- αισθητοποιεί την έννοια του χρονο-τόπου
- εκπαιδεύει στην οπτικοποίηση της σκέψης
- ενδείκνυται για επικοινωνία της γνώσης στην ψηφιακή εποχή

**Σε ποιους απευθύνεται;** Σε όλους όσους διαθέτουν βασικό ψηφιακό γραμματισμό.

**Υλικό που απαιτείται:**

- Η/Υ
- Online Πρόσβαση στην ελεύθερη εφαρμογή StoryMap (Knight Lab, Παν/μιο Northwestern. ΗΠΑ)
- Λογαριασμός Google
- Φύλλα εργασίας

**Τι περιλαμβάνει (περιγραφή-στάδια);** Ακολουθούν τα βήματα – στάδια που προτείνονται από την έναρξη ως την ολοκλήρωση της δραστηριότητας.

Ανάλογα με τη μέση ηλικία των εκπαιδευόμενων, μπορεί <u>το πρώτο μέρος</u> (προετοιμασία) να γίνει χωρίς τη χρήση Η/Υ (unplugged) και στη συνέχει με χρήση PowerPoint, Word ή εγγράφων Google.

**Πρώτο Μέρος: Προετοιμασία**

Επιλογή της ιστορίας με βάση δύο **κριτήρια:**

Η ιστορία να είναι ευσύνοπτη: να ολοκληρώνεται σε μέχρι 15 διαφάνειες

Η δράση να εκτυλίσσεται σε πάνω από έναν τόπους

**Δεύτερο Μέρος:** Χρήση της εφαρμογής StoryMap$^{JS}$

1. Πάτα το κουμπί «Make a StoryMap»

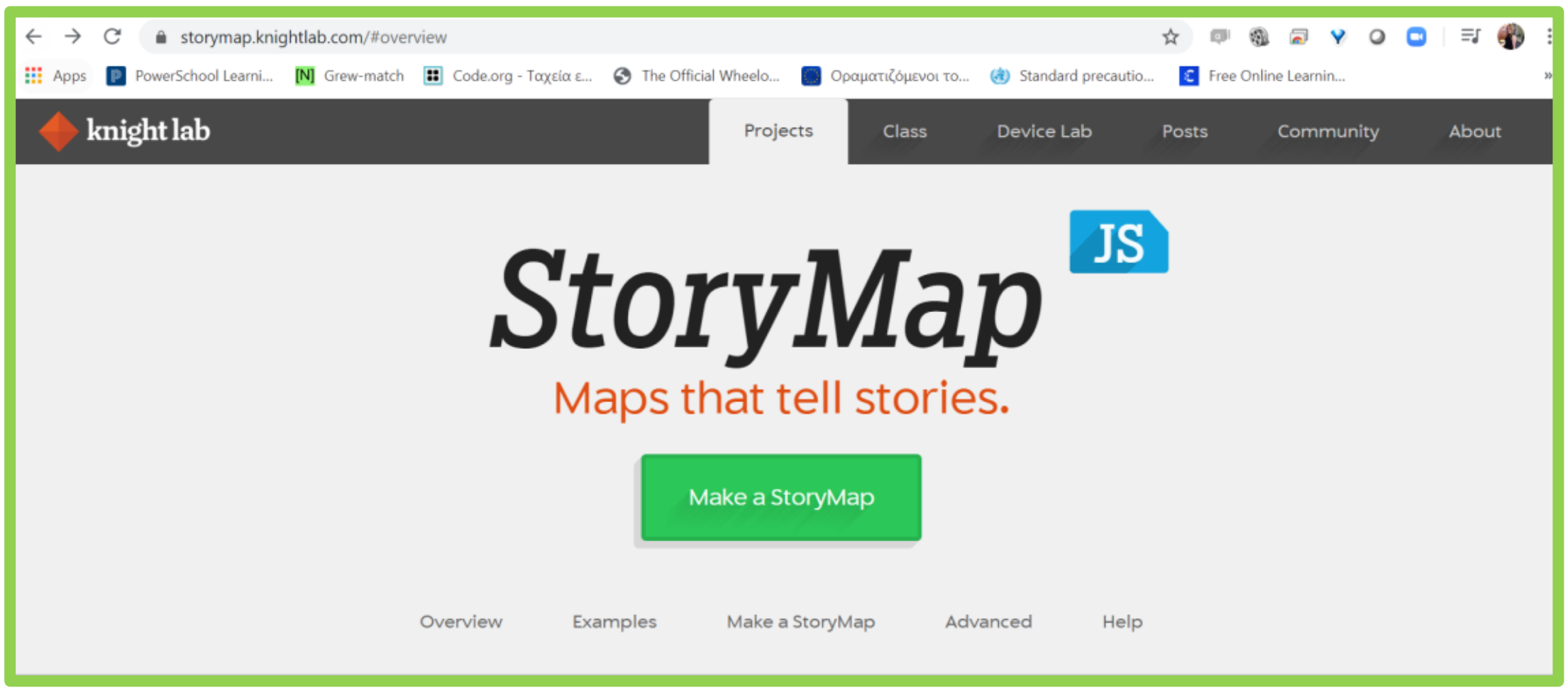

2. Κάνε εγγραφή μέσω του λογαριασμού σου Google

3. Δώσε όνομα στην ψηφιακή σου ιστορία

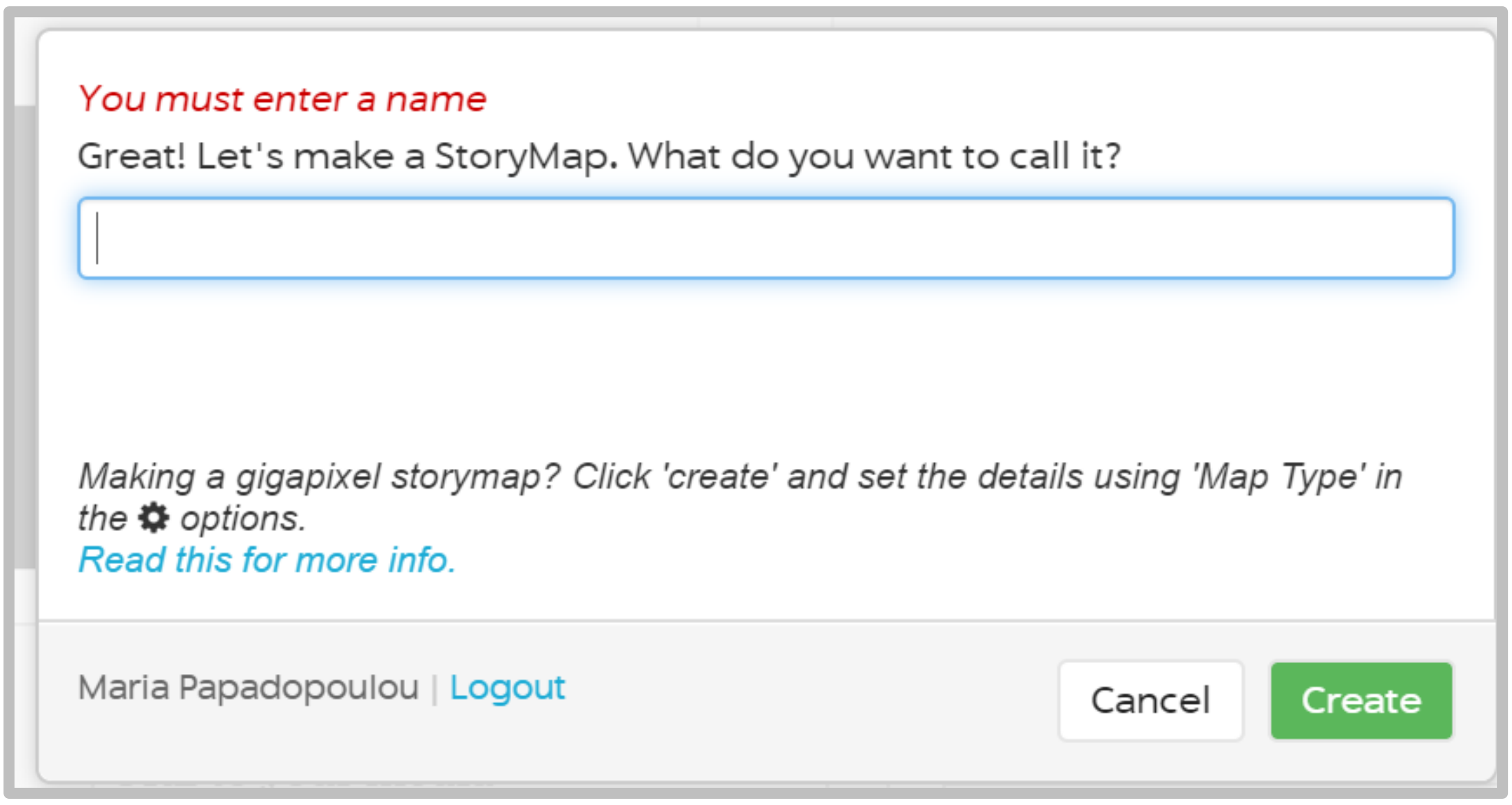

4. Θα βρεθείς στο περιβάλλον που βλέπεις παρακάτω

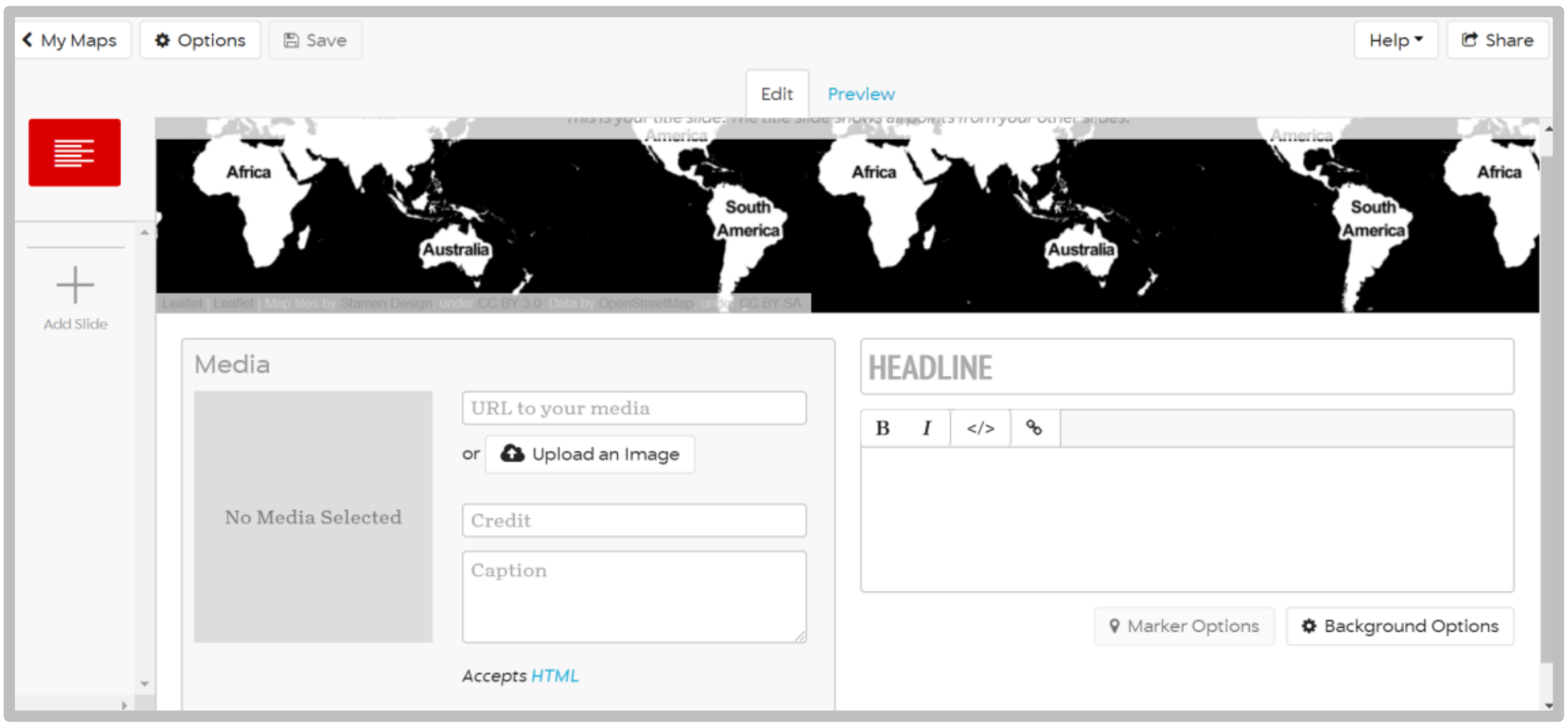

5. Εξερεύνησε τα βασικά χαρακτηριστικά του

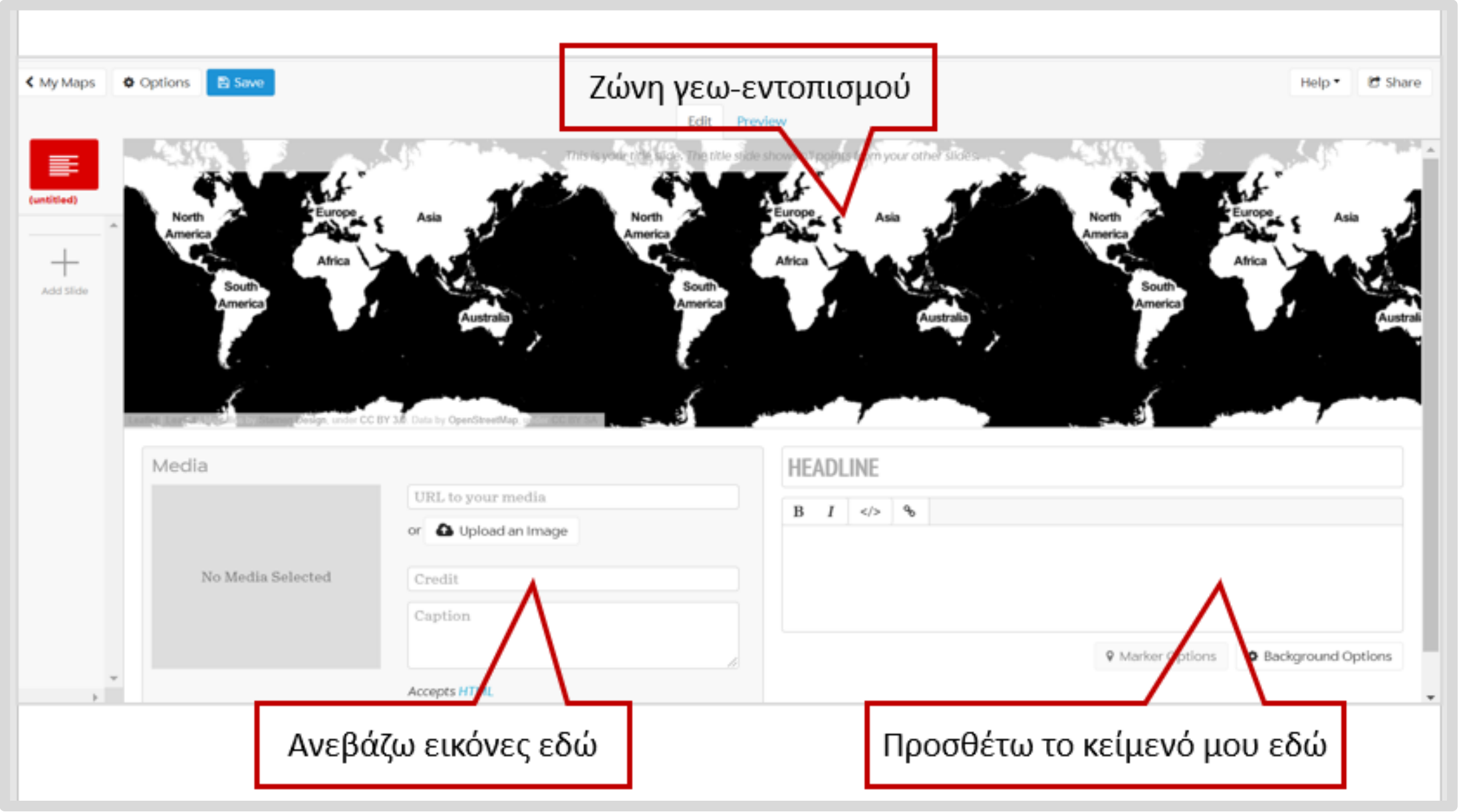

6. Δημιούργησε την πρώτη σου διαφάνεια κάνοντας κλικ εκεί που βλέπεις στην εικόνα

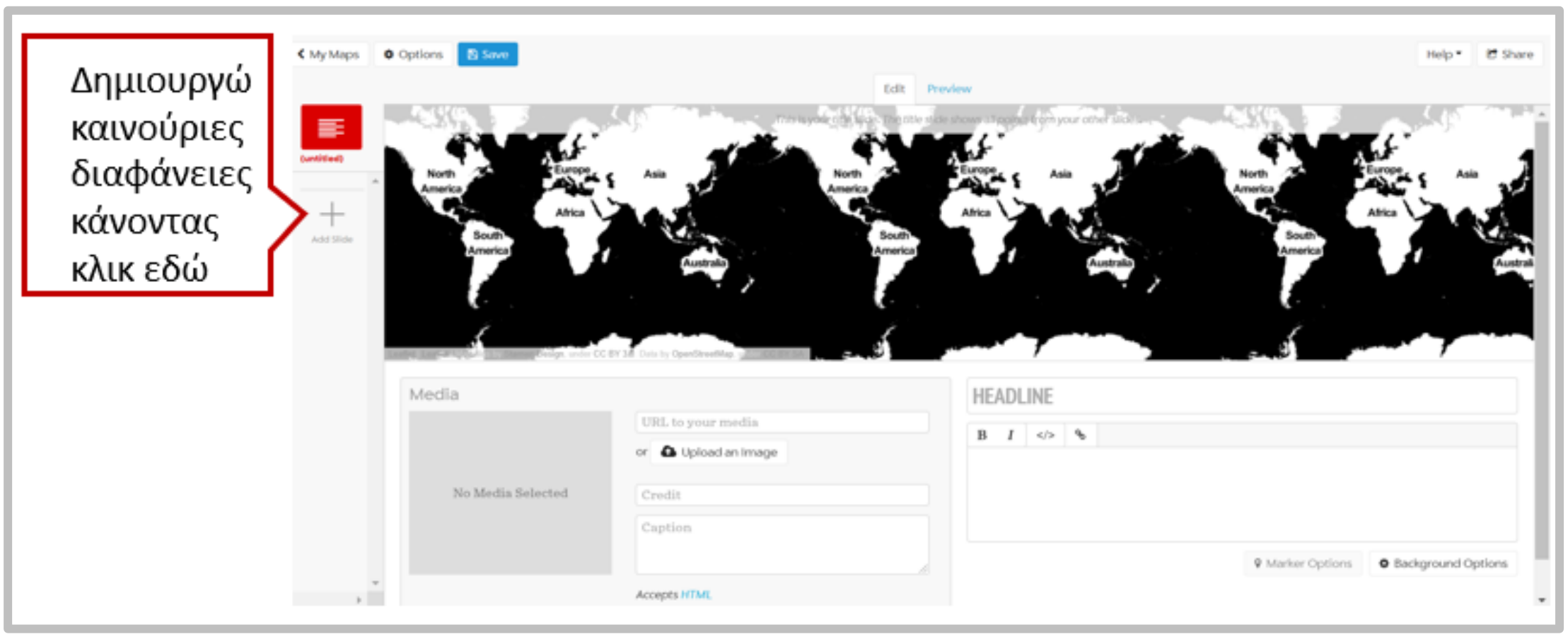

7. Χωροεντόπισε τη δράση αυτού του μέρους της ιστορίας, πληκτρολογώντας τον τόπο, και διαλέγοντας από το drop-down menu, όπως βλέπεις στην εικόνα:

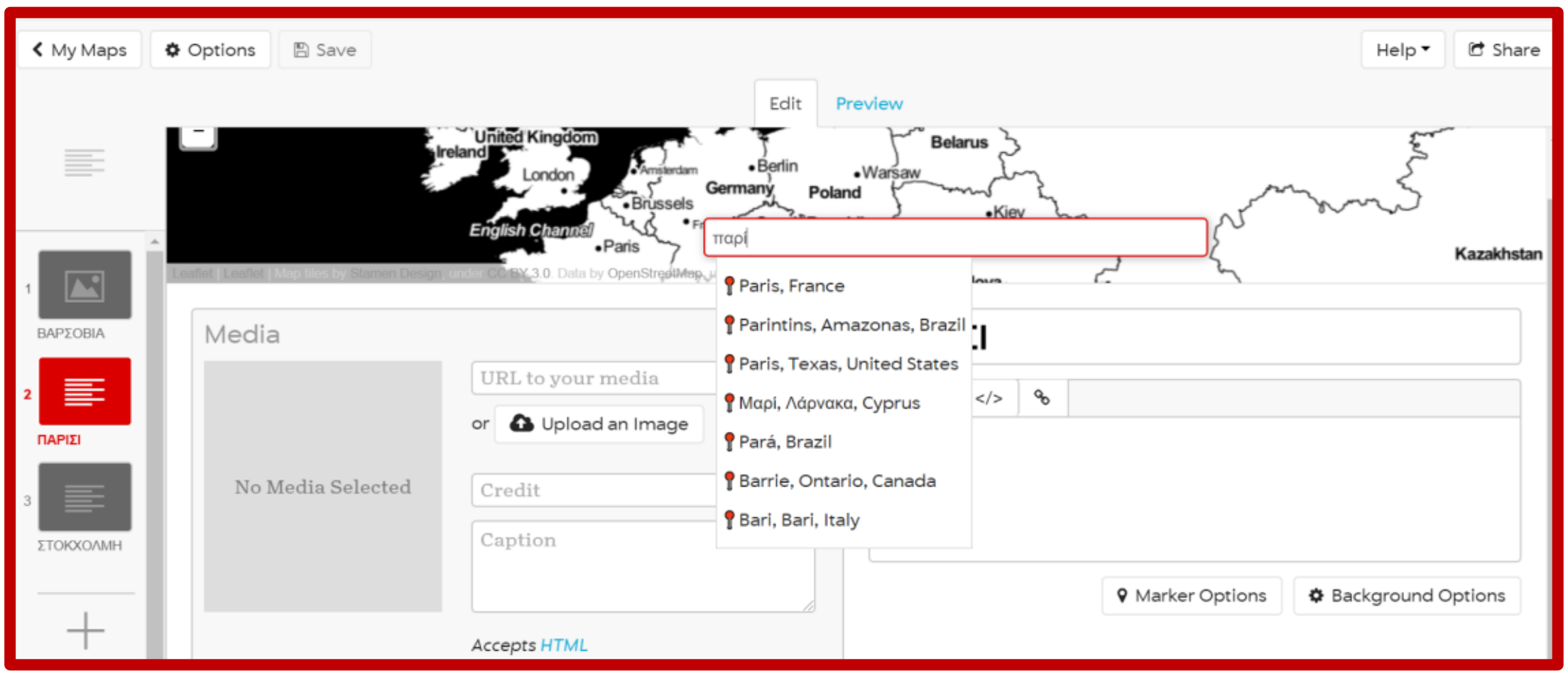

8. Επανάλαβε τα βήματα 6 και 7 ώσπου να ολοκληρώσεις την αφήγηση της ιστορίας

9. Ανά πάσα στιγμή μπορείς να πατήσεις το κουμπί «Preview», για να δεις ολοκληρωμένη την ιστορία σου, ως το σημείο εκείνο.

10. Μπορείς, επίσης, να μοιράσεις τη δουλειά σου, πατώντας το κουμπί «Share».

11. Θυμήσου, ανά τακτά διαστήματα να πατάς το κουμπί «Save», ώστε να σώζεις τη δουλειά σου. Καλή αφήγηση!

---

**Παραδείγματα:**

**1. Εικονομαχία,**
https://uploads.knightlab.com/storymapjs/91acd6afea3c13ad239a11f740744c85/eikonoma

[khia/index.html](khia/index.html)

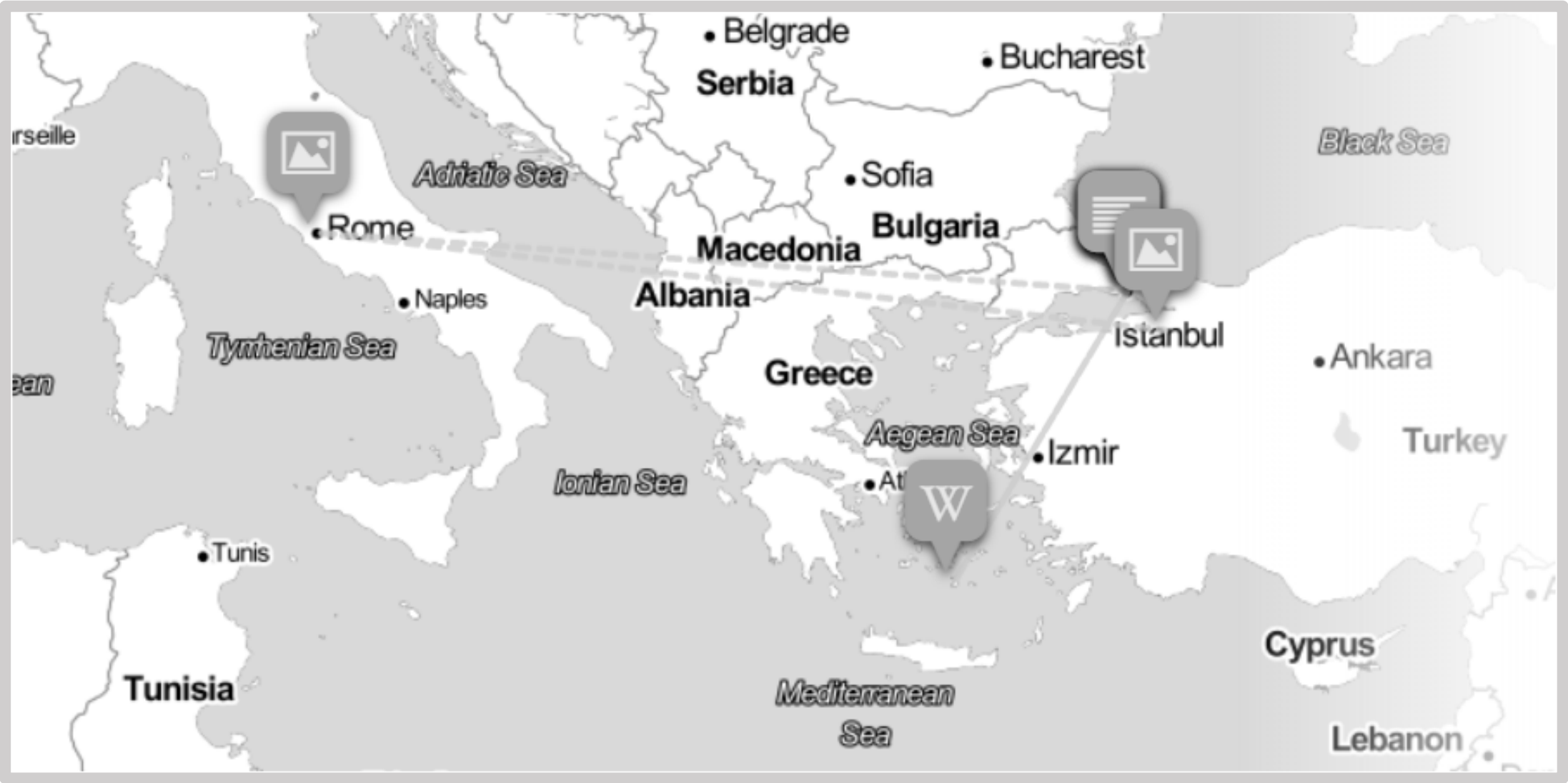

**2. Το Ναπολεόντειο άνοιγμα,** [https://uploads.knightlab.com/storymapjs/35d128260d404d75026c2bdc64bffad3/mpla/index.html](https://uploads.knightlab.com/storymapjs/35d128260d404d75026c2bdc64bffad3/mpla/index.html)

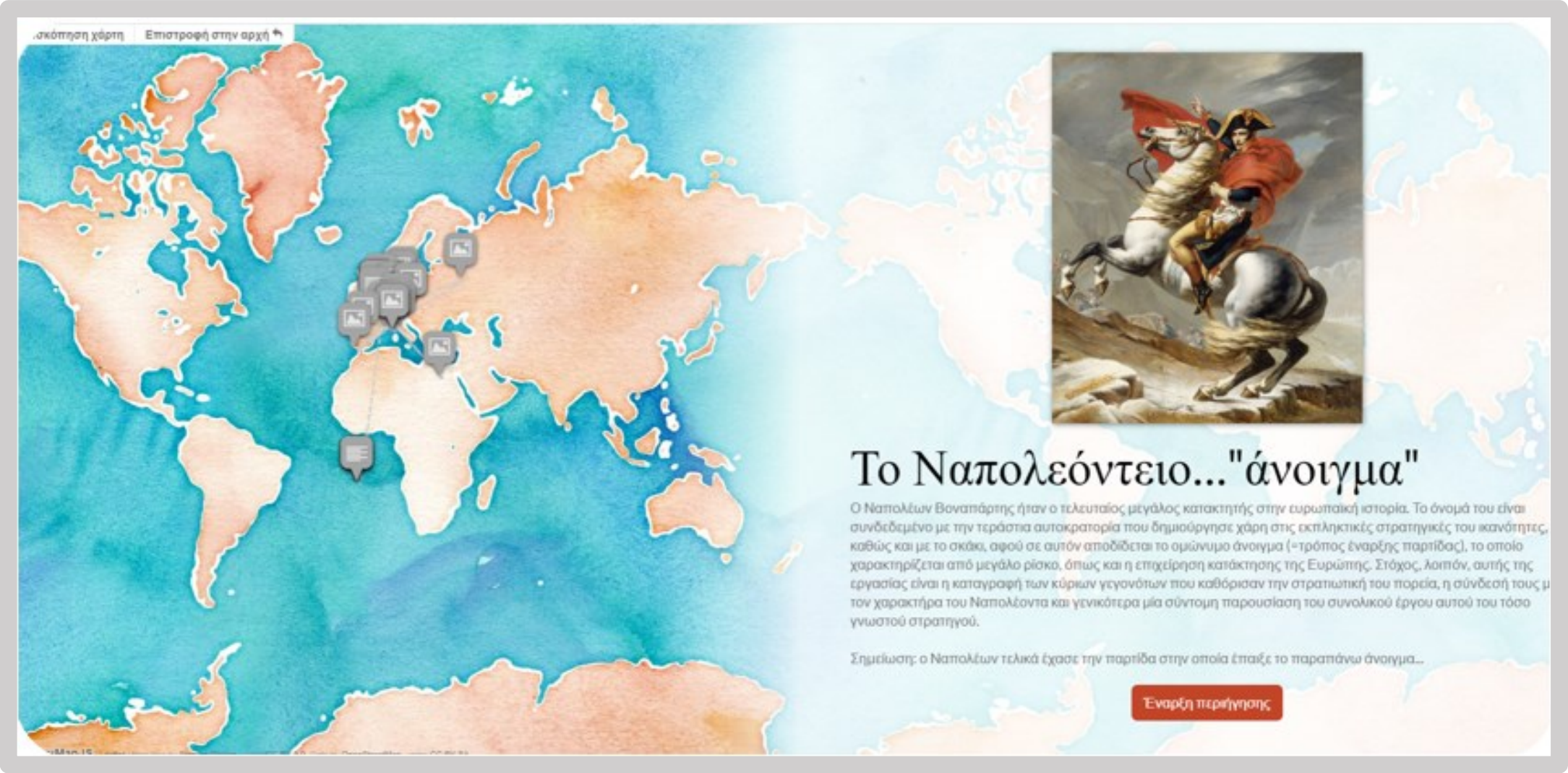

**3. Γεωγραφικές Ανακαλύψεις**
[https://uploads.knightlab.com/storymapjs/2c44c4241a368b484eb2a31518336930/geographik](https://uploads.knightlab.com/storymapjs/2c44c4241a368b484eb2a31518336930/geographik)

es-anakalupseis-1482-1521/index.html

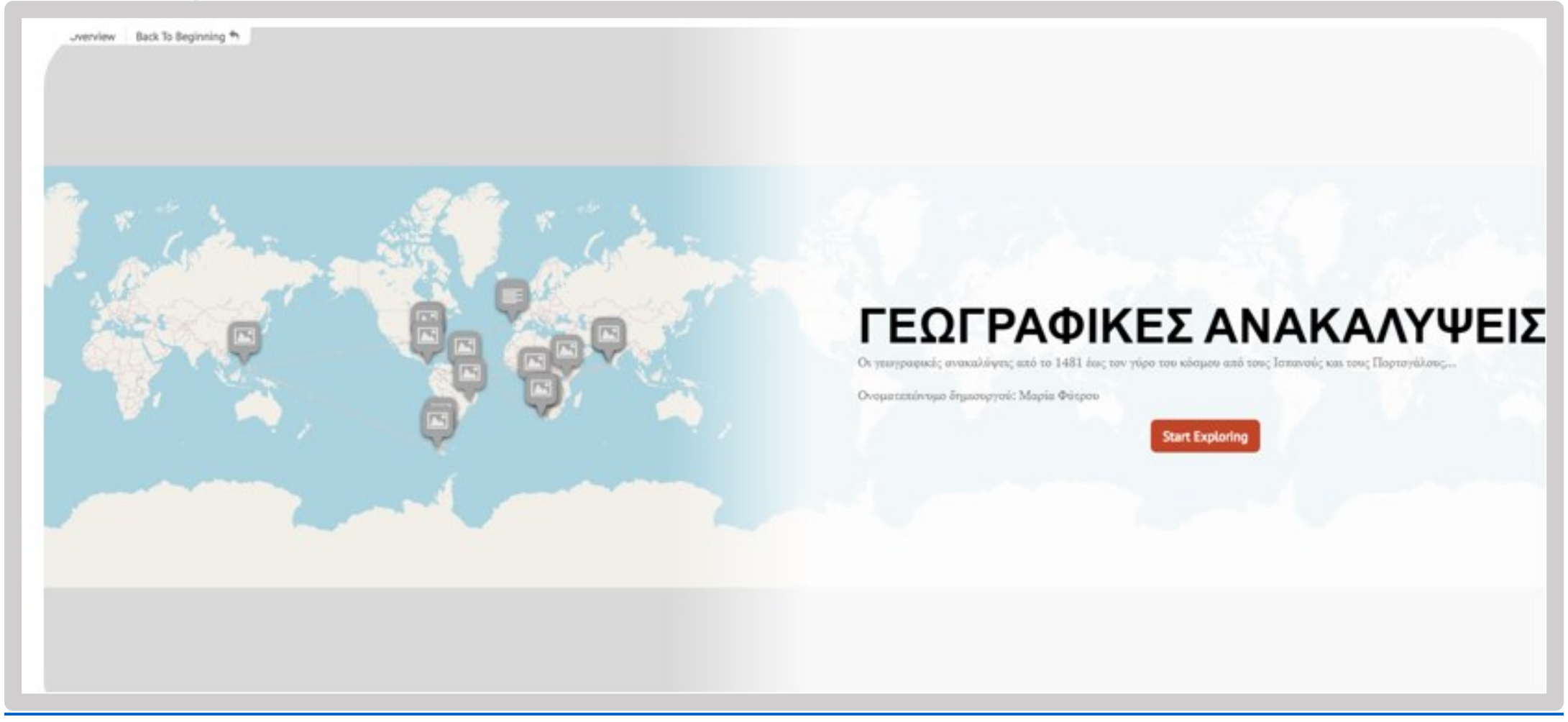

**4. Μαρία Κιουρί. Η γυναίκα που πήρε δύο Νόμπελ: Μαρία Κιουρί (1867-1934)**

**Βασικές πηγές:**

- Η βιογραφία της Μ. Κιουρί γραμμένη από την κόρη της, Εύα. Στα Ελληνικά από τις εκδόσεις Ωκεανίδα.
- Marie Curie – Facts. NobelPrize.org. Nobel Media AB 2020. Tue. 5 May 2020. https://www.nobelprize.org/prizes/physics/1903/marie-curie/facts/

Η δραστηριότητα αφορά ψηφιακή αφήγηση με χρήση χάρτη της ζωής και του έργου της Marie Curie. Χρησιμοποιείται η εφαρμογή StoryMap[JS].

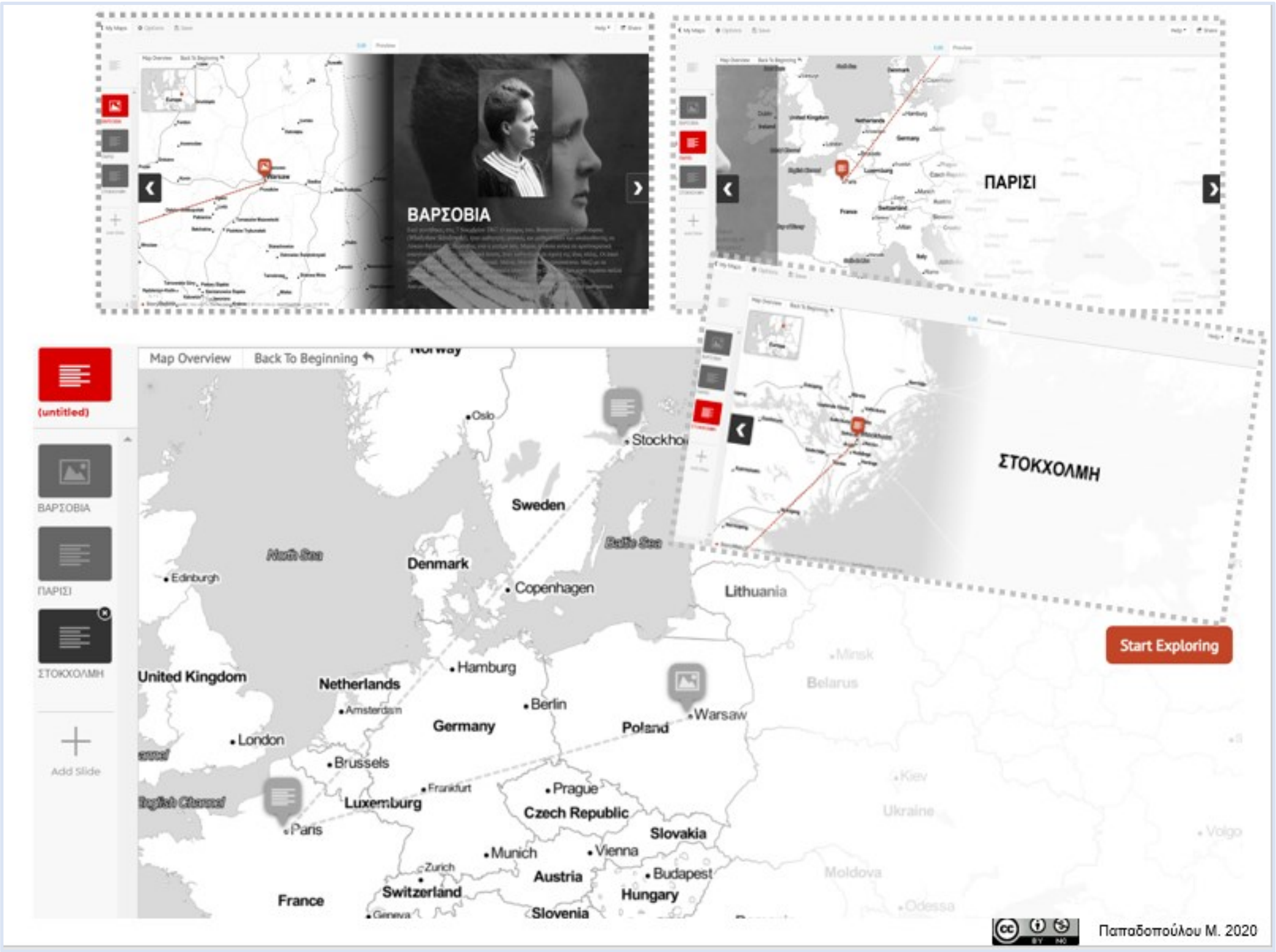

## Φύλλο Εργασίας 2: Επεξεργασία κειμένου με τα εργαλεία Voyant

**ΑΠΑΙΤΟΥΜΕΝΟΣ ΧΡΟΝΟΣ:** δύο (2) διδακτικές ώρες

**ΔΙΔΑΚΤΙΚΗ ΕΝΟΤΗΤΑ** π.χ. Ξενοφών, Ελληνικά 2.2.16-23 (σελ. 70-7 του σχολικού εγχειριδίου «Αρχαίοι Έλληνες Ιστοριογράφοι» της Α' τάξης Λυκείου7)

Έχουμε ήδη διδάξει την ενότητα. Μέσω της παρούσας δραστηριότητας θα δούμε πώς μπορούμε να προσεγγίσουμε την ενότητα αυτή με την στοχοθεσία που ακολουθεί.

**ΠΡΩΤΗ ΔΙΔΑΚΤΙΚΗ ΩΡΑ**
**ΜΑΘΗΣΙΑΚΟΙ ΣΤΟΧΟΙ**

- Να εξοικειωθούν οι μαθητές/τριες με τη χρήση του διαδραστικού βιβλίου μαθητή.
- Να ασκηθούν στο να εντοπίζουν τις τροπικότητες (modalities) του κειμένου σε έντυπη μορφή και τις τροπικότητες του ψηφιακού κειμένου με υπερσυνδέσμους (hyperlinks).
- Να γνωρίσουν την ΨΗΦΙΑΚΗ ΒΙΒΛΙΟΘΗΚΗ Perseus (Παν/μιο Tufts)
- Να εξοικειωθούν με την επεξεργασία και την οπτικοποίηση κειμενικών δεδομένων μέσω της χρήσης των ψηφιακών εργαλείων ανοιχτής πρόσβασης Voyant
- Να έχουν μια πρώτη επαφή με την γλώσσα XML (eXtensible Markup Language). Η XML είναι γλώσσα σήμανσης (mark-up language), που περιέχει ένα σύνολο κανόνων για την ηλεκτρονική κωδικοποίηση κειμένων κατά τρόπο ώστε να είναι αναγνώσιμα και 'κατανοητά' από τους ηλεκτρονικούς υπολογιστές. Η κωδικοποίηση κειμένων των ιστοσελίδων στη γλώσσα αυτή θέτει τις βάσεις για την επόμενη, πιο 'ευφυή' γενιά του Ιστού.

**ΠΡΟΑΠΑΙΤΟΥΜΕΝΕΣ ΕΝΝΟΙΕΣ & ΔΕΞΙΟΤΗΤΕΣ:**

- δεξιότητες βασικού ψηφιακού εγγραμματισμού - γνώσεις πλοήγησης σε Web 2.0

**ΚΑΙΝΟΥΡΙΕΣ ΕΝΝΟΙΕΣ:**

- Τροπικότητα (modality)
- Γλώσσα σήμανσης (markup language)
- Οπτικοποίηση δεδομένων (data visualization)
- Επεξεργασία Φυσικής Γλώσσα (Natural Language Processing-NLP)
- Λέξεις τερματισμού (stop words)

**ΒΗΜΑ 1** Πήγαινε στην ιστοσελίδα του βιβλίου

ΑΡΧΑΙΟΙ ΕΛΛΗΝΕΣ ΙΣΤΟΡΙΟΓΡΑΦΟΙ **http://ebooks.edu.gr/modules/ebook/show.php/DSGL-A108/225/1640,5208/**

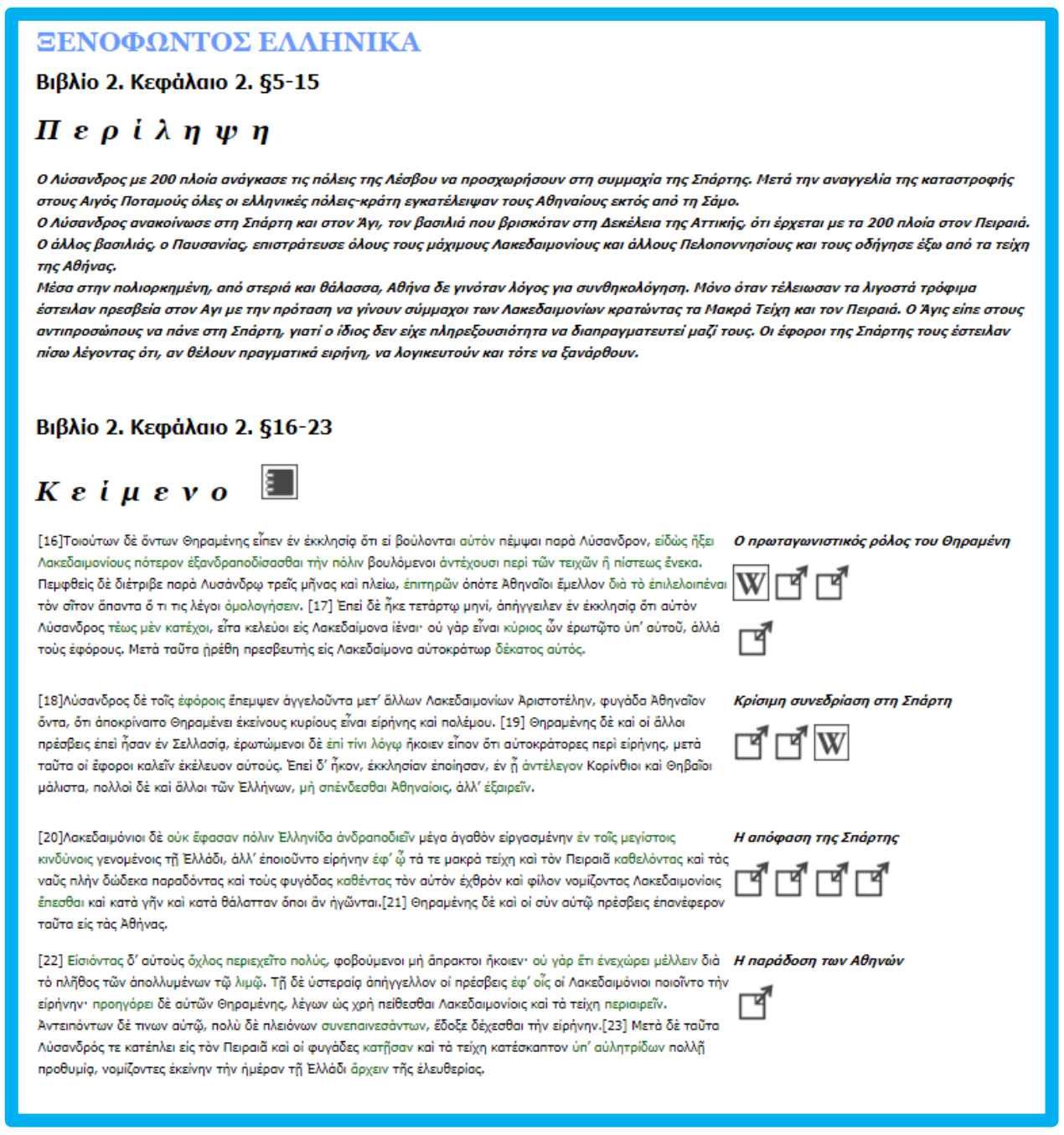

ΞΕΝΟΦΩΝΤΟΣ ΕΛΛΗΝΙΚΑ

Βιβλίο 2. Κεφάλαιο 2. §5-15

*Π ε ρ ί λ η ψ η*

***Ο Λύσανδρος με 200 πλοία ανάγκασε τις πόλεις της Λέσβου να προσχωρήσουν στη συμμαχία της Σπάρτης. Μετά την αναγγελία της καταστροφής στους Αιγός Ποταμούς όλες οι ελληνικές πόλεις-κράτη εγκατέλειψαν τους Αθηναίους εκτός από τη Σάμο.***
***Ο Λύσανδρος ανακοίνωσε στη Σπάρτη και στον Άγι, τον βασιλιά που βρισκόταν στη Δεκέλεια της Αττικής, ότι έρχεται με τα 200 πλοία στον Πειραιά. Ο άλλος βασιλιάς, ο Παυσανίας, επιστράτευσε όλους τους μάχιμους Λακεδαιμονίους και άλλους Πελοποννησίους και τους οδήγησε έξω από τα τείχη της Αθήνας.***
***Μέσα στην πολιορκημένη, από στεριά και θάλασσα, Αθήνα δε γινόταν λόγος για συνθηκολόγηση. Μόνο όταν τέλειωσαν τα λιγοστά τρόφιμα έστειλαν πρεσβεία στον Αγι με την πρόταση να γίνουν σύμμαχοι των Λακεδαιμονίων κρατώντας τα Μακρά Τείχη και τον Πειραιά. Ο Άγις είπε στους αντιπροσώπους να πάνε στη Σπάρτη, γιατί ο ίδιος δεν είχε πληρεξουσιότητα να διαπραγματευτεί μαζί τους. Οι έφοροι της Σπάρτης τους έστειλαν πίσω λέγοντας ότι, αν θέλουν πραγματικά ειρήνη, να λογικευτούν και τότε να ξανάρθουν.***

Βιβλίο 2. Κεφάλαιο 2. §16-23

*Κ ε ί μ ε ν ο*

[16]Τοιούτων δὲ ὄντων Θηραμένης εἶπεν ἐν ἐκκλησίᾳ ὅτι εἰ βούλονται αὐτὸν πέμψαι παρὰ Λύσανδρον, εἰδὼς ἥξει Λακεδαιμονίους πότερον ἐξανδραποδίσασθαι τὴν πόλιν βουλόμενοι ἀντέχουσι περὶ τῶν τειχῶν ἢ πίστεως ἕνεκα. Πεμφθεὶς δὲ διέτριβε παρὰ Λυσάνδρῳ τρεῖς μῆνας καὶ πλείω, ἐπιτηρῶν ὁπότε Ἀθηναῖοι ἔμελλον διὰ τὸ ἐπιλελοιπέναι τὸν σῖτον ἅπαντα ὅ τι τις λέγοι ὁμολογήσειν. [17] Ἐπεὶ δὲ ἧκε τετάρτῳ μηνί, ἀπήγγειλεν ἐν ἐκκλησίᾳ ὅτι αὐτὸν Λύσανδρος τέως μὲν κατέχοι, εἶτα κελεύοι εἰς Λακεδαίμονα ἰέναι· οὐ γὰρ εἶναι κύριος ὧν ἐρωτῷτο ὑπ᾽ αὐτοῦ, ἀλλὰ τοὺς ἐφόρους. Μετὰ ταῦτα ᾑρέθη πρεσβευτὴς εἰς Λακεδαίμονα αὐτοκράτωρ δέκατος αὐτός.

***Ο πρωταγωνιστικός ρόλος του Θηραμένη***

[18]Λύσανδρος δὲ τοῖς ἐφόροις ἔπεμψεν ἀγγελοῦντα μετ᾽ ἄλλων Λακεδαιμονίων Ἀριστοτέλην, φυγάδα Ἀθηναῖον ὄντα, ὅτι ἀποκρίναιτο Θηραμένει ἐκείνους κυρίους εἶναι εἰρήνης καὶ πολέμου. [19] Θηραμένης δὲ καὶ οἱ ἄλλοι πρέσβεις ἐπεὶ ἦσαν ἐν Σελλασίᾳ, ἐρωτώμενοι δὲ ἐπὶ τίνι λόγῳ ἥκοιεν εἶπον ὅτι αὐτοκράτορες περὶ εἰρήνης, μετὰ ταῦτα οἱ ἔφοροι καλεῖν ἐκέλευον αὐτούς. Ἐπεὶ δ᾽ ἧκον, ἐκκλησίαν ἐποίησαν, ἐν ᾗ ἀντέλεγον Κορίνθιοι καὶ Θηβαῖοι μάλιστα, πολλοὶ δὲ καὶ ἄλλοι τῶν Ἑλλήνων, μὴ σπένδεσθαι Ἀθηναίοις, ἀλλ᾽ ἐξαιρεῖν.

***Κρίσιμη συνεδρίαση στη Σπάρτη***

[20]Λακεδαιμόνιοι δὲ οὐκ ἔφασαν πόλιν Ἑλληνίδα ἀνδραποδιεῖν μέγα ἀγαθὸν εἰργασμένην ἐν τοῖς μεγίστοις κινδύνοις γενομένοις τῇ Ἑλλάδι, ἀλλ᾽ ἐποιοῦντο εἰρήνην ἐφ᾽ ᾧ τά τε μακρὰ τείχη καὶ τὸν Πειραιᾶ καθελόντας καὶ τὰς ναῦς πλὴν δώδεκα παραδόντας καὶ τοὺς φυγάδας καθέντας τὸν αὐτὸν ἐχθρὸν καὶ φίλον νομίζοντας Λακεδαιμονίοις ἕπεσθαι καὶ κατὰ γῆν καὶ κατὰ θάλατταν ὅποι ἂν ἡγῶνται.[21] Θηραμένης δὲ καὶ οἱ σὺν αὐτῷ πρέσβεις ἐπανέφερον ταῦτα εἰς τὰς Ἀθήνας.

***Η απόφαση της Σπάρτης***

[22] Εἰσιόντας δ᾽ αὐτοὺς ὄχλος περιεχεῖτο πολύς, φοβούμενοι μὴ ἄπρακτοι ἥκοιεν· οὐ γὰρ ἔτι ἐνεχώρει μέλλειν διὰ τὸ πλῆθος τῶν ἀπολλυμένων τῷ λιμῷ. Τῇ δὲ ὑστεραίᾳ ἀπήγγελλον οἱ πρέσβεις ἐφ᾽ οἷς οἱ Λακεδαιμόνιοι ποιοῖντο τὴν εἰρήνην· προηγόρει δὲ αὐτῶν Θηραμένης, λέγων ὡς χρὴ πείθεσθαι Λακεδαιμονίοις καὶ τὰ τείχη περιαιρεῖν. Ἀντειπόντων δέ τινων αὐτῷ, πολὺ δὲ πλειόνων συνεπαινεσάντων, ἔδοξε δέχεσθαι τὴν εἰρήνην.[23] Μετὰ δὲ ταῦτα Λύσανδρός τε κατέπλει εἰς τὸν Πειραιᾶ καὶ οἱ φυγάδες κατῇσαν καὶ τὰ τείχη κατέσκαπτον ὑπ᾽ αὐλητρίδων πολλῇ προθυμίᾳ, νομίζοντες ἐκείνην τὴν ἡμέραν τῇ Ἑλλάδι ἄρχειν τῆς ἐλευθερίας.

***Η παράδοση των Αθηνών***

**ΒΗΜΑ 2** Βρες το «βιβλίο 2, κεφάλαιο 2, παρ. 16-23» των Ελληνικών του Ξενοφώντα

**ΒΗΜΑ 3** Πάτησε το εικονίδιο

Τι παρατηρείς;

**ΒΗΜΑ 4** Πάτησε έναν-έναν τους υπόλοιπους υπερσυνδέσμους. Κάνε την παρακάτω άσκηση.

ΑΣΚΗΣΗ: Ποιες από τις πιο κάτω τροπικότητες έχει η έντυπη εκδοχή του διδακτικού εγχειριδίου; Διάλεξε από μία (1) έως πέντε (1) και δώσε τη σελίδα για καθεμία που επιλέγεις.

☐ κείμενο ☐ εικόνα ☐ χρονολόγιο ☐ λεξιλόγιο ☐ χάρτης

**ΒΗΜΑ 5** Δες την αρχή των Ελληνικών του Ξενοφώντα στον ιστότοπο της Ψηφιακής Βιβλιοθήκης Perseus:

**ΒΗΜΑ 6** Κάνε πλοήγηση και βρες στον Perseus την τελευταία παράγραφο της ενότητας (**2.2.23)**

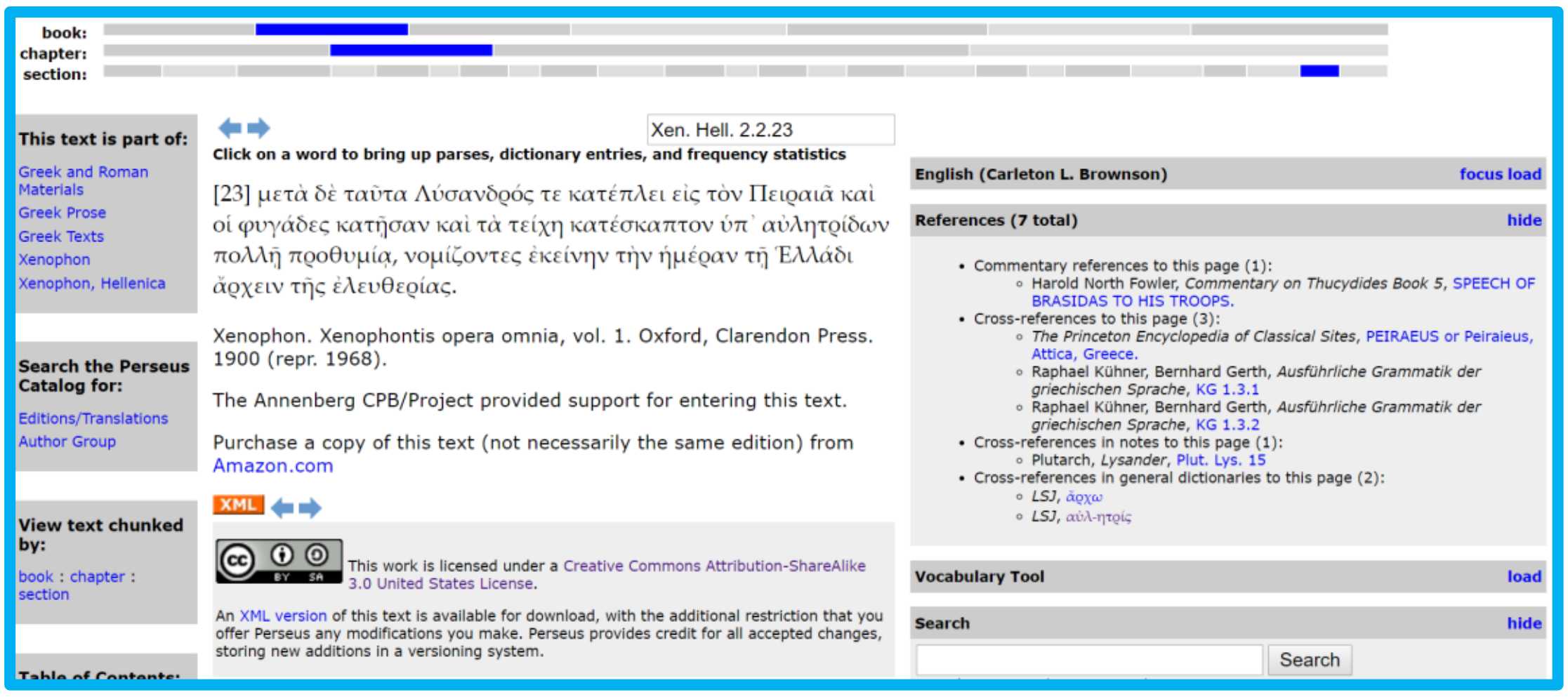

**ΒΗΜΑ 7** Ποιες γλώσσες (φυσικές και τεχνητές) συναντάς στην ψηφιακή εκδοχή του Ξενοφώντειου αποσπάσματος της βιβλιοθήκης Perseus;

☐ αρχαίο Ελληνικό κείμενο
☐ μετάφραση στα αγγλικά
☐ HTML
☐ XML

**ΒΗΜΑ 8** Πήγαινε στην ιστοσελίδα των εργαλείων ψηφιακής οπτικοποίησης κειμενικών δεδομένων **Voyant**.

**ΒΗΜΑ 9** Αντίγραψε το κείμενο της παραγράφου 23 από την ιστοσελίδα του Perseus

- πρόσθεσέ το στο παράθυρο της ιστοσελίδας του Voyant
- πάτα Reveal (Αποκάλυψε)

**ΒΗΜΑ 10** Τι παρατηρείς στο παράθυρο Cirrus; Ποια πληροφορία παίρνεις;

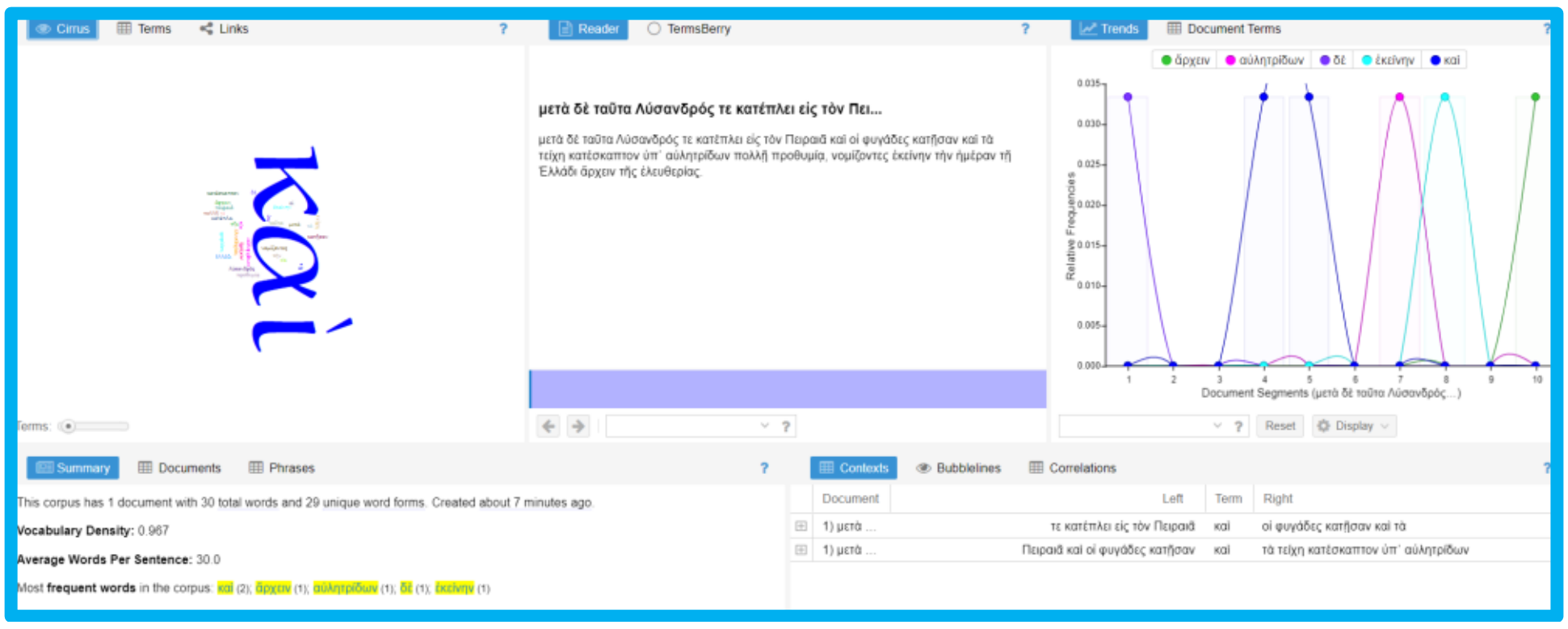

**ΒΗΜΑ 11** Άνοιξε το παράθυρο του εργαλείου TermsBerry. Τι παρατηρείς;

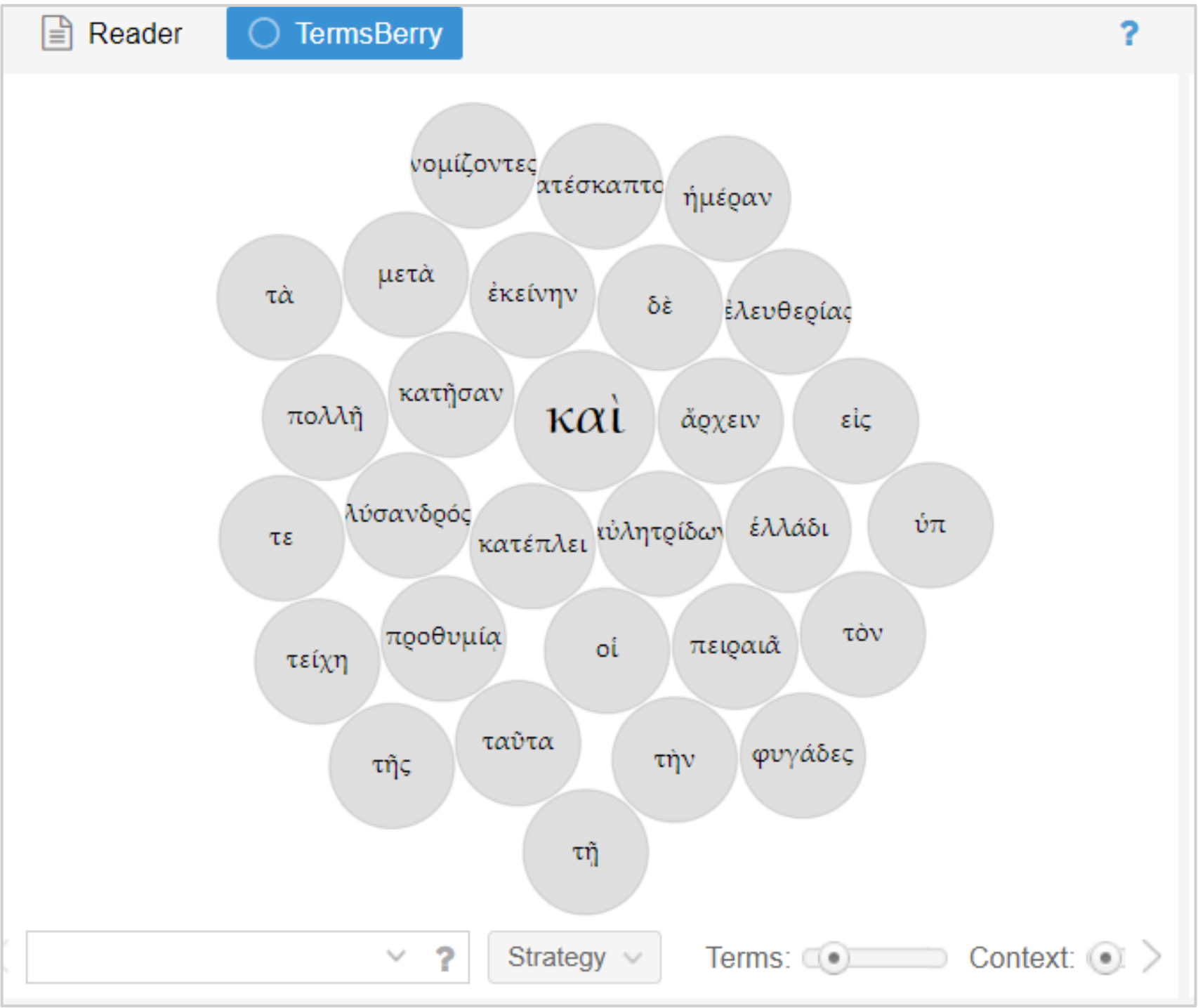

**ΒΗΜΑ 12** Πήγαινε πίσω στο κείμενο του Perseus. Πάτα την λέξη ***καὶ***. Τι παρατηρείς ως προς τη συχνότητα της λέξης;

Λέξεις όπως αυτή λέγονται stopwords (λέξεις τερματισμού) σύμφωνα με την ορολογία του επιστημονικού πεδίου Επεξεργασία Φυσικής Γλώσσας (Natural Language Processing – NLP), δεν θεωρούνται φορείς εννοιών και αφαιρούνται στη φάση του καθαρισμού των δεδομένων (data cleaning).

Μπορείς να βρεις έναν κατάλογο λέξεων τερματισμού της Αρχαίας Ελληνικής εδώ (GitHub) και εδώ (Digital Classicist wiki).

Καλή πλοήγηση!

**ΔΕΥΤΕΡΗ ΔΙΔΑΚΤΙΚΗ ΩΡΑ**

**ΒΗΜΑ 1**

Το στιγμιότυπο οθόνης (screenshot) που ακολουθεί δημιουργήθηκε με το λογισμικό ανοικτού κώδικα Voyant. Αναγνωρίζετε το κείμενο;

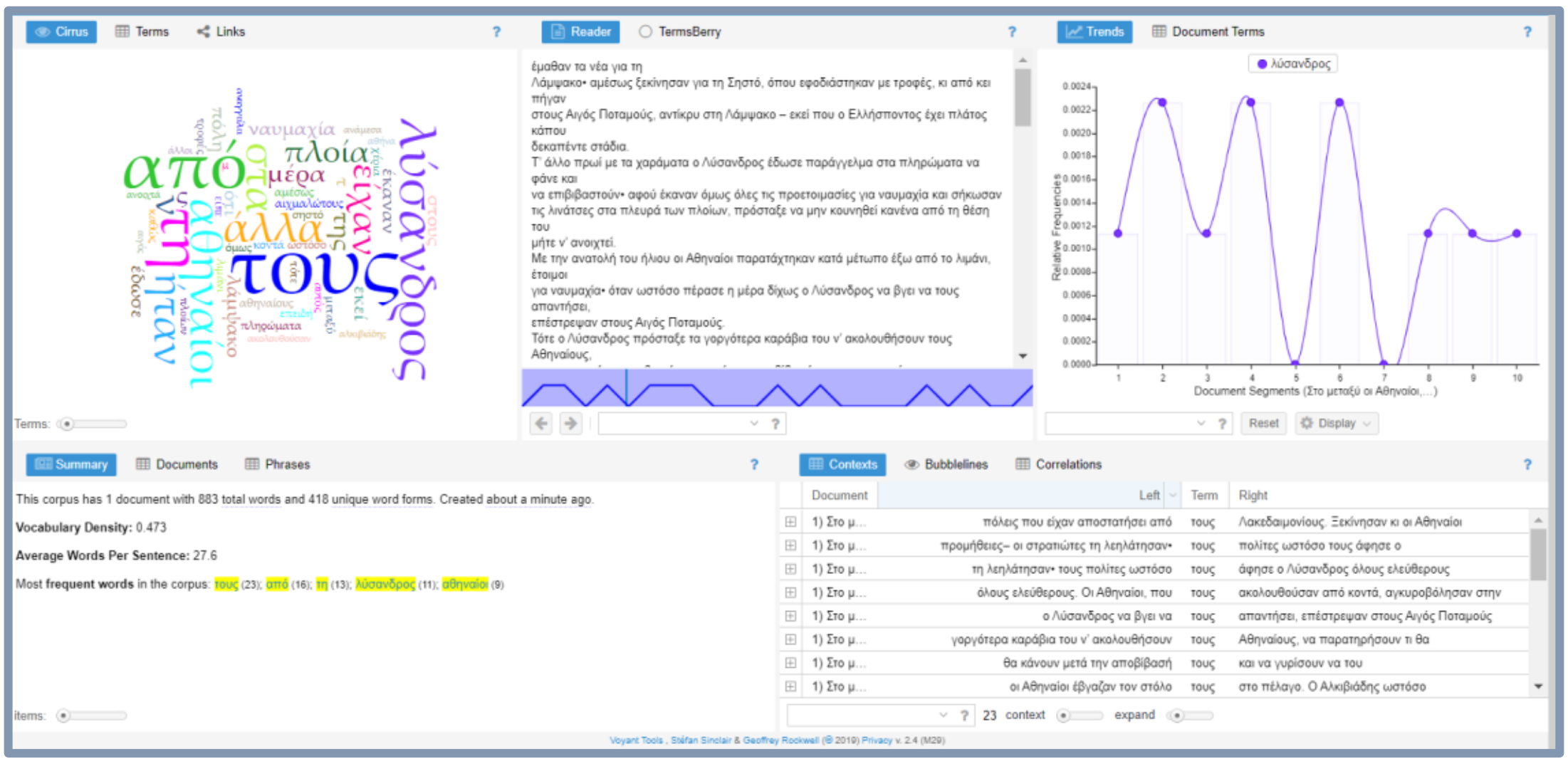

Αν ναι, γράψτε εδώ..............................................................................................................................

Για πρόσβαση στο κείμενο πατήστε [εδώ] και για πρόσβαση στο στιγμιότυπο πατήστε [εδώ].

## ΒΗΜΑ 2

Πατήστε πάνω στη λέξη **Λύσανδρος**. Τι παρατηρείτε;.............................................

## ΒΗΜΑ 3

Οπτικοποιήστε το σε νέφος λέξεων με το εργαλείο Cirrus. Σε τι χρησιμεύει το εργαλείο αυτό;..........................................................................................................

Δείτε τις ρυθμίσεις του Cirrus. Δοκιμάστε να βρείτε τη ρύθμιση που φαίνεται στο στιγμιότυπο.

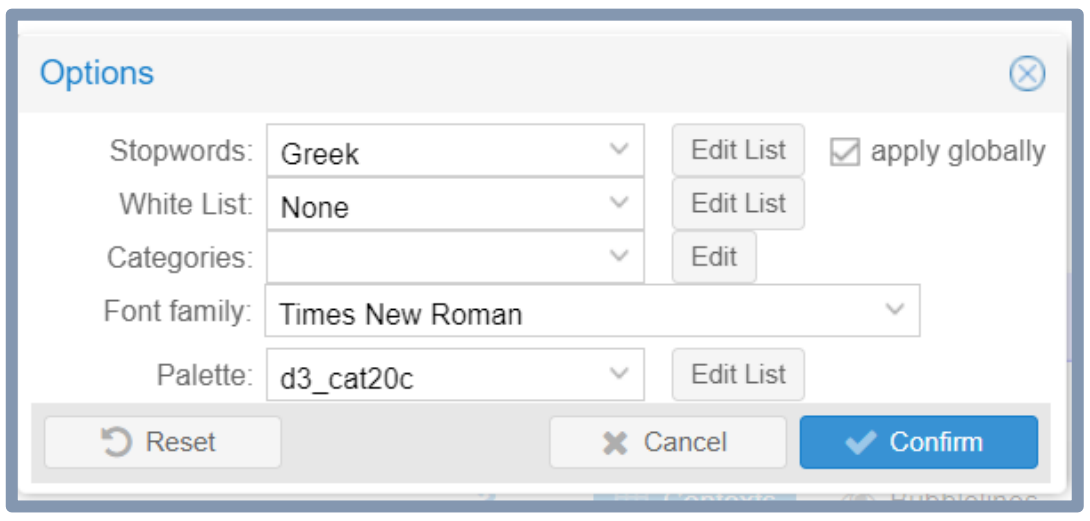

## ΒΗΜΑ 4

Άνω δεξιά στο παράθυρο του Cirrus μπορείτε να εμφανίσετε τις Επιλογές –Options. Θυμηθείτε όσα μάθατε την προηγούμενη ώρα για τις λέξεις τερματισμού. Ρυθμίστε σε Αρχαία Ελληνικά.

Τι πετυχαίνει κανείς με αυτή τη ρύθμιση;...........................................................................

## ΒΗΜΑ 5

Δείτε τα εργαλεία Trends, Collocates, Document Terms.

Σε τι χρησιμεύει το καθένα;......................................................................

**ΒΗΜΑ 6**

Το Voyant προσφέρει πάρα πολλά εργαλεία, όπως φαίνεται και στο στιγμιότυπο. Ποιο σας φαίνεται πιο χρήσιμο; Για ποιο σκοπό;

.........................................................................................................................................

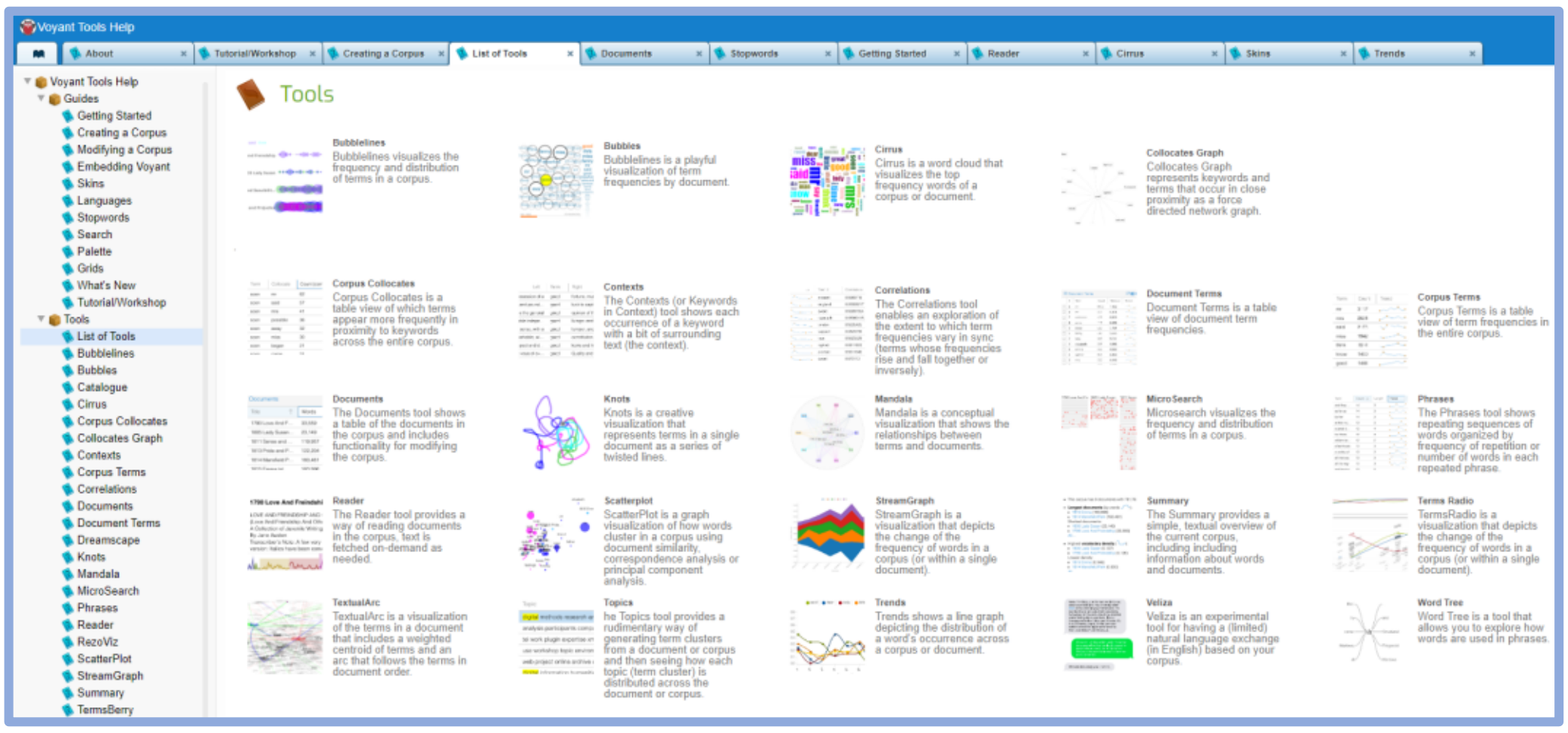

**ΒΗΜΑ 7**

Το Voyant μπορεί να χρησιμοποιηθεί για τη δημιουργία σώματος κειμένων (corpus = σώμα < Λατινικά, πληθ. corpora).

Να φτιάξεις ένα σώμα 2 κειμένων (άνω των 20 λέξεων το καθένα) και να τα ανεβάσεις στην ηλεκτρονική μας τάξη. Μπορείς να χρησιμοποιήσεις κείμενα από οποιοδήποτε ψηφιακό σχολικό βιβλίο.

## Φύλλο Εργασίας 3: Επισημειώνω διαλειτουργικά στο περιβάλλον Recogito

**ΤΙ ΘΑ ΚΑΝΟΥΜΕ:**

Σε ομάδες, θα επισημειώσουμε διαλειτουργικά αποσπάσματα από το 3ο βιβλίο του Θουκυδίδη. Διαλειτουργική επισημείωση θα πει: προσθέτω ένα στρώμα στοιχείων που είναι δομημένα δεδομένα επεξεργάσιμα από τον Η/Υ.

**ΒΗΜΑ 1**

Πήγαινε στον Ιστότοπο της εφαρμογής Recogito: https://recogito.pelagios.org/

Κάνε εγγραφή.

Θα βρεθείς σε ένα παρόμοιο περιβάλλον με αυτό:

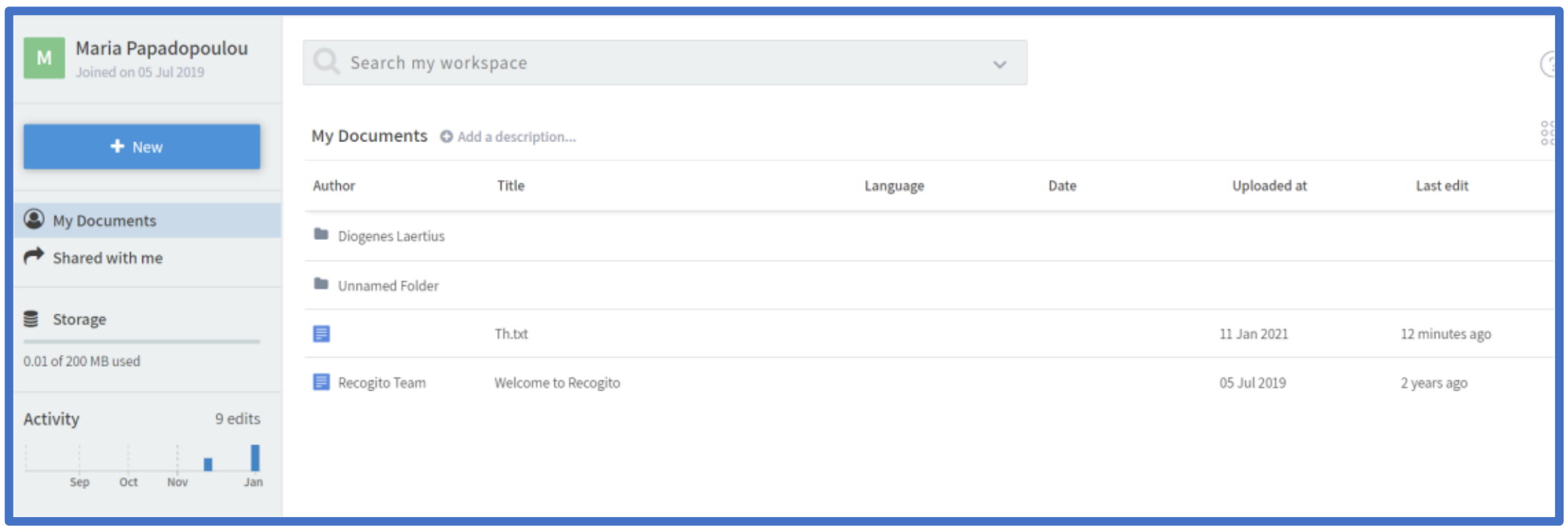

**ΒΗΜΑ 2**

Πήγαινε στον ιστότοπο του εμπλουτισμένου σχολικού βιβλίου: [http://ebooks.edu.gr/ebooks/v/html/8547/2662/Archaioi-Ellines-Istoriografoi_A-Lykeiou_html-empl/](http://ebooks.edu.gr/ebooks/v/html/8547/2662/Archaioi-Ellines-Istoriografoi_A-Lykeiou_html-empl/)

Αντίγραψε (και σώστε) σε ένα αρχείο .txt κείμενα του Θουκυδίδη από το σχολικό βιβλίο.

Μπορείς να χρησιμοποιήσεις τον αγαπημένο σου εκδότη κειμένου. Σου προτείνω τον εκδότη κειμένου sublime text. Η δωρεάν έκδοση δεν απαιτεί εγγραφή.

**ΒΗΜΑ 3**

Στο περιβάλλον του Recogito που φαίνεται στην εικόνα μπορείς να εισαγάγεις το αρχείο .txt

Τι συμβαίνει, αφού το κάνεις;

**ΒΗΜΑ 4**

Στο περιβάλλον Recogito μπορείς να επισημειώσεις το κείμενό σου ως προς τύπο οντότητας (Πρόσωπο, Τόπο, Γεγονός). Δοκίμασέ το.

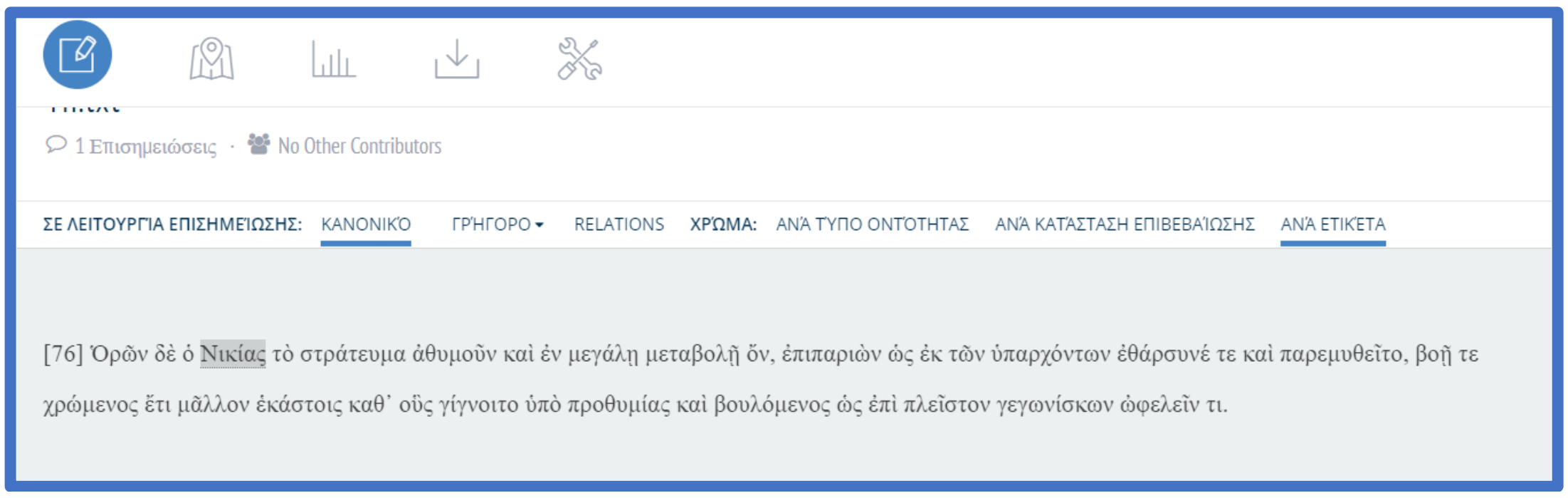

**ΒΗΜΑ 5**

Στο περιβάλλον Recogito μπορείς να δουλέψεις συνεργατικά.

Ερώτηση: Πώς θα καλέσω τα μέλη της ομάδας μου να μοιραστούν το ίδιο έγγραφο με μένα;

Απάντηση:

Θα πρέπει πρώτα να μεταβείς στο Document Settings:

Πρέπει να επιλέξεις την άδεια με την οποία θα δημοσιεύσεις το επισημειωμένο κείμενό σου στον Ιστό μέσω του drop down menu Select a license. Σου προτείνω να διαλέξεις την άδεια CC Attribution-Non-Commercial 4.0 International:

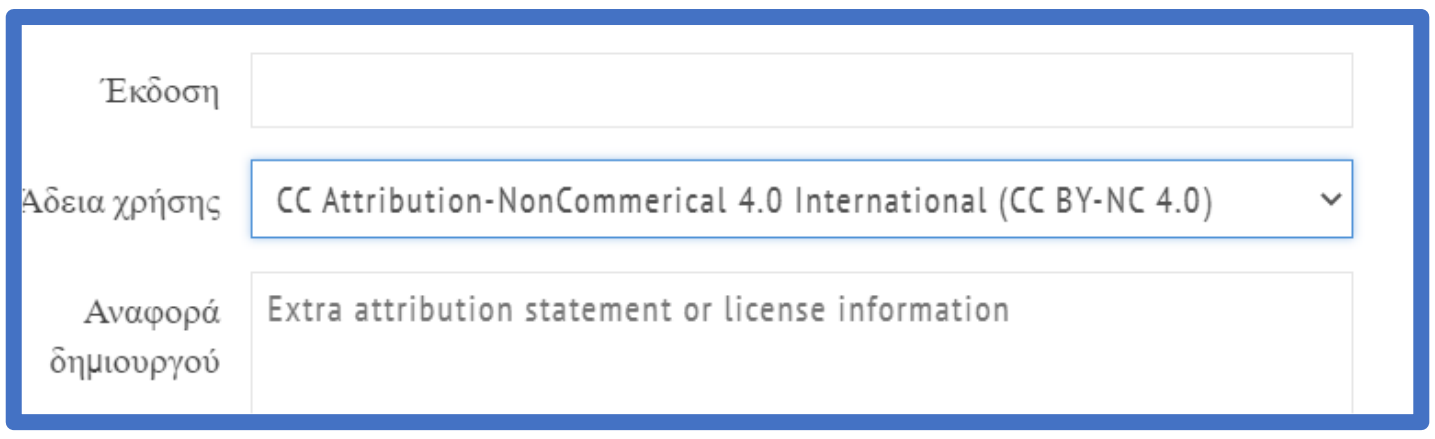

Ερώτηση: Τι άλλο μπορώ να κάνω με το recogito;

Απάντηση:

- Μπορώ να οπτικοποιήσω σε χάρτη, πατώντας το κουμπί που είναι πρώτο από αριστερά.
- Μπορώ να εξαγάγω τις επισημειώσεις σε διάφορους μορφότυπους, πατώντας το κουμπί που είναι πρώτο από δεξιά.
- Το κουμπί που βρίσκεται στη μέση δίνει τα στατιστικά στοιχεία σχετικά τις συνεισφορές κάθε μέλους της ομάδας και το πότε έγιναν οι επισημειώσεις.

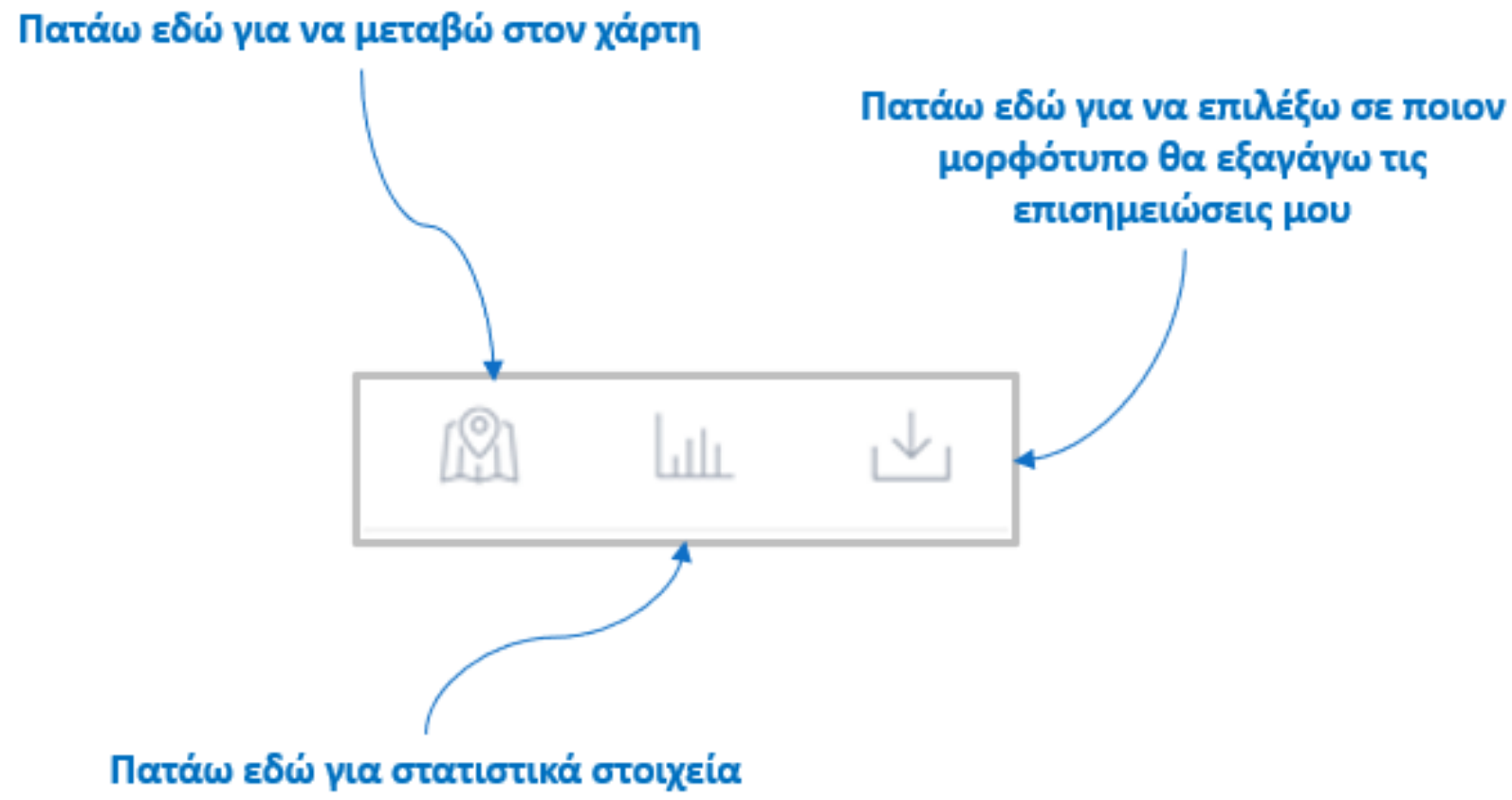

Pelagios network

Το δίκτυο Πελάγιος (Pelagios Network https://pelagios.org/ σας ευχαριστεί για την εθελοντική σας συνεισφορά στη διαλειτουργική επισημείωση αρχαίων κειμένων!